\documentstyle[12pt]{article}
\newlength{\pubnumber} 

\catcode`\@=11
\@addtoreset{equation}{section}

\def\section{\@startsection{section}{1}{\z@}{3.5ex plus 1ex minus .2ex}
 {2.3ex plus .2ex}{\large\bf}}
\def\subsection{\@startsection{subsection}{2}{\z@}{2.3ex plus .2ex}
 {2.3ex plus .2ex}{\bf}}
\newcommand\Appendix[1]{\def\thesection{Appendix \Alph{section}}
 \section{\label{#1}}\def\thesection{\Alph{section}}}

\begin{document}

\begin{titlepage}
\samepage{
\setcounter{page}{1}
\rightline{IASSNS-HEP-94/113}
\rightline{\tt hep-th/9505046}
\rightline{published in {\it Nucl.\ Phys.}\/ {\bf B457} (1995) 409}
\rightline{May 1995}
\vfill
\begin{center}
 {\Large \bf Gauge Coupling Unification in \\
     Realistic Free-Fermionic String Models\\}
\vfill
\vfill
 {\large Keith R. Dienes\footnote{
   E-mail address: dienes@sns.ias.edu}
   $\,$and$\,$ Alon E. Faraggi\footnote{
   E-mail address: faraggi@sns.ias.edu}\\}
\vspace{.12in}
 {\it  School of Natural Sciences, Institute for Advanced Study\\
  Olden Lane, Princeton, N.J.~~08540~ USA\\}
\end{center}
\vfill
\begin{abstract}
  {\rm
  We discuss the unification of gauge couplings within the framework
  of a wide class of realistic free-fermionic string models which have appeared
  in the literature, including the flipped
  $SU(5)$, $SO(6)\times SO(4)$, and various $SU(3)\times SU(2)\times U(1)$
  models.
  If the matter spectrum below the string scale is that of
  the Minimal Supersymmetric Standard Model (MSSM), then string unification is
  in disagreement
  with experiment.
  We therefore examine several effects that may modify the
  minimal string predictions.  First, we develop a systematic procedure for
  evaluating
  the one-loop heavy string threshold corrections in free-fermionic string
  models,
  and we explicitly evaluate these corrections for each of the realistic
  models.
  We find that these string threshold corrections are small,
  and we provide general arguments explaining why such threshold corrections
  are suppressed in string theory.
  Thus heavy thresholds cannot resolve the disagreement with experiment.
  We also study the effect of non-standard hypercharge normalizations,
  light SUSY thresholds,
  and intermediate-scale gauge structure, and similarly conclude that these
  effects
  cannot resolve
  the disagreement with low-energy data.  Finally, we examine
  the effects of additional
  color triplets and electroweak doublets beyond the MSSM.
   Although not required in
  ordinary grand unification scenarios, such states generically
  appear within the context of
   certain realistic free-fermionic string models.
  We show that if these states exist at the appropriate thresholds,
  then the gauge couplings will indeed unify at the string scale.
  Thus, within these string models, string unification can be in
  agreement with low-energy data.  }
\end{abstract}
\smallskip}
\end{titlepage}

\setcounter{footnote}{0}

\def\beq{\begin{equation}}
\def\eeq{\end{equation}}
\def\beqn{\begin{eqnarray}}
\def\eeqn{\end{eqnarray}}

\def\ie{{\it i.e.}}
\def\eg{{\it e.g.}}
\def\half{{\textstyle{1\over 2}}}
\def\third{{\textstyle {1\over3}}}
\def\quarter{{\textstyle {1\over4}}}
\def\m{{\tt -}}
\def\p{{\tt +}}

\def\slash#1{#1\hskip-6pt/\hskip6pt}
\def\slk{\slash{k}}
\def\GeV{\,{\rm GeV}}
\def\TeV{\,{\rm TeV}}
\def\y{\,{\rm y}}
\def\SM{Standard-Model }
\def\SUSY{supersymmetry }
\def\SSM{supersymmetric standard model}
\def\vev#1{\left\langle #1\right\rangle}
\def\l{\langle}
\def\r{\rangle}

\def\Htw{{\tilde H}}
\def\chibar{{\overline{\chi}}}
\def\qbar{{\overline{q}}}
\def\ibar{{\overline{\imath}}}
\def\jbar{{\overline{\jmath}}}
\def\Hbar{{\overline{H}}}
\def\Qbar{{\overline{Q}}}
\def\abar{{\overline{a}}}
\def\alphabar{{\overline{\alpha}}}
\def\betabar{{\overline{\beta}}}
\def\tautwo{{ \tau_2 }}
\def\thetatwo{{ \vartheta_2 }}
\def\thetathree{{ \vartheta_3 }}
\def\thetafour{{ \vartheta_4 }}
\def\ttwo{{\vartheta_2}}
\def\tthree{{\vartheta_3}}
\def\tfour{{\vartheta_4}}
\def\ti{{\vartheta_i}}
\def\tj{{\vartheta_j}}
\def\tk{{\vartheta_k}}
\def\calF{{\cal F}}
\def\smallmatrix#1#2#3#4{{ {{#1}~{#2}\choose{#3}~{#4}} }}
\def\ab{{\alpha\beta}}
\def\Minv{{ (M^{-1}_\ab)_{ij} }}
\def\bone{{\bf 1}}
\def\ii{{(i)}}
\def\V{{\bf V}}
\def\b{{\bf b}}
\def\N{{\bf N}}
\def\t#1#2{{ \Theta\left\lbrack \matrix{ {#1}\cr {#2}\cr }\right\rbrack }}
\def\C#1#2{{ C\left\lbrack \matrix{ {#1}\cr {#2}\cr }\right\rbrack }}
\def\tp#1#2{{ \Theta'\left\lbrack \matrix{ {#1}\cr {#2}\cr }\right\rbrack }}
\def\tpp#1#2{{ \Theta''\left\lbrack \matrix{ {#1}\cr {#2}\cr }\right\rbrack }}
\def\l{\langle}
\def\r{\rangle}


\def\inbar{\,\vrule height1.5ex width.4pt depth0pt}

\def\IC{\relax\hbox{$\inbar\kern-.3em{\rm C}$}}
\def\IQ{\relax\hbox{$\inbar\kern-.3em{\rm Q}$}}
\def\IR{\relax{\rm I\kern-.18em R}}
 \font\cmss=cmss10 \font\cmsss=cmss10 at 7pt
\def\IZ{\relax\ifmmode\mathchoice
 {\hbox{\cmss Z\kern-.4em Z}}{\hbox{\cmss Z\kern-.4em Z}}
 {\lower.9pt\hbox{\cmsss Z\kern-.4em Z}}
 {\lower1.2pt\hbox{\cmsss Z\kern-.4em Z}}\else{\cmss Z\kern-.4em Z}\fi}

\def\AEF{A.E. Faraggi}
\def\NPB#1#2#3{{\it Nucl.\ Phys.}\/ {\bf B#1} (19#2) #3}
\def\PLB#1#2#3{{\it Phys.\ Lett.}\/ {\bf B#1} (19#2) #3}
\def\PRD#1#2#3{{\it Phys.\ Rev.}\/ {\bf D#1} (19#2) #3}
\def\PRL#1#2#3{{\it Phys.\ Rev.\ Lett.}\/ {\bf #1} (19#2) #3}
\def\PRT#1#2#3{{\it Phys.\ Rep.}\/ {\bf#1} (19#2) #3}
\def\MODA#1#2#3{{\it Mod.\ Phys.\ Lett.}\/ {\bf A#1} (19#2) #3}
\def\IJMP#1#2#3{{\it Int.\ J.\ Mod.\ Phys.}\/ {\bf A#1} (19#2) #3}
\def\nuvc#1#2#3{{\it Nuovo Cimento}\/ {\bf #1A} (#2) #3}
\def\etal{{\it et al\/}}

\hyphenation{su-per-sym-met-ric non-su-per-sym-met-ric}
\hyphenation{space-time-super-sym-met-ric}
\hyphenation{mod-u-lar mod-u-lar--in-var-i-ant}


\setcounter{footnote}{0}
\section{Introduction}

LEP precision data provides remarkable confirmation of the Standard Model
of particle physics. However, many fundamental problems are not addressed
in the context of the Standard Model, leading to the expectation that
a more fundamental theory must exist
in which the Standard Model appears as an effective low-energy limit. While
many possible extensions of the Standard Model are highly constrained
or ruled out by experiment, supersymmetric theories are in agreement with all
available data. Furthermore, the top-quark mass range
required in supersymmetric
scenarios of electroweak symmetry breaking \cite{SUSYreviews}
is in agreement with that suggested
by CDF/D0 direct observation \cite{CDF} and LEP precision data \cite{LEP}.
In recent years it has also been suggested
that the success of gauge coupling unification in the Minimal
Supersymmetric Standard Model (MSSM) \cite{gcumssm}
provides evidence for the validity of supersymmetric grand unified
theories (SUSY GUT's).

While SUSY GUT's provide a useful parametrization of the sparticle spectrum
at low energies and of the boundary conditions at the GUT scale, they are
incomplete
theories. First, they do not explain the origin of the Standard Model spectrum.
Second, in order to evade proton-lifetime constraints, some {\it ad hoc}\/
global
symmetries must be imposed, and a doublet-triplet splitting mechanism is
required. Finally, despite the fact that the unification scale is
just one or two orders of magnitude below the Planck scale, a consistent
treatment of quantum gravity is lacking.

Remarkably, superstring theories \cite{Sreviews},
the only available candidates for a
consistent theory of quantum gravity, accommodate $N=1$ supersymmetric theories
as their low-energy effective theories. Even more remarkably, heterotic string
theories \cite{heterotic}
provide a general framework in which the origin of the observed
particle spectrum and interactions may be understood \cite{CHSW}
and the doublet-triplet splitting problem may be resolved \cite{DTSM}.

Among the string models constructed to date, the most phenomenologically
realistic have been
formulated within the so-called ``free-fermionic''
construction \cite{KLT,ABK,KLST}.
Indeed, within this construction,
many three-generation models can be obtained.
While this may be an accident, it is more likely to be a reflection of some
fundamental properties of string compactification. The free-fermionic
construction is formulated at a highly symmetric point in the compactification
space at which spacetime symmetries are maximally enhanced.
It turns out that the realistic
free-fermionic models admit a $Z_2\times Z_2$ orbifold structure with
standard embedding which is
realized in this construction through
the so-called ``NAHE set'' \cite{FOC} of fermionic
boundary-condition basis vectors.
Such $Z_2\times Z_2$
orbifolds possess a structure which can naturally accommodate three generations
due to the existence of exactly three twisted sectors. These
result from the action of the $Z_2\times Z_2$ twist on a six-dimensional
compactified space.  In general, the $Z_2\times Z_2$
orbifold does not produce a number of fixed points that can be reduced
to three generations. However, precisely at the free-fermionic
point in the toroidal compactification space, the number of fixed points
is such that a reduction to three generations can be achieved.
Indeed, the number of fixed points from each twisted sector
is reduced to one. Thus,
each twisted sector produces one of the chiral generations of the
Standard Model.

The realistic free-fermionic models achieve remarkable success in trying
to explain the different features of the Standard Model spectrum. In
addition to naturally producing three generations with Standard Model gauge
group, they also provide a plausible explanation for the heavy top-quark mass
and
the fermion mass hierarchy. Indeed, the realistic standard-like
models
suggest that only the top quark mass term is obtained at the cubic level of the
superpotential, while the lighter fermion mass terms are obtained from
non-renormalizable terms which are naturally suppressed relative to
the leading cubic-level mass term. In fact, in this way a successful prediction
of the
mass of the top quark was obtained in Ref.~\cite{TOP}, three years
prior to its experimental observation.
Furthermore, an analysis of non-renormalizable terms up to eighth order then
reveals
how the fermion masses and mixing angles may also be generated \cite{CKM}.

The realistic free-fermionic models are therefore very appealing from a
theoretical point of view, and may successfully explain the different features
of the
Standard Model. However, string theory in general, and the free-fermionic
models in  particular, predict that the string unification scale
is related to the Planck scale, and should be numerically of the order of
${\cal O}(5\cdot10^{17})$ GeV. Thus, a factor of approximately twenty
separates the string unification scale from the usual MSSM unification
scale extrapolated from low-energy data.
This discrepancy is one of major problems confronting string model-building.

One possible solution is to construct string GUT models in which
the GUT symmetry is broken in the effective low-energy field
theory \cite{Lewellen,Fermilab}.
However, no realistic models of this sort have yet been constructed.
Moreover, in this scenario, the problems with proton-lifetime
constraints reemerge. Indeed, this problem is more severe in string GUT's
due to the possible appearance of baryon- and lepton-number-violating
dimension-four operators \cite{DTSM} and the anticipated difficulty in
implementing the
GUT doublet-triple splitting mechanism in string models.

Another solution is provided if additional
thresholds exist in the desert between the electroweak scale and
the string unification scale. In fact, the availability of such additional
thresholds in realistic free-fermionic models has been demonstrated in
Ref.~\cite{GCU}.
In this respect, imposing the restriction that the spectrum below the
string scale is just that of the MSSM is {\it ad hoc}\/, and may be too
restrictive.
The successful unification of gauge couplings
within the MSSM would then indicate
that the
MSSM is only an approximation to the complete theory between the weak scale
and the Planck scale.

A third suggestion, due to Ib\'a\~nez \cite{ibanez}, is that in string models
the normalization of the weak hypercharge is, in general, different
from that in grand-unified theories, and may just have the right
value to allow string unification, even if the spectrum below the string
scale is that of the MSSM.
It is found that if $k_1$ is in the range $1.2 \leq k_1 \leq 1.4$,
then unification at the string scale can be consistent with low-energy data.
However, whether such values of $k_1$ can be achieved within consistent
string models remains an open question. Moreover, we will argue that in string
models
we must have $k_1\ge5/3$.

A final possibility is that the
contributions arising from the infinite towers of heavy string states
can modify the above string tree-level predictions.  Within this scenario,
one would hope that  these heavy string states can give rise to
additional heavy string threshold corrections which shift the unification
point down to ${\cal O}(10^{16})$ GeV.

In this paper, we systematically re-examine the problem of gauge
coupling unification within the context of a wide variety of realistic
free-fermionic models which have appeared in the literature.
There are various reasons why this is an important undertaking.
For example, while
many of the possible effects that we consider have been
discussed previously, each was treated in an abstract setting and in isolation.
However, within the tight constraints of a given realistic
string model, the mechanisms giving rise to three
generations and the MSSM gauge group
may ultimately prove inconsistent with, for example,
large threshold corrections or extra non-MSSM matter.
Moreover, the increased complexity of the known realistic string models
may substantially alter previous expectations
based on simplified or idealized scenarios.
It is therefore important to rigorously calculate all
of these effects simultaneously,
within the context of a wide variety of actual
realistic string models,
in order to determine which path to successful gauge
coupling unification (if any) such models actually take.

Our analysis proceeds in several steps.
First, we study the role of threshold corrections due to
the infinite tower of heavy string states.
In this we follow the definition of threshold corrections as given by
Kaplunovsky \cite{Kaplunovsky}, and explicitly calculate these corrections
for a wide range of realistic
free-fermionic models including the flipped
$SU(5)$ ``revamped'' model \cite{revamp},
several $SU(3)\times SU(2)\times U(1)$ models \cite{EU,TOP,SLM,custodial},
an $SO(6)\times SO(4)$ model \cite{ALR}, and even models with
non-supersymmetric spacetime spectra.
Our evaluation of these threshold corrections is done
in two stages. First, we analytically evaluate the traces
over the entire Fock space corresponding to a given
string model, with various relevant combinations of gauge charge
operators inserted into the trace.
Then, for each combination of gauge charge insertions, these results are
expanded level-by-level,
and the contributions from the states at each string energy level
are integrated over the modular-group fundamental domain.
In this manner, we can obtain results to any desired accuracy.
Moreover, as we shall see, various non-trivial consistency checks can be
performed.

Our results show that threshold corrections due to the massive
string states are small in free-fermionic string models. This result is
 {\it a priori}\/ surprising, given the infinite numbers of heavy
string states which potentially contribute
to the string threshold corrections,
but we are able to provide a general argument
which explains why the threshold corrections from the massive string
states are naturally suppressed in string theory.
Our argument also explains why
threshold corrections can grow large only
for large values of the string moduli.

Given these results, we then proceed to
systematically examine the other effects
which might potentially alleviate the discrepancy between the GUT
and string scales.  As discussed above, these include the
effects of stringy non-standard $U(1)$
hypercharge normalizations, light SUSY thresholds, intermediate gauge
structure, and additional matter beyond that predicted by the MSSM.
We find that the effects of hypercharge normalizations, light SUSY
thresholds, and intermediate gauge structure are not sufficient
to remove the discrepancy.  By contrast, we surprisingly find that only
the presence of additional matter in these models
(in particular, certain color triplets and electroweak doublets with
special hypercharge assignments) has a profound effect on
the running of the gauge couplings, and precisely this matter
appears naturally in a variety of the realistic string models.
Indeed, within these models,
we show the gauge couplings can indeed unify at the string scale when all
of the above effects are taken into account.
Thus, for such models, the disagreement between the GUT scale
and the string scale can be  naturally resolved.

It is remarkable that string theory, which predicts an unexpectedly high
unification scale $M_{\rm string}$, in many cases
compensates by simultaneously also predicting
precisely the extra exotic particles needed to reconcile
this higher scale with low-energy data.
Moreover, our analysis shows that while some string models
naturally contain the extra matter needed to resolve this
disagreement, other models do not and can actually be ruled out on this basis.
Thus, the appearance of such extra matter becomes a
low-energy prediction of these models which may be accessible to
present-day experiments.

We stress that this is the first time that such an exhaustive examination
of all possible effects has been performed within the context
of actual realistic string models,
and within the constraints that these models impose.
Hence, one of the main results of our analysis
is the new observation that only the appearance of extra exotic matter
in particular representations can possibly resolve the experimental
discrepancies.
Indeed, because we have been able to rule out all other possible
within these models, the appearance of such extra exotic matter
becomes a {\it prediction}\/ of successful string-scale unification.
It is of course an old idea that the presence of extra matter
can resolve the discrepancy between the GUT and string unification
scales.  What is highly non-trivial, however, is that this now appears
to be the {\it only}\/ way in which realistic string models can solve the
problem.

This paper is organized as follows.  In Sect.~2 we review the
general features of the realistic free-fermionic models,
and in Sect.~3 we discuss how heavy string threshold corrections
may be evaluated within the context of such models.
In Sect.~4 we then explicitly calculate the threshold corrections within
the flipped $SU(5)$, $SO(6)\times SO(4)$, and $SU(3)\times SU(2)\times U(1)$
string models, and in Sect.~5 we test whether these models are
in agreement with low-energy data  by
systematically analyzing these results along with the effects
due to
light SUSY thresholds,
intermediate gauge and matter thresholds,
as well as two-loop and
Yukawa-coupling effects.
It is here that we find, contrary to naive
expectations
based on comparing $M_{\rm GUT}$ and $M_{\rm string}$,
that certain string models may be in full agreement with low-energy
data and grand unification.
Sect.~6 then contains a general argument based
on the modular properties of the threshold corrections
which explains why they must be small in the
free-fermionic models, and  Sect.~7 contains our conclusions.
Finally, an Appendix contains an explicit listing of all of the
string models that we will be considering in this paper.
A short summary of some of the main results of this paper
can also be found in Ref.~\cite{summ}.

\setcounter{footnote}{0}
\section{Realistic free-fermionic models}

In the free-fermionic formulation of the heterotic string \cite{heterotic},
all of the worldsheet degrees of freedom needed to cancel the conformal anomaly
are
represented in terms of internal free fermions propagating on the string
worldsheet.
In four dimensions, this requires 44 real left-moving (Majorana-Weyl)
fermions
and  20 real right-moving Majorana-Weyl fermions (or equivalently, half
as many complex fermions,
or any such consistent combinations of real and complex fermions).
Under parallel transport around a non-contractible loop on the toroidal
string worldsheet,
these fermionic fields can generally accrue a phase, and each set of
specified phases for all worldsheet fermions
around all such non-contractible loops is called the
``spin structure'' of the model.
Such spin structures are usually given in the form of boundary-condition
``vectors'',
with each element of the vector specifying the phase of a corresponding
worldsheet fermion.
The possible spin structures which can be used in the construction
of string models are constrained by various string consistency requirements
({\it e.g.}, the existence of a proper worldsheet supercurrent, proper
spacetime
spin-statistics assignments, physically sensible projections, and modular
invariance).
A model is therefore constructed by choosing
a set of boundary condition vectors which satisfy these constraints.
In general,
these basis vectors $\b_k$ must span a finite
additive group $\Xi=\sum_k{{n_k}{\b_k}}$
where $n_k=0,\cdots,{{N_{z_k}}-1}$.
The physical massless states in the Hilbert space of a given sector
$\alpha\in{\Xi}$ are then obtained by acting on the vacuum state of
that sector  with the worldsheet bosonic and fermionic mode operators, and
by subsequently applying the generalized GSO projections. The $U(1)$
charges ${\bf Q}(f)$ with respect to the unbroken Cartan generators of the
four-dimensional gauge group are in one-to-one correspondence with the $U(1)$
currents ${f^*}f$ for each complex worldsheet fermion $f$, and are given by:
\beq
       {\bf Q}(f) ~=~ {1\over 2}\,\alpha(f) ~+~ F(f)~.
\label{uonecharges}
\eeq
Here $\alpha(f)$ is the boundary condition of the worldsheet fermion $f$
in the sector $\alpha$;  $\alpha(f)=0$ for Neveu-Schwarz boundary conditions,
and $\alpha(f)=1$ for Ramond.
Likewise, $F_\alpha(f)$ is a fermion-number operator counting $+1$
for each mode of $f$ (and $-1$ for each mode of $f^\ast$, if $f$ is complex).
For periodic complex fermions  [{\it i.e.}, for $\alpha(f)=1$],
the vacuum is a spinor
with two degenerate vacuum states ${\vert +\rangle}$ and ${\vert -\rangle}$.
These states are respectively annihilated by the zero modes $f_0$ and
${{f_0}^\ast}$ which obey the Clifford algebra, and have
fermion numbers  $F(f)=0,-1$ respectively.

In the realistic free-fermion models that we will be considering,
the worldsheet fermions are as follows:
\begin{itemize}

\item   a complex right-moving fermion, denoted $\psi^\mu$, formed from the two
        real fermionic superpartners of the coordinate
         boson $X^\mu$;

\item   six real right-moving fermions denoted $\chi^{1,...,6}$,
         often paired to form three complex right-moving fermions
         denoted $\chi^{12}$, $\chi^{34}$, and $\chi^{56}$;

\item   12 real right-moving fermions, denoted $y^{1,...,6}$
          and $\omega^{1,...,6}$;

\item   12 corresponding real left-moving fermions, denoted
$\overline{y}^{1,...,6}$
          and $\overline{\omega}^{1,...,6}$; and

\item   16 remaining complex left-moving fermions, denoted
          $\overline{\psi}^{1,...,5}$,  $\overline{\eta}^{1,...,3}$,
          and $\overline{\phi}^{1,...,8}$.

\end{itemize}
The realistic models in the free-fermionic formulation are then generated
by specifying a special basis of boundary-condition vectors
\cite{revamp,FNY,ALR,EU,TOP,SLM,LNY}
for these worldsheet fermions.
This basis is constructed in two stages. The first stage consists
of introducing the so-called NAHE set \cite{SLM}, which is a set
of five boundary condition basis
vectors denoted $\lbrace{\bf 1},S,\b_1,\b_2,\b_3\rbrace$.
With `0' indicating Neveu-Schwarz boundary conditions
and `1' indicating Ramond boundary conditions, these vectors are as follows:
\beqn
 &&\begin{tabular}{c|c|ccc|c|ccc|c}
 ~ & $\psi^\mu$ & $\chi^{12}$ & $\chi^{34}$ & $\chi^{56}$ &
        $\overline{\psi}^{1,...,5} $ &
        $\overline{\eta}^1 $&
        $\overline{\eta}^2 $&
        $\overline{\eta}^3 $&
        $\overline{\phi}^{1,...,8} $ \\
\hline
\hline
      {\bf 1} &  1 & 1&1&1 & 1,...,1 & 1 & 1 & 1 & 1,...,1 \\
          $S$ &  1 & 1&1&1 & 0,...,0 & 0 & 0 & 0 & 0,...,0 \\
\hline
  ${\bf b}_1$ &  1 & 1&0&0 & 1,...,1 & 1 & 0 & 0 & 0,...,0 \\
  ${\bf b}_2$ &  1 & 0&1&0 & 1,...,1 & 0 & 1 & 0 & 0,...,0 \\
  ${\bf b}_3$ &  1 & 0&0&1 & 1,...,1 & 0 & 0 & 1 & 0,...,0 \\
\end{tabular}
   \nonumber\\
   ~  &&  ~ \nonumber\\
   ~  &&  ~ \nonumber\\
     &&\begin{tabular}{c|cc|cc|cc}
 ~&      $y^{3,...,6}$  &
        $\overline{y}^{3,...,6}$  &
        $y^{1,2},\omega^{5,6}$  &
        $\overline{y}^{1,2},\overline{\omega}^{5,6}$  &
        $\omega^{1,...,4}$  &
        $\overline{\omega}^{1,...,4}$   \\
\hline
\hline
    {\bf 1} & 1,...,1 & 1,...,1 & 1,...,1 & 1,...,1 & 1,...,1 & 1,...,1 \\
    $S$     & 0,...,0 & 0,...,0 & 0,...,0 & 0,...,0 & 0,...,0 & 0,...,0 \\
\hline
${\bf b}_1$ & 1,...,1 & 1,...,1 & 0,...,0 & 0,...,0 & 0,...,0 & 0,...,0 \\
${\bf b}_2$ & 0,...,0 & 0,...,0 & 1,...,1 & 1,...,1 & 0,...,0 & 0,...,0 \\
${\bf b}_3$ & 0,...,0 & 0,...,0 & 0,...,0 & 0,...,0 & 1,...,1 & 1,...,1 \\
\end{tabular}
\label{nahe}
\eeqn
As can be seen, the vector {\bf 1}
has periodic boundary conditions for all the worldsheet fermions.
The vector $S$ generates the spacetime supersymmetry, and the
sectors $\b_1$, $\b_2$, and $\b_3$ correspond to the three twisted sectors
of a $Z_2\times Z_2$ orbifold.
Corresponding to this set of boundary-condition vectors is the following
choice of phases which define how the generalized GSO projections are to
be performed in each sector of the theory:
\beq
      C\left\lbrack \matrix{\b_i\cr \b_j\cr}\right\rbrack~=~
      C\left\lbrack \matrix{\b_i\cr S\cr}\right\rbrack ~=~
      C\left\lbrack \matrix{\bone \cr \bone \cr}\right\rbrack ~= ~ -1~.
\label{nahephases}
\eeq
The remaining projection phases can be determined from those above through
the self-consistency constraints.
The precise rules governing the choices of such vectors and phases, as well
as the procedures for generating the corresponding spacetime particle
spectrum, are given in Refs.~\cite{KLT,ABK}.
The mapping between the notation used here \cite{ABK} and
the notation used in Ref.~\cite{KLT} is given in the Appendix.

After imposing the NAHE set,
the resulting model has gauge group $SO(10)\times SO(6)^3\times E_8$
and $N=1$ spacetime supersymmetry.
The vector $S$ is the supersymmetry generator and the superpartners of
the states from any given sector $\alpha$ are obtained from the sector
$S+\alpha$. The spacetime vector bosons that generate the gauge group
arise from the Neveu-Schwarz sector and from the sector
$I \equiv {\bf 1}+\b_1+\b_2+\b_3$.
The Neveu-Schwarz sector produces the generators of
$SO(10)\times SO(6)^3\times SO(16)$. The sector ${\bf 1}+\b_1+\b_2+\b_3$
produces the spinorial {\bf 128} of $SO(16)$ and completes the hidden
gauge group to $E_8$.
The vectors $\b_1$, $\b_2$, and $\b_3$ correspond to the three twisted
sectors in the corresponding orbifold formulation and produce
48 spinorial {\bf 16}'s of $SO(10)$, sixteen from each of the sectors $\b_1$,
$\b_2$, and $\b_3$.

As can be seen from (\ref{nahe}),
the NAHE set divides the 44 left-moving and 20 right-moving real internal
fermions in the following way: the complex fermions ${\bar\psi}^{1,\cdots,5}$
produce the observable $SO(10)$ symmetry;  the complex fermions
${\bar\phi}^{1,\cdots,8}$
produce the hidden $E_8$ gauge group;
and the fermions $\{{\bar\eta}^1,{\bar y}^{3,\cdots,6}\}$,
$\{{\bar\eta}^2,{\bar y}^{1,2},
 {\bar\omega}^{5,6}\}$, and $\{{\bar\eta}^3,{\bar\omega}^{1,\cdots,4}\}$
give rise to the three horizontal $SO(6)$ symmetries. The left-moving
fermions $\{y,\omega\}$
are divided as well, in groups
$\{{y}^{3,\cdots,6}\}$, $\{{y}^{1,2},{\omega}^{5,6}\}$,
and $\{{\omega}^{1,\cdots,4}\}$. The left-moving
fermions $\chi^{12},\chi^{34},\chi^{56}$
carry the supersymmetry charges.
Each sector $\b_1$, $\b_2$, and $\b_3$ imposes periodic boundary conditions
for the fermions $(\psi^\mu\vert{\bar\psi}^{1,\cdots,5})$ and for
$(\chi_{12},\{y^{3,\cdots,6}\vert{\bar y}^{3,\cdots6}\},{\bar\eta}^1)$,
$(\chi_{34},\{y^{1,2},\omega^{5,6}\vert{\bar y}^{1,2}{\bar\omega}^{5,6}\},
 {\bar\eta}^2)$, or $(\chi_{56},\{\omega^{1,\cdots,4}\vert{\bar\omega}^{1,
\cdots4}\},{\bar\eta}^3)$ respectively.
This division of the internal fermions is a
reflection of the equivalent underlying $Z_2\times Z_2$ orbifold
compactification \cite{FOC}. The set of internal fermions
    ${\{y,\omega\vert{\bar y},{\bar\omega}\}^{1,\cdots,6}}$
corresponds to the left/right symmetric conformal field theory of the
heterotic string, or equivalently to the six-dimensional compactified
manifold in a bosonic formulation.
This set of left/right symmetric internal fermions
plays a fundamental role  in the determination of the low-energy properties
of the realistic free-fermionic models.

The second stage in the construction of the realistic models consists
of adding three additional basis vectors to the above NAHE set.
These three additional basis vectors, which are often called
$\lbrace \alpha,\beta, \gamma\rbrace$,
correspond to ``Wilson lines'' in the orbifold construction.
The allowed fermion boundary conditions in these additional basis vectors are
of course also constrained by the string consistency
constraints, and must preserve modular invariance and
worldsheet supersymmetry.
The choice of these additional basis vectors
$\lbrace \alpha,\beta,\gamma\rbrace$ nevertheless distinguishes
between different models and determine their low-energy properties.
For example, three additional vectors are
needed to reduce the number of massless
generations to three, one from each sector $\b_1$, $\b_2$, and $\b_3$,
and the choice of their boundary conditions for the internal fermions
${\{y,\omega\vert{\bar y},{\bar\omega}\}^{1,\cdots,6}}$
also determines the Higgs doublet-triplet splitting and
the Yukawa couplings. These
low-energy phenomenological requirements therefore impose strong
constraints \cite{SLM} on the possible assignment of boundary conditions to the
set of internal world-sheet fermions
${\{y,\omega\vert{\bar y},{\bar\omega}\}^{1,\cdots,6}}$.

One then finds a variety of possibilities, with
the $SO(10)$ gauge symmetry
broken to one of its subgroups, either $SU(5)\times U(1)$,
$SO(6)\times SO(4)$ or
$SU(3)\times SU(2)\times U(1)_{B-L}\times U(1)_{T_{3_R}}$.
These breakings are achieved by the assignment of the following
boundary conditions to the fermion set
${\bar\psi}^{1\cdots5}_{1/2}$.
To achieve a breaking to $SU(5)\times U(1)$,
we need to assign boundary conditions $1/2$
to all of these fermions simultaneously.
By contrast, to achieve a breaking
to $SO(6)\times SO(4)$, we assign boundary condition $1$
to only the first three fermions,
while assigning boundary condition $0$
to the remaining two.
Finally, to break the $SO(10)$ symmetry to
$SU(3)\times SU(2)\times
U(1)_C\times U(1)_L$,
we impose {\it both}\/ of these breakings
via two separate basis vectors.\footnote{Recall
      that $U(1)_C={3\over2}U(1)_{B-L}$ and
      that $U(1)_L=2U(1)_{T_{3_R}}.$}
The complete listings of boundary conditions and GSO phases
for each of the models we will be considering in this paper can be found
in the Appendix.

Note that all of these realistic free-fermionic models
have three $U(1)$ symmetries,
denoted by $U(1)_{r_j}$
$(j=1,2,3)$; these are respectively generated by the
left-moving world-sheet
currents ${\bar\eta}^1_{1/2}{\bar\eta}^{1^*}_{1/2}$,
${\bar\eta}^2_{1/2}{\bar\eta}^{2^*}_{1/2}$, and
${\bar\eta}^3_{1/2}{\bar\eta}^{3^*}_{1/2}$. These three $U(1)$
symmetries arise due to the $Z_2\times Z_2$ twist with standard embedding.
Additional horizontal $U(1)$ symmetries, denoted by $U(1)_{r_j}$
$(j=4,5,...)$,
arise by pairing two real fermions from the sets
 $\{{\bar y}^{3,\cdots,6}\}$, $\{{\bar y}^{1,2},{\bar\omega}^{5,6}\}$, and
$\{{\bar\omega}^{1,\cdots,4}\}$. The final observable gauge group depends on
the number of such pairings. Indeed, the rank of the entire gauge group can
vary
between $r=16$ to $r=22$. To every one of the horizontal gauged
right-moving $U(1)$ symmetries
corresponds a horizontal left-moving global $U(1)$ symmetry.
Finally, the realistic free-fermionic models also contain Ising-model
operators that are obtained by pairing a right-moving real fermion with a
left-moving real fermion \cite{KLN}.

The hidden sector in the free-fermionic standard-like models is determined
by the boundary condition of the internal left-moving fermions
${\bar\phi}^{1,\cdots,8}$.
In the NAHE set, the contribution to the hidden $E_8$
gauge group  comes from the Neveu-Schwarz sector
and from the sector $I\equiv {\bf 1}+\b_1+\b_2+\b_3$,
which produces the adjoint and spinorial
representations of $SO(16)$ respectively.
The final hidden gauge group
is determined by the assignment of boundary conditions for
the worldsheet fermions
${\bar\phi}^{1,\cdots,8}$ from the vectors
$\{{\alpha,\beta,\gamma}\}$,
and by the choices of generalized GSO projection phases.

The realistic free-fermionic models contain three chiral generations
from the sectors $\b_1$, $\b_2$, and $\b_3$. At the level of the NAHE
set alone,
there are 48 generations, 16 from each of these sectors. However,
the vector $\gamma$
reduces the number
of generations by a factor of two
by fixing the charge under $U(1)_{R_j}$ for each of the sectors
$\b_1$, $\b_2$, and $\b_3$ to be either $+1/2$ or $-1/2$.
A further reduction to three generations is then obtained by carefully
choosing the $\{\alpha,\beta,\gamma\}$-boundary conditions for the
real fermions $\{y,\omega\vert{\bar y},{\bar\omega}\}^{1,\cdots,6}$.
Each one of the vectors $\{\alpha,\beta,\gamma\}$ reduces
by a factor of two the number
of degenerate vacua that arise from the real fermions
$\{y,\omega\vert{\bar y},{\bar\omega}\}^{1,\cdots,6}$
in each of the sectors $\b_1$, $\b_2$, and $\b_3$.
After the GSO projections from the vectors $\{\alpha,\beta,\gamma\}$,
each sector $\b_1$, $\b_2$, and $\b_3$ produces one chiral generation.
Thus, the assignment of boundary conditions for the set
$\{y,\omega\vert{\bar y},{\bar\omega}\}^{1,\cdots,6}$ from the vectors
$\{\alpha,\beta,\gamma\}$ is constrained by requiring three
light generations. Further constraints on the assignment of boundary conditions
for the set $\{y,\omega\vert{\bar y},{\bar\omega}\}^{1,\cdots,6}$
from the vectors $\{\alpha,\beta,\gamma\}$ are imposed by requiring
other phenomenological criteria, such as the presence of Higgs doublets in the
massless spectrum, the projection of colored Higgs triplets,
the existence of a phenomenologically realistic supersymmetric vacuum, and
the existence of non-vanishing Yukawa couplings that may produce a realistic
fermion mass spectrum. Satisfying all of these phenomenological
criteria simultaneously is a highly non-trivial task, and it is indeed
 a remarkable feat that models which  successfully incorporate all
of these features have been constructed.

\setcounter{footnote}{0}
\section{Calculating Threshold Corrections in Free-Fermionic
         Models}

In this section we first provide a general model-independent summary of how
threshold corrections are defined in string theory \cite{Kaplunovsky}.  We
then
outline our specific procedure for calculating threshold
corrections within the context of free-fermionic string models.

\bigskip
\noindent{\it  Preliminary Comments Concerning String Unification and
               Mass Scales}
\bigskip

We begin, however, with some preliminary comments concerning
the unification of gauge couplings and mass scales within string theory.

In order to relate string dynamics to low-energy experimental observables,
we are interested in the effective field theories of the light
(massless) string excitations.
In particular, we are interested in the
effective
gauge couplings of the four-dimensional gauge group, and their beta functions.
Unlike the case in field theory, however, in string theory all couplings
and mass scales are dynamical variables, and are ultimately set by
the values of certain moduli fields.  Consequently, determination
of the gauge couplings, and especially their ``unification scale'',
becomes a highly non-trivial issue.

Perhaps the most important factor in this regard is the presence of
moduli fields in the string spectrum.
Such moduli fields, which are massless gauge-neutral Lorentz scalar fields
whose effective potential is classically and perturbatively flat,
arise as follows.
Recall that the consistency
of a heterotic string theory requires 22 internal bosonic
degrees of freedom in the non-supersymmetric sector (or more generally,
an internal conformal field theory of total central charge 22).
Sixteen of these bosonic degrees of freedom
are compactified on a flat torus with fixed radius,
and generically give rise to the internal symmetries of the theory
which are interpreted as the gauge symmetries of the low-energy
field theory.
The remaining six bosonic degrees of freedom in the non-supersymmetric sector,
combined with six compactified bosonic
degrees of freedom in the supersymmetric sector,
are compactified on an internal manifold that can be
a Calabi-Yau manifold or an orbifold.
The moduli fields are then precisely those massless fields which parametrize
the size and shape of this six-dimensional internal manifold.

In general, the (left-moving) gauge symmetry that one
obtains for generic points
in the moduli space of four-dimensional strings is $U(1)^{22}$.
At special points in the moduli space, however, the gauge symmetries may be
enhanced, with additional vector states becoming massless.
Indeed, such points with enhanced symmetries can often be
realized as the results of compactifications
of higher-dimensional string theories, with the enhancement of
the gauge symmetry arising for special choices of the compactification
moduli (or equivalently, special choices of background fields) \cite{NSW}.
In this paper, we are interested in such points
at which the effective gauge symmetry is enlarged to contain at
least the MSSM gauge group $SU(3)_C\times SU(2)_L\times U(1)_Y$.
Similarly, as one moves around in moduli space, there also exist special points
at which the internal  $(c_{\rm right},c_{\rm left})=(9,22)$
worldsheet conformal field theory effectively decomposes into a product of
smaller worldsheet conformal field theories.
For example, in the class of models that we study in this paper,
all of the internal degrees of freedom can be represented
in terms of free fermions propagating on the string worldsheet.  Hence,
each of the phenomenological models that we study in this paper
corresponds to a special set of fixed values for the moduli fields.

The gauge couplings of the four-dimensional gauge group factors are all
generally
related to the gauge coupling of the string theory in ten dimensions,
and are consequently related to each other at tree level.  Thus, a natural
unification of gauge couplings occurs in string theory regardless of the
presence of any grand-unified gauge group.  At tree level, the size of
these couplings at unification is determined \cite{dilvev} by the VEV of a
special moduli field, the dilaton $\phi$, via
$g_i\sim e^{-\langle \phi\rangle}$.
Unfortunately, the presence of a classically flat dilaton potential
implies that one does not know, {\it a priori}, the value of the dilaton
VEV.  Hence, at the classical level, one does not know the general size of
the string coupling constant at unification.

This observation is not just limited to the gauge couplings, but
applies to the gravitational coupling as well.
Indeed, because string theories naturally incorporate quantum gravity
and contain a massless graviton in their spectra,
the gauge coupling in the effective four-dimensional gauge group
can ultimately be related to the gravitational coupling by calculating the
scattering amplitude between two gauge bosons and the graviton \cite{Ginsparg}.
Since the string coupling constant
is universal at the classical level (modulo Kac-Moody levels),
and since this coupling is determined by the vacuum expectation value
of the four-dimensional dilaton field, the scale for the gravitational coupling
is therefore also set by the value of the dilaton VEV.
Thus {\it all}\/ of the string couplings are related at tree
level to a single string coupling constant whose value
is {\it a priori}\/ unknown.

Even more importantly, the string ``unification scale''
$M_{\rm string}$ is also not a fixed quantity,
for the specific choice of $M_{\rm string}$ is ultimately a
matter of definition which depends on the chosen renormalization scheme.
This will be discussed below.
In general, this scale is related to the string tension (or the
phenomenological Planck scale) via
$M_{\rm string}\sim 1/\sqrt{\alpha^\prime}\sim g_{\rm string}M_{\rm Planck}$.
It is an important feature that $M_{\rm string}$ and $M_{\rm Planck}$
be of roughly the same order of magnitude;  otherwise, the effective
four-dimensional theory will not be weakly coupled \cite{scales},
and our subsequent perturbative one-loop analysis will not be valid.

The above considerations apply at tree level in string perturbation theory.
However, at the one-loop level, this situation becomes even more non-trivial,
for in general the {\it quantum corrections}\/ are sensitive to
not only the massless string modes, but also the entire infinite
tower of massive string modes,
and these in turn generally depend critically on the expectation values
of all of the moduli fields.  In particular, the quantum
corrections to the gauge couplings --- {\it i.e.}, the so-called
``threshold corrections'' to be discussed below --- are complicated functions
of the moduli fields, and hence the running of the gauge couplings of
the four-dimensional group factors
in the effective low-energy theory generally depends on
the expectation values of all of these moduli fields.
This important dependence has been studied by several groups
\cite{louis,moduli,ILR},
and is crucial for understanding how
an effective potential might be generated which not only selects a preferred
string vacuum (thereby lifting the degeneracy of string vacua corresponding
to different values of the moduli VEV's), but which might also
provide a means of dynamical supersymmetry breaking ({\it e.g.}, through
the formation of gaugino condensates).
Furthermore, understanding the dependence of the threshold corrections on
the moduli fields gives great insight into the structure and symmetries of the
string effective low-energy supergravity Lagrangian \cite{moduli}.

It should be noted, however, that most of these studies of the moduli
dependence of the threshold corrections make some simplifying
assumptions that do not hold for the realistic string models
that we will be studying here.
For example, the moduli dependence in those studies is usually extracted for
the case of heterotic string theories with $(2,2)$ worldsheet
supersymmetry,\footnote{
   Recent results concerning the moduli-dependence of threshold
   corrections in $(2,0)$ compactifications can be found
   in Ref.~\cite{modulitwozero}.  }
and for restricted classes of toroidal compactifications.
However, the models that we will be studying here correspond to $(2,0)$
compactifications in which the moduli fields are not simple to identify.
Similarly, this class of models contains twisted
moduli and sectors that correspond to Wilson lines in an orbifold formulation.
Therefore, the previous results on the moduli dependence of the gauge couplings
are in general not applicable to the realistic string models we will
be studying,
and we will need to develop a different strategy.
Of course, as discussed above, in our setting the moduli fields are fixed at
particular values in the moduli space.  Thus our results apply
only at (or sufficiently near) those points in the moduli space.

\bigskip
\noindent{\it Heavy String Threshold Corrections}
\bigskip

We now discuss the heavy string threshold corrections in further detail.

As we have discussed above, a natural unification of couplings occurs
in string theory through which (at tree level in string perturbation theory)
the gravitational (Newton) coupling constant $G_N$
and all of the gauge couplings $g_i$ are related to one
fundamental string coupling $g_{\rm string}$.
The precise relation is \cite{Ginsparg}
\beq
     g^2_{\rm string}~=~8\pi \,{G_N\over\alpha'}~=~ g_i^2\,k_i
       ~~~~{\rm for ~all}~i~,
\label{treelevelrelation}
\eeq
where $\alpha'$ is the Regge slope, and $g_i$ and $k_i$ are respectively
the gauge coupling and Ka\v{c}-Moody level of the gauge-group factor $G_i$.
However, at the one-loop level, the above tree-level relations are modified to
\beq
     {{16\pi^2}\over{g_i^2(\mu)}}~=~k_i{{16\pi^2}\over{g_{\rm string}^2}}~+~
     b_i\,\ln{ M^2_{\rm string}\over\mu^2}~+~\tilde \Delta_i
\label{onelooprunning}
\eeq
where $b_i$ are the one-loop beta-function coefficients, and $\tilde \Delta_i$
are the quantum corrections which reflect the contributions from
the infinite tower of massive string states.
These heavy string states are usually
neglected in an analysis of the purely massless ({\it i.e.}, observable)
string spectrum, but they nevertheless contribute to the running
of the gauge couplings.  The threshold correction terms $\tilde \Delta_i$
therefore represent these contributions.
Because these terms $\tilde \Delta_i$ enter these equations in the
way that threshold corrections do in field theory,
they are typically referred to as heavy string threshold corrections.

There are several things to note about these threshold corrections.
First, as we have already indicated,
these threshold correction terms $\tilde \Delta_i$
are complicated functions of the moduli fields,
and hence their values vary as one moves around in moduli space
and considers different string vacua.
However, our goal is to examine the role of these threshold corrections
within a particular set of realistic free-fermionic models, and
thus  we are essentially evaluating these
corrections at particular fixed points in moduli space.
Consequently, for our purposes, these threshold corrections are merely numbers,
and our conclusions concerning these numbers will apply only at
(or sufficiently near) those points in moduli space.  For our examination
of gauge coupling unification within these string models, however, this
will be sufficient.

A second point concerns the renormalization scheme-dependence of
these threshold corrections.  Clearly the gauge couplings themselves
are physical quantities, and thus their values completely scheme-independent.
Consequently their unification is indeed a general property.
By contrast, the values for the unification scale $M_{\rm string}$ and
the threshold corrections $\Delta_i$ are scheme-dependent,
and consequently their values can only be specified
within a particular scheme.
For example, it is possible to choose a definition for $M_{\rm string}$ which
entirely absorbs the threshold dependence, leading to a definition of $M_{\rm
string}$
which depends on the VEV's of the string moduli.
However, such a definition proves inconvenient,
and for supersymmetric theories it is
preferable to work within the supersymmetric $\overline{DR}$ scheme.  Thus,
by selecting such a scheme, we have intrinsically selected not only certain
definition for
$M_{\rm string}$, but also a certain corresponding definition for the $\tilde
\Delta_i$ which
appear in the renormalization group equations (\ref{onelooprunning}).
We shall discuss these choices for the $\overline{DR}$ scheme below.

Our final point concerns the dependence of these threshold corrections
$\tilde \Delta_i$ on the gauge group ({\it i.e.}, their dependence on the
subscript $i$).
In general, there are typically two terms that arise in a complete
calculation of $\tilde \Delta_i$, one of which depends on the gauge
group in a complicated fashion (and which will be denoted $\Delta_i$ without
the tilde), and the other of which depends on the gauge group only
through the Ka\v{c}-Moody level $k_i$ at which it is realized:
\beq
        \tilde \Delta_i ~=~ \Delta_i~+~ k_i\,Y~.
\label{gaugedependence}
\eeq
Here $Y$ denotes the contribution which is independent of the gauge
group \cite{Yterm}.
However, in our subsequent calculations, we shall only be interested
in the {\it relative}\/ running of the coupling constants $g_i$,
and this means
that we shall only need to evaluate the {\it differences}\/
between the threshold corrections $\tilde \Delta_i$ for
different gauge group factors.
We shall see this explicitly in Sect.~4.
Moreover, since all of the realistic string models we will
be examining have non-abelian gauge group factors realized
at levels $k_i=1$, the value of the group-independent quantity $Y$ will be
irrelevant for our purposes.
We shall therefore need to evaluate only the simpler quantities $\Delta_i$.

Towards this end,
the most important objects that we need to calculate
are the one-loop string partition function
$Z(\tau)$ and the so-called ``modified'' one-loop string partition functions
$B_G(\tau)$
corresponding to each gauge group factor $G$.
Let us first focus on the partition function $Z(\tau)$.
In general, the partition function $Z(\tau)$ of a given theory takes the
form
\beq
    Z(\tau) ~=~  \sum_{\alpha}
      (-1)^F\, {\rm Tr}\,({\alpha})~.
\label{genform}
\eeq
Here the sum over $\alpha$ represents the sum over all sectors
in the theory,
the overall factor of $(-1)^F$ ensures that spacetime bosonic and fermionic
states contribute with opposite signs, and
${\rm Tr}(\alpha)$ indicates a trace
over the Fock space of mode excitations  of the worldsheet fields:
\beq
       {\rm Tr}\,(\alpha) ~\equiv~
      {\rm Tr}\, q^{H_\alpha}\, \qbar^{\overline{H}_\alpha}~,~~~~~~
	    q\equiv e^{2\pi i \tau}~.
\eeq
Here $H_\alpha$ and $\Hbar_\alpha$
are respectively the right- and left-moving
Hamiltonians for the worldsheet degrees of freedom in the $\alpha$-sector,
and thus this trace simply counts the number of string states
at each worldsheet energy $(H_\alpha,\Hbar_\alpha)$, as expected for a
partition function.
In string models this trace is generically realized as the result
of a GSO projection between subsectors of the theory:
\beq
	{\rm Tr}\,(\alpha)~=~ {1\over g}\,\sum_{\beta}
      \,c({\alpha},{\beta})\, {\rm Tr}\,({\alpha},{\beta})~.
\label{GSO}
\eeq
Here the $\beta$-sum
implements the GSO projection,
$c(\alpha,\beta)$ are the chosen GSO phases,
$g$ is a normalization factor,
and ${\rm Tr}(\alpha,\beta)$ indicates a restricted trace
over the appropriate $(\alpha,\beta)$ subsector.

The ``modified'' partition functions $B_G$ which are needed in the calculation
of the gauge coupling threshold corrections
are then defined in a manner similar to the partition function,
since they too must weigh the contributions from infinite towers of states.
Indeed, they take the same form as $Z$ in (\ref{genform}) except that
they are multiplied by two factors of $\tau_2\equiv {\rm Im}\,\tau$, and
their corresponding traces are modified through insertions of the square of the
spacetime helicity operator $\overline{Q}_H$ and the square of the gauge group
generator $Q_G$:
\beq
     {\rm Tr}\,(\alpha)~\to~
	   {\rm Tr}_G\,(\alpha)~\equiv ~ {\tau_2}^2\,
       {\rm Tr}~  \overline{Q}_H^2 \, Q_G^2 ~q^{H_\alpha}
\qbar^{\overline{H_\alpha}}~
\label{modtracegen}
\eeq
[with identical corresponding insertions into ${\rm Tr}\,(\alpha,\beta)$].
Thus, these modified partition functions also count the numbers of states
at each string energy level, but as expected each such degeneracy is multiplied
by the gauge charge carried by the corresponding state.

Finally, just as the one-loop vacuum energy is defined as the integral
of the ordinary partition function over the modular-group fundamental domain
$\calF$,
the threshold correction contribution from the infinite tower of massive
string states is given \cite{Kaplunovsky} by an analogous integral:
\beq
      \Delta_G ~\equiv~ \int_\calF
	   {d^2\tau\over {\tau_2}^2}
     \, \biggl\lbrack B_G(\tau) - \tau_2 \,b_G \biggr\rbrack ~.
\label{Deltadef}
\eeq
Note that since the contributions of the massless string states
have already been included the calculation of the beta-function
coefficient $b_G$ appearing in (\ref{onelooprunning}),
they must be subtracted from the modified partition function $B_G$
so that $\Delta_G$ records only the {\rm extra}\/ contributions
from the infinite towers of {\it massive}\/ string states.
This subtraction which excludes the massless states
also renders the integral (\ref{Deltadef}) finite.
Note that while the measure of integration in (\ref{Deltadef}) is
modular-invariant, the integrand is not.
We shall discuss the modular properties of this expression in Sect.~6.

As we have indicated at the beginning of this section,
there are various assumptions which enter the derivation
of the one-loop result (\ref{Deltadef}), and consequently there
are various conditions under which (\ref{Deltadef}) may be used.
We shall merely list them here for completeness;  a full discussion can be
found
in Ref.~\cite{Kaplunovsky}.
First, as mentioned above, this result is scheme-dependent, and has in fact
been derived in the so-called $\overline{DR}$ renormalization
scheme;  this is the scheme that is typically used for supersymmetric
field theories and renormalization-group equations.
Therefore, as we shall see in Sect.~5, it will be necessary to include explicit
scheme-conversion corrections when comparing with low-energy data
obtained through other schemes (such as the usual $\overline{MS}$ scheme).
Second, as we have discussed above, this choice of scheme in turn
fixes the choice of string unification scale \cite{Kaplunovsky},
and amounts to the definition
\beq
     M_{\rm string} ~\equiv~ {e^{(1-\gamma)/2} \,3^{-3/4}
                        \over 4\pi} \,g_{\rm string}\, M_{\rm Planck}
\eeq
where $\gamma\approx 0.577$ is the Euler constant, and
where $g_{\rm string}\equiv \sqrt{32 \pi/(\alpha' M_{\rm Planck}^2)}$ is
the string coupling.
Numerically, this yields
\beq
   M_{\rm string}~ \approx ~ g_{\rm string}\,\times\,5\,\times\,10^{17}~
         {\rm GeV}~.
\label{Mstringvalue}
\eeq
Third, we reiterate that the result (\ref{Deltadef}) is only a partial
result which neglects
the additional contributions such as the term $Y$ in (\ref{gaugedependence})
that are the same for all gauge group factors in a given model.
Consequently, by using (\ref{Deltadef}) we cannot consider absolute values
of the threshold corrections for a given gauge group factor, but rather only
the {\it relative
differences}\/ of the threshold corrections $\tilde \Delta_{G_1}-\tilde
\Delta_{G_2}
 =\Delta_{G_1}-\Delta_{G_2}$ between two
gauge group factors $G_1$ and $G_2$ realized at equal Ka\v{c}-Moody
levels $k_i$.\footnote{
    We point out, however, that there has recently appeared
   an alternative procedure for calculating threshold
   corrections \cite{Kiritsis} which includes
   these gauge-independent constant terms.
   We will discuss this briefly at the end of Sect.~6.
   In any case, these constant terms will not be necessary for our analysis,
   so it will be sufficient for our purposes
    to use the definition (\ref{Deltadef}).}
However, beyond these restrictions,
the one-loop expression in (\ref{Deltadef}) is
completely general.  In particular, it makes no additional assumptions
about the structure of a given heterotic string model,
or the values of various string moduli.
Thus this expression (\ref{Deltadef}) will be our starting point
when evaluating the threshold corrections in the realistic free-fermionic
models.  In particular, since these models have $(2,0)$ [rather than
$(2,2)$] worldsheet supersymmetry, we cannot use various
moduli-dependent expressions \cite{louis,moduli} which are derived from this.

\bigskip
\noindent{\it Threshold Corrections in Free-Fermionic Models}
\bigskip

We now concentrate on how this general procedure for calculating
threshold corrections applies to the case of
string models built through the free-fermionic construction \cite{KLT,ABK}.
As briefly discussed in Sect.~2, in the free-fermionic construction of
four-dimensional heterotic
string models, the light-cone gauge worldsheet field content
consists of two transverse spacetime coordinate bosons, their two right-moving
fermionic
superpartners, and an additional set of
62 purely internal fermions of which 18 are right-moving and 44 are
left-moving.
Collectively these 64 Majorana-Weyl fermions may be denoted $\psi_{\ell}$
($\ell=1,...,64$),
with an ordering such that $\ell=1,2$ correspond to the right-moving
superpartner fermions
carrying spacetime Lorentz indices, $3\leq \ell \leq 20$ correspond to
the purely-internal right-moving fermions,
and $\ell\geq 21$ correspond to the purely-internal left-moving fermions.
A string model is then realized by specifying, for each sector of the theory,
the boundary conditions of these 64 fermions as they traverse
the two non-contractible loops of the torus, as well as a set of phases which
specify
the generalized GSO projections which are to be applied in that sector.
These parameters are not all independent, however, and must be chosen in
such a way that certain self-consistency conditions (guaranteeing a proper
worldsheet
supercurrent, proper spacetime spin-statistics assignments,
physically sensible projections, and modular invariance)
are satisfied.
The rules governing the construction of such models are
given in Refs.~\cite{KLT,ABK,KLST}.
In particular, it is shown
that the parameters describing the 64 fermionic
boundary conditions in all sectors can ultimately be described via a simple
``basis set'' of
$(N+1)$ different 64-component vectors ${\bf V}_i$ ($0\leq i \leq N$), and that
the GSO phases
in all sectors can be similarly described through a single matrix $k_{ij}$
($0\leq i,j\leq N$).
The number of basis vectors $N$ which are necessary is of course
model-dependent,
increasing with the complexity of the model desired,
and the complete constraint equations which
relate self-consistent choices for the $\V_i$ vectors and $k_{ij}$ matrix
are given in Ref.~\cite{KLST}.

In the free-fermionic construction,
each of the $\alpha$-sectors discussed above corresponds to a different set of
fermionic
worldsheet boundary conditions around the {\it spacelike}\/ cycle of the torus,
whereas the GSO projection is realized through
the $\beta$-summation over different fermionic
boundary conditions around the {\it timelike}\/ cycle of the torus.
The corresponding GSO phases and traces are then unambiguously defined
once a particular model ({\it i.e.}, a set of
$\V_i$ and $k_{ij}$ parameters) is specified.
For completeness we now write these phases and traces explicitly
in terms of the $\V_i$ and $k_{ij}$.
Note that henceforth we will be following the
notation and conventions of Ref.~\cite{KLST}, where
the definitions of any unexplained symbols below can be found.
First, in the free-fermionic models, the
traces ${\rm Tr}\,(\alpha,\beta)$
in (\ref{GSO}) are given by
\beqn
     {\rm Tr}\,(\alpha,\beta)&\equiv& {\rm Tr}\, \left\lbrack
	\exp\left\lbrace -2\pi i \beta\V\cdot \N'_{\alpha\V}\right\rbrace
       \,q^{H_{\alpha\V}} \,\qbar^{\Hbar_{\alpha\V}} \right\rbrack \nonumber\\
	 &=& {\tau_2}^{-1}\, {1\over \overline{\eta}^{12}\eta^{24}} \,
	 \left\lbrace \prod_{\ell=1}^{20}
	   \sqrt{\t{-a_\ell}{+b_\ell}^\ast}
	 \right\rbrace \,
	 \left\lbrace \prod_{\ell=21}^{64}
	   \sqrt{\t{-a_\ell}{+b_\ell}}
	 \right\rbrace
\label{trace1}
\eeqn
where $N'_{\alpha V}$ indicates the worldsheet fermionic number
operator in the $\alpha \V$ sector with the Ramond zero modes excluded,
the asterisk indicates complex conjugation, and where
\beqn
	    && a_\ell\equiv (\overline{\alpha\V})^\ell~,~~~~~
	       b_\ell\equiv (\overline{\beta\V})^\ell~\nonumber\\
	  && \eta(\tau)\equiv q^{1/24}\,\prod_{n=0}^\infty \,(1-q^n)\nonumber\\
	  && \t{a}{b}(\tau)\equiv \sum_{n= -\infty}^\infty \,e^{2\pi i b n}\,
q^{(n+a)^2/2}~.
\label{trace2}
\eeqn
Note that the ${\tau_2}^{-1}$ factor in (\ref{trace1}) represents
the contribution to the trace from the zero-modes of the two transverse
coordinate bosons $X^i$, and that the sign of the square roots in
(\ref{trace1})
is  not of practical importance for our calculation because there will always
be an even number of $\sqrt{\Theta}$ factors for each type of
$\Theta$-function.
This latter property is guaranteed by the so-called ``cubic constraint'' of
Ref.~\cite{KLST}.

With the definition given in (\ref{trace1}), the
corresponding GSO phases and normalization factor in (\ref{GSO}) are then given
by
\beqn
      c(\alpha,\beta)&=& \Gamma^{\overline{\alpha\V}}_{\overline{\beta\V}}\,
	\exp\left\lbrace 2\pi i \, \sum_{i=0}^N \beta_i \left(
	  k_{ij} \alpha_j + s_i - \V_i\cdot \overline{\alpha\V}
\right)\right\rbrace\nonumber\\
       g &=&  \prod_{i=0}^N\, m_i
\label{phases}
\eeqn
where $\Gamma^{\overline{\alpha\V}}_{\overline{\beta\V}}=\pm 1$ represents the
additional
phase contribution which can arise due to permutations of
the Ramond fermionic zero-mode gamma-matrices.
The definition of
$\Gamma^{\overline{\alpha\V}}_{\overline{\beta\V}}$ is given
explicitly\footnote{
    We point out, however, that when calculating
    $\Gamma^{\overline{\alpha\V}}_{\overline{\beta\V}}$,
    one must include $\gamma$-matrix factors from
    only those Majorana-Weyl worldsheet fermions
    which are {\it necessarily real}, {\it i.e.},
    those fermions which cannot be globally paired in all sectors with any
    other Majorana-Weyl fermion in
    all sectors to form a complex Weyl fermion.
    This crucial restriction is not stated
    in the definition given in Ref.~\cite{KLST},
    but must be incorporated in the cases
    of models which simultaneously contain not only
    necessarily real fermions but also
    complex fermions with twisted ``multi-periodic'' boundary conditions.
    Most free-fermionic models of phenomenological interest fall into
    this class.  This restriction was also independently pointed out in
    Ref.~\cite{Fermilab}.  }
in Eq.~(3.6) of Ref.~\cite{KLST}.  Note that
$\Gamma^{\overline{\alpha\V}}_{\overline{\beta\V}}= \pm 1$,
since for a given sector $(\alpha\V,\beta\V)$
it can be shown that
$\Gamma^{\overline{\alpha\V}}_{\overline{\beta\V}}$
always contains an even number of factors of $\gamma_\ell$ for each necessarily
real fermion $\ell$.
The quantities $m_i$ in (\ref{phases}) are defined in Ref.~\cite{KLST}.

The calculation of the ``modified'' partition functions $B_G$ is similar
to that described above for $Z$;
indeed, the GSO phases remain unchanged, and
the only change is that the traces are modified as in
(\ref{modtracegen}) due to the insertions of the charge operators and two
extra factors of $\tau_2$.
Now, these two extra factors of $\tau_2$ are simple overall factors which
merely increase by two the power of $\tau_2$ in (\ref{trace1}).
Let us therefore focus first on the gauge group generator $Q_G^2$.
In general, this object may be defined as
\beq
	Q_G^2 ~\equiv~ {1\over R}\,\sum_{r=1}^{R} \,\left\lbrace
         Q^{(r)}\right\rbrace^2
\label{squaredgenerator}
\eeq
where $R$ is the rank of the group $G$ and where $Q^{(r)}$ are the
normalized charge elements of the Cartan subalgebra.
In free-fermionic models, these charges $Q^{(r)}$ generally appear
as linear combinations of the individual charge operators
corresponding to individual worldsheet fermions;  such
linear combinations describe how the gauge group $G$ is
ultimately embedded or realized through free fermions.
Note that in the free-fermionic construction, the gauge group can
be realized only through those worldsheet real Majorana-Weyl fermions
which can be consistently paired in all sectors to form complex Weyl
fermions, and it is indeed only for such pairs of real fermions
--- or equivalently for such complex Weyl fermions --- that these
fermionic charge operators can be defined.
Let us write the indices of such complex Weyl fermions with capital
letters, so that $L\equiv(\ell_1,\ell_2)$ represents a pair of
real-fermion indices with corresponding charge operator $Q_L$.
We find, then, that one can generically expand $Q_G^2$ as a polynomial in terms
of the
underlying fermionic charge operators $Q_L$:
\beq
	Q_G^2 ~=~ \sum_{L,M}\, c_{LM}^{(G)}\, Q_L Q_M
\label{Qexpansion}
\eeq
where $c^{(G)}_{LM}$ are the model-specific coefficients which
describe the gauge group embeddings.

Having related the gauge-group generator $Q_G^2$
to the charge operators $Q_L$ of the individual (complex) fermions,
it is now straightforward to calculate the effect of the insertion
of $Q_G^2$ into the trace.
Let us first consider the effect of a single term $Q_L$ on the
contribution to the trace from the corresponding $L^{\rm th}$ complex fermion,
where $L\equiv(\ell_1,\ell_2)$.  Without the insertion, this isolated
contribution from the two real fermions $(\ell_1,\ell_2)$ would have
been
\beqn
      \eta^{-1}\, \sqrt{
	\t{-a_{\ell_1}}{+b_{\ell_1}} \t{-a_{\ell_2}}{+b_{\ell_2}}  }
      &=& \eta^{-1}\, \t{-a_{\ell_1}}{+b_{\ell_1}} \nonumber\\
      &=& \eta^{-1}\, \sum_{n= -\infty}^\infty \,
	 \exp\left\lbrace 2\pi i b_{\ell_1} n\right\rbrace \,
	 q^{(n-a_{\ell_1})^2/2} ~.
\label{thetanoinsertion}
\eeqn
The first equality follows since the $\ell_1$ and $\ell_2$ fermions
cannot be joined to form a complex fermion $L$ unless
they have the same boundary conditions in all sectors, so that
$a_{\ell_1}=a_{\ell_2}$ and  $b_{\ell_1}=b_{\ell_2}$.
The second line is just the expansion of the single $\Theta$ function.
Let us recall, however, that we may equivalently bosonize this complex fermion,
in which case each term in this sum represents
the contribution to the single-boson trace from a state in the corresponding
bosonic
one-dimensional momentum lattice {\bf P}
with lattice coordinate $P= n-a_{\ell_1}$ (so that the power of
$q$ in the expansion is the usual worldsheet energy contribution $H=P^2/2$).
But this bosonic momentum lattice {\bf P} is nothing but
the corresponding fermionic charge lattice {\bf Q} \cite{KLT}.  Thus we
immediately recognize
that the charge of each state contributing to the $n^{\rm th}$
term in (\ref{thetanoinsertion}) is simply $n-a_{\ell_1}$, so that
the trace with a single $Q_L$ insertion is simply
\beqn
      \eta^{-1}\, \tp{-a_{\ell_1}}{+b_{\ell_1}} ~\equiv~
      \eta^{-1}\, \sum_{n= -\infty}^\infty \, (n-a_{\ell_1})\,
	 \exp\left\lbrace 2\pi i b_{\ell_1} n\right\rbrace \,
	 q^{(n-a_{\ell_1})^2/2} ~.
\label{thetaprime}
\eeqn
Likewise, an insertion of $Q_L^2$ yields
\beqn
      \eta^{-1}\, \tpp{-a_{\ell_1}}{+b_{\ell_1}} ~\equiv~
      \eta^{-1}\, \sum_{n= -\infty}^\infty \, (n-a_{\ell_1})^2\,
	 \exp\left\lbrace 2\pi i b_{\ell_1} n\right\rbrace \,
	 q^{(n-a_{\ell_1})^2/2} ~.
\label{thetaprimeprime}
\eeqn
Thus for each term $Q_L Q_M$
in (\ref{Qexpansion}), the modified trace is the same as
(\ref{trace1}) except that certain $\Theta$ functions are
replaced\footnote{
    In performing this replacement, there is in fact a subtle
    sign ambiguity: although $\t{-a}{b}=\t{a}{-b}$ as an algebraic identity,
    we find $\tp{-a}{b}= - \tp{a}{-b}$.  The question
    then arises as to whether $\tp{-a}{b}$ or $\tp{a}{-b}$ is to
    be used in the modified trace, amounting to an ambiguity in the overall
    sign of the
    charge $Q_L$.  However, after the GSO projections are performed,
    the resulting charge lattice {\bf Q} must be always be invariant under the
    inversion
    ${\bf Q}\to -{\bf Q}$ (this is tantamount to CPT invariance).
    Thus, one must ensure only that consistent {\it relative}\/ signs are used
    between GSO-related sectors.  This amounts to choosing the same sign
    for $a_\ell$ in all sectors with identical $\alpha$-boundary conditions but
    different $\beta$-boundary conditions.}
by either $\Theta'$ or $\Theta''$ as defined above:
for $L=M$ we replace the corresponding single factor of $\Theta$
with $\Theta''$,
and for $L\not =M$ we replace each of the two corresponding factors
of $\Theta$ with $\Theta'$.

The helicity operator insertion $\overline{Q}_H^2$
can be handled in precisely the same way, since the spacetime helicity
operator is essentially the same as the gauge charge operator, only
applied to the two real {\it right-moving}\/ fermions
$L=(\ell_1,\ell_2)=(1,2)$
which carry spacetime Lorentz indices, and with an overall minus
sign (to take into account that we are dealing with the corresponding
complex conjugate $\overline{\Theta}$-functions).
Thus, the insertion of $\overline{Q}_H^2$ into the trace simply amounts to
an analogous replacement of the corresponding $\overline{\Theta}$ functions
with $-\overline{\Theta}''$, as discussed above.

The final issue we discuss is a practical matter concerning
the integration which is then necessary to obtain a value
of $\Delta_G$ as in (\ref{Deltadef}).
It is of course possible to calculate $b_G$ independently from
explicit knowledge of the massless spectrum of the particular
string model in question.  However, there is an important
self-consistency check which can be performed, since
if we expand the corresponding modified partition function $B_G$
as a power series of the form
\beq
	B_G ~=~ \tau_2\, \sum_{m,n}\, b_{mn}^{(G)}\, \qbar^m q^n~,
\label{Bexpand}
\eeq
we see that the coefficient $b_{00}^{(G)}$, which tallies
the contributions from the massless states with total (left,right)
worldsheet energies $(n,m)=(0,0)$
respectively, should be equal to $b_G$:
\beq
	b_{00}^{(G)}~=~ b_G~.
\label{selfcheck}
\eeq
Thus, the subtraction of $\tau_2 b_G$ within the integrand of
(\ref{Deltadef}) can be most easily achieved by expanding
$B_G$ in the form (\ref{Bexpand}), and then simply setting
the coefficient $b_{00}^{(G)}$ to zero.  In any case,
as remarked earlier,
a non-zero value of $b^{(G)}_{00}$ will result in a divergent
integral for $\Delta_G$, so setting the $b^{(G)}_{00}$ coefficient
to zero is an efficient analytic way of removing what would
otherwise be a logarithmic divergence.
The resulting integral can then be performed by computing the sum
\beq
     \Delta_G ~=~ \sum_{m,n} \, b_{mn}^{(G)}\, I^{(1)}_{mn}
\label{Deltasum}
\eeq
where the general integrals
\beq
      I^{(s)}_{mn} ~\equiv ~\int_\calF {d^2\tau \over {\tau_2}^s}
                \,\qbar^m q^n~.
\label{integraldef}
\eeq
can each be numerically evaluated.
This procedure has the dual advantages of incorporating the cancellations
inherent in the GSO projections at an early stage
(by calculating the coefficients $b_{mn}^{(G)}$ prior
to integration), and of analytically removing the divergence from
the massless physical states $b^{(G)}_{00}$.
Furthermore, the integrals (\ref{integraldef})
then need to be evaluated only once,
for any change in the particular string model under discussion merely amounts
to a change in the corresponding coefficients $b_{mn}^{(G)}$.
Note that inherent in our procedure is an exchange in the order
of the summation over $(m,n)$ and the integration over $\tau$.
As discussed in Ref.~\cite{louis}, this procedure is valid
provided that the sum in (\ref{Bexpand}) is convergent as
$\tau\to i\infty$.  However, this is necessarily the case for
all string theories lacking charged physical tachyonic states.
Indeed, any divergences arising from different sectors in the theory
will have already cancelled in the sum over all sectors, so that
the total coefficients $b_{mn}$ which we use necessarily have
$b_{nn}=0$ for all $n<0$.  Hence, once the contributions from all sectors
have been added together (as we have done here), the interchange is valid.

Finally, we point out that in removing the logarithmic
divergence from the massless states, we have defined $b_{00}$ through the
expansion (\ref{Bexpand}).  In particular, we have {\it not}\/ followed
Ref.~\cite{Kaplunovsky} in defining
\beq
      b_{00} ~ {\buildrel {?}\over {=}} ~ \lim_{\tau\to i\infty}\,
       \biggl\lbrace \tau_2^{-1} \,B(\tau)\biggr\rbrace~,
\label{Kapwrong}
\eeq
as there can be cases for which such an identification is incorrect.
For example, if $B(\tau)$ has non-zero coefficients $b_{mn}$ with $m+n<0$
and $m\not=n$ ({\it i.e.}, contributions from charged ``unphysical
tachyons''), then
the $\tau\to i\infty$ limit of $B(\tau)$ diverges, and (\ref{Kapwrong})
is in error.
We shall see explicit examples of this sort of occurrence in Sect.~4,
where we shall be considering certain models which are free of {\it physical}\/
charged tachyons ({\it i.e.}, which have $b_{nn}=0$ for all $n<0$),
but which nevertheless have non-zero
coefficients $b_{mn}$ for negative unequal values of $m$ and $n$.
Although we shall find that $m+n>0$ for such coefficients, the appearance of
such contributions with negative and unequal values of $m$ and $n$ shows that
these
sorts of unphysical tachyons generically appear, and that unphysical tachyons
with
$m+n<0$ cannot be generically ruled out.
In any case, however, such unphysical tachyons do not lead to the sorts of
divergences in the threshold corrections $\Delta$ that would arise
from physical tachyons with $m=n$, since
the contributions from {\it unphysical}\/ tachyons
are projected out of the integral in the $\tau\to\infty$ region.
We shall discuss the implications of such unphysical tachyons in Sect.~6.

\setcounter{footnote}{0}
\section{Threshold Corrections in Particular Models}

In the previous section we discussed how threshold corrections
$\Delta_G$ can be calculated in general free-fermionic models.
In this section we now apply this procedure to the cases of
the realistic free-fermionic models described in Sect.~2.
In particular, we shall explicitly calculate the threshold
corrections $\Delta_G$ for each gauge group factor
in each realistic free-fermionic model which has appeared
in the literature to date.  This includes the ``revamped''
flipped $SU(5)$ model, several $SU(3)\times SU(2)\times U(1)$ models,
and an $SO(6)\times SO(4)$ model.  Furthermore, to test the dependence
of our results on the existence of spacetime supersymmetry
in these models, we shall also consider the case of a certain
 {\it non}\/-supersymmetric $SU(3)\times SU(2)\times U(1)$ model.

As we discussed in the previous section, our procedure for each
model is essentially the same, and can be broken down into
several distinct steps.

First, for each model, we must determine the embedding of each
of its gauge group factors within the free worldsheet fermions ---
 {\it i.e.}, we must essentially determine the coefficients $c_{LM}^{(G)}$
which appear in (\ref{Qexpansion}).  We will refer to this
procedure as obtaining the appropriate {\it charge polynomial}\/
for each factor.  In general, this requires not only analyzing
how each gauge group factor arises within the free-fermionic
construction for each model, but also calculating the appropriate
overall normalizations.  These issues will be discussed for each
model below.

Second, for each model, we must then actually perform the
threshold calculations, and calculate the sum over {\it all}\/ of the
contributing sectors that appear in the model.  For each of
the realistic models that
we will be analyzing, this typically amounts to summing the separate
contributions
from several thousand individual sectors for each gauge group factor.

Third, for each model, we then perform a set of important self-consistency
checks by verifying that the relation (\ref{selfcheck}) holds for each
gauge group factor in the model.
In order to verify this relation, we therefore have to
do two separate calculations:  we must calculate the $b_{00}^{(G)}$
coefficient that arises from an explicit expansion of the modified partition
function $B^{(G)}(\tau)$ for each gauge group factor $G$, and
we must also independently calculate the beta-function coefficient $b_G$
using our prior knowledge of the massless spectrum of the model
in question.  Ensuring that these two quantities agree for all gauge
group factors thus guarantees that our calculations of both the massless
spectrum of
the model and the threshold corrections from its massive states are not in
error
(or inconsistent with each other).
This also provides a non-trivial check of our charge polynomials,
which in turn verifies our determinations of both the gauge group embeddings,
and their non-trivial normalizations.

Finally, for each model, we then put all of these results together to
calculate the resulting threshold corrections $\Delta_i$.
The effects of these threshold corrections
on gauge coupling unification will be analyzed in Sect.~5.

\subsection{Flipped SU(5) model}

We first analyze the threshold correction
in the ``revamped'' flipped $SU(5)$ model \cite{revamp}.
The defining parameters (worldsheet fermion boundary conditions
and GSO projection phases) which yield this model
are listed in the Appendix.
The analysis of heavy threshold corrections for this model was already
performed
in Ref.~\cite{AELN}, but we shall use this example to set the procedure for the
analysis in subsequent models. Furthermore, as we shall see,
our results differ
substantially from the numerical results of Ref.~\cite{AELN}.
However, we believe that this difference is ultimately a reflection
of our improved method of calculating the modified partition functions
$B(\tau)$, and of numerically integrating these functions
$B(\tau)$ over the modular-group fundamental domain.  In particular, as
discussed in the
previous section,  our procedure {\it analytically}\/
removes the logarithmic divergence from the massless states, and
in so doing explicitly verifies
that $b_G$ and $b_{00}^{(G)}$ agree (as required for self-consistency).
Furthermore, we will see that the size of our results
is more in line with those from the other realistic free-fermionic
string models we will be examining, as well as from previous
string threshold calculations in various
orbifold \cite{Kaplunovsky,MNS} and Type-II \cite{DL} models.
We therefore believe that our results are more reliable.

\bigskip
\noindent{\it Determining the Charge Polynomials}
\bigskip

This model has the observable gauge symmetry $SU(5)\times U(1)$, which is
realized
through the following five complex (left-moving) worldsheet fermions:
\beqn
       L=1 ~&\Longleftrightarrow&~ \ell_i=(33,49) ~~~\longleftarrow~{\rm
fermion}~\overline{\psi}^1
                  \nonumber\\
       L=2 ~&\Longleftrightarrow&~ \ell_i=(34,50) ~~~\longleftarrow~{\rm
fermion}~\overline{\psi}^2
                  \nonumber\\
       L=3 ~&\Longleftrightarrow&~ \ell_i=(35,51) ~~~\longleftarrow~{\rm
fermion}~\overline{\psi}^3
                  \nonumber\\
       L=4 ~&\Longleftrightarrow&~ \ell_i=(36,52) ~~~\longleftarrow~{\rm
fermion}~\overline{\psi}^4
                  \nonumber\\
       L=5 ~&\Longleftrightarrow&~ \ell_i=(37,53) ~~~\longleftarrow~{\rm
fermion}~\overline{\psi}^5~.
\label{firstfive}
\eeqn
Explicitly, this means that we are labelling as $L=1,...,5$
the five complex fermions which can be formed, as indicated, from the
real fermions $33,...,37$ and $49,...,53$.  This latter numbering
reflects the ordering of the real fermions as listed in
the Appendix.
These five complex fermions $L=1,...,5$
are respectively the complex
fermions called $\overline{\psi}^{1,...,5}$ in Sect.~2.

As discussed in Ref.~\cite{AELN}, the four traceless $SU(5)$
generators for this model can then be written in terms of the
charges $Q_L$ corresponding to these complex fermions,
\beqn
  Q_{SU(5)}^{(1)}&=&{1\over{\sqrt2}}\,(Q_1-Q_2)\nonumber\\
  Q_{SU(5)}^{(2)}&=&{1\over{\sqrt6}}\,(Q_1+Q_2-2Q_3)\nonumber\\
  Q_{SU(5)}^{(3)}&=&{1\over{\sqrt{12}}}\,(Q_1+Q_2+Q_3-3Q_4)\nonumber\\
  Q_{SU(5)}^{(4)}&=&{1\over{\sqrt{20}}}\,(Q_1+Q_2+Q_3+Q_4-4Q_5)~,
\label{SU5generators}
\eeqn
while the remaining single orthogonal $U(1)$ generator (essentially
the trace of the original larger $U(5)$ symmetry) is therefore
\beq
      Q^{(1)}_{U(1)}~=~ {1\over{\sqrt5}}\,(Q_1+Q_2+Q_3+Q_4+Q_5)~.
\label{SU5U1generator}
\eeq
Note that each of these generators is normalized so as to produce
conformal dimension one for the massless states.

Given these individual generators $Q^{(r)}$
for the $SU(5)$ and $U(1)$ group factors,
we then compute the corresponding squared polynomials $Q_G^2$ for each factor
according to (\ref{squaredgenerator}).
Note that since the complex fermions $L=1,...,5$ all
have identical boundary conditions in all sectors of this
model, their charges $Q_L$ will be identical in all sectors.
It is therefore possible to replace
any occurrence of products of charges of the form $Q_L Q_M$ with
${Q_1}^2$ if $L=M$, and with $Q_1 Q_2$ if $L\not=M$.\footnote{
      Note, however, that we must continue to distinguish
      between the squared charge for a single fermion,
      and the product of single charges for two separate fermions.
      These two charge combinations have different effects
      within the trace, as discussed below  (\ref{thetaprimeprime}).}
The resulting squared polynomials then take the simple form
\beqn
     Q^2_{SU(5)}&=&Q_1^2~-~Q_1Q_2~\nonumber\\
     Q^2_{U(1)} &=&Q_1^2~+~4\,Q_1Q_2~.
\label{SU5poly}
\eeqn
Thus, the only two ``basis''
charge insertions that we need to consider
are ${Q_1}^2$ and $Q_1 Q_2$.

Note that since we will ultimately be interested in only the
 {\it difference}\/ of  the threshold corrections
for $SU(5)$ and $U(1)$, we could at this stage proceed
to consider only the difference of their corresponding
charge polynomials in (\ref{SU5poly}).
Indeed, this would cancel the ${Q_1}^2$ term completely,
and leave us with a single ``basis'' insertion $Q_1 Q_2$.
However, this simplification would later rob us of an important
self-consistency check on the corresponding beta-function coefficients, for
we would be able to compare only the difference of these coefficients,
rather than each one separately.  For this reason we shall retain
both charge polynomials in our basis.

\bigskip
\leftline{\it Threshold Contributions:  Results of Calculation}
\medskip

Following the procedure outlined in the previous section,
we now calculate the separate contributions to the threshold
corrections that arise from each of these ``basis'' insertions.
Specifically, for each basis insertion $Q_L Q_M$ ({\it i.e.},
either ${Q_1}^2$ or $Q_1 Q_2$), we calculate the corresponding
trace $B(\tau)$ over mass levels and all sectors, its massless
expansion coefficient $b_{00}$, and
its corresponding integral $\Delta$ with its massless
coefficient $b_{00}$ set to zero (to analytically remove the
divergence from the massless states).
Our results are as follows:
\beq
\begin{tabular}{c|cc}
 ~ & $b_{00}$ & $\Delta$  \\
\hline
${Q_1}^2$ &   $3.5$ &   $  12.331       $ \\
$Q_1 Q_2 $ &  $    4.5 $  &  $     1.53625 $  \\
\end{tabular}
\label{results1}
\eeq

In evaluating $\Delta$, we have followed the procedure indicated
in (\ref{Deltasum}) and (\ref{integraldef}),
analytically $q$-expanding each modified partition
function $B(\tau)$ to the fourth excited energy level, and
numerically evaluating the convergent integrals $I_{mn}^{(1)}$
to the precision indicated in the results quoted above.
This procedure thus minimizes any numerical error, so that
essentially none remains to this level of accuracy.

\bigskip
\leftline{\it Self-Consistency Checks}
\medskip

Next we perform our self-consistency checks by comparing
these values of $b_{00}$ for each charge insertion with
the one-loop beta-function coefficients expected
from the (known) spectrum of this model.
In general, the one-loop beta-function coefficients are given by
\beqn
    b_{SU(5)}&=&-15~+~2\,N_g~+~N_h\nonumber\\
    b_{U(1)}&=&2\,N_g~+~N_h~+~{5\over4}\,N_4~,
\label{SU5betafun}
\eeqn
where $N_g$ is the number of generations and antigenerations
[with a generation consisting
of a full ${\bf 16}$ representation of $SO(10)$ decomposed under
$SU(5)\times U(1)$],
where $N_h$ is the number of ${\bf5}+{\bf{\bar5}}$ Higgs pairs,
where $N_4$ is the number of hidden $SU(4)$ pairs ${\bf4}+{\bf{\bar4}}$ that
   carry $U(1)$ charge $\pm{5/4}$.
In the revamped flipped $SU(5)$ model, there are four
${\bf{16}}$ representations and one ${\bf{\overline{16}}}$
representation of $SO(10)$, $N_h=4$ and $N_4=6$. Thus, the expected
one-loop beta-function coefficients are
\beq
          b_{SU(5)}=-1~,~~~~~~b_{U(1)}={21.5}.
\label{SU5betafuncoef}
\eeq

We now compare these numbers with the values of $b_{00}$ in
(\ref{results1}).
Using the charge polynomials given in (\ref{SU5poly}),
we indeed find
\beqn
        b_{SU(5)}&=& 3.5 ~-~ 4.5~=~ -1\nonumber\\
        b_{U(1)}&=&  3.5 ~+~ 4\,(4.5)~=~ 21.5~,
\eeqn
in perfect agreement with (\ref{SU5betafuncoef}).

\bigskip
\leftline{\it Final Threshold Corrections}
\medskip

Given these results, we now compute the final relative threshold correction
$\Delta_{U(1)}-\Delta_{SU(5)}$.
Since $Q^2_{U(1)}-Q^2_{SU(5)}= 5 Q_1 Q_2$,
the difference $\Delta_{U(1)}-\Delta_{SU(5)}$
will simply be five times the value of $\Delta$
in (\ref{results1}) for $Q_1 Q_2$, or
\beq
     \Delta_{U(1)}-\Delta_{SU(5)}~=~7.68125~.
\label{su5result}
\eeq
Note that this result is smaller by approximately a factor of three
from the value that was found in Ref.~\cite{AELN}.

Given this relative threshold correction,
we can then compute its effect on the
string unification scale $M_{\rm string}$.  From the renormalization-group
equations (\ref{onelooprunning})
for the $SU(5)$ and $U(1)$ couplings,
and by taking the difference between
the $SU(5)$ and $U(1)$ one-loop RGE's, we find that
the corrected unification scale is in general given by
\beq
          M_{\rm string}^{\rm (corrected)}~=~M_{\rm string}^{\rm
(uncorrected)}\,
      \exp\left\lbrace {\Delta_1-\Delta_5\over 2(b_1-b_5)} \right\rbrace~.
\label{su5xu1unificationscale}
\eeq
Since $b_{U(1)}-b_{SU(5)}=22.5$, we find using (\ref{su5result})
that
\beq
    M_{\rm string}^{\rm (corrected)}~\approx~5.93 \,\times\,10^{17}~ {\rm
GeV}~
\eeq
for this model.

We remind the reader that in the realistic free-fermionic models, one
combination of the
$U(1)$ factors is anomalous. This anomalous $U(1)$ combination arises due to
the
fact that
these free-fermionic models are $(2,0)$ compactifications rather than $(2,2)$.
The anomalous
$U(1)$ gives rise to a Fayet-Iliopoulos $D$-term that breaks supersymmetry and
destabilizes
the vacuum. The models must therefore choose
non-zero VEV's for some of the scalar fields (twisted moduli)
so as to cancel the anomalous $U(1)$ $D$-term \cite{DSW}.
This corresponds to a shift in the string vacuum, and one can then ask whether
this shift
can modify the estimate of $\delta\Delta$, and consequently our estimate of
$M_{\rm string}^{\rm (corrected)}$.
However, the contributions to the threshold corrections come from string
states weighing $\geq g_{\rm string}M_{\rm Planck}/\sqrt{8\pi}$,
whereas the (extra) masses acquired from shifting
the string vacuum are of higher order in $g_{\rm string}$.
Hence, these extra masses only affect the
value of $M_{\rm string}^{\rm (corrected)}$ at higher
order, and can be ignored in
our analysis \cite{AELN}.
The same also holds true for the other models we will be examining.

\subsection{First SU(3) $\times$ SU(2) $\times$ U(1) model}

This model \cite{EU} is listed in the Appendix.
The observable gauge group of this model is $SU(3)\times SU(2)\times U(1)^2$,
with one of the $U(1)$ factors
associated with the $SU(3)_{\rm color}$
group factor (arising as the trace of the original larger $U(3)$ color
symmetry),
and the other associated with the $SU(2)_{\rm left}$
group factor (and arising as the trace of the original larger $U(2)$ weak
isospin symmetry).
We shall denote these two $U(1)$ factors as $U(1)_C$ and $U(1)_L$
respectively.
The electroweak hypercharge $U(1)$ is then a linear combination
of these two $U(1)$ factors.

\bigskip
\noindent{\it Determining the Charge Polynomials}
\medskip

All of these group factors are realized through
the same five complex fermions
as in (\ref{firstfive}) for the flipped $SU(5)$ model.
Unlike the case of the flipped $SU(5)$ model, however,
these five fermions no longer share the same boundary conditions
(indeed, it is for this reason that the gauge group is altered).
Rather, only the $L=\lbrace 1,2,3\rbrace$ fermions share the same boundary
conditions;  likewise, the remaining $L=\lbrace 4,5\rbrace$ fermions
share the same set of {\it different}\/ boundary conditions.
Therefore, we expect that our needed charge polynomials must be built from
the larger set of {\it five}\/ elementary ``basis'' charge bilinears
$\lbrace {Q_1}^2, Q_1 Q_2, Q_1 Q_4, {Q_4}^2, Q_4 Q_5\rbrace$.

Explicitly, the generators of the
$SU(3)\times SU(2)\times U(1)^2$ group are as follows.
The traceless $SU(3)$ generators are given by
\beqn
         Q_{SU(3)}^{(1)}&=& {1\over{\sqrt2}}\,(Q_1-Q_2)\nonumber\\
         Q_{SU(3)}^{(2)}&=& {1\over{\sqrt6}}\,(Q_1+Q_2-2Q_3)~,
\label{SU3gene}
\eeqn
and the orthogonal $U(1)_C$ generator corresponding to this
color group factor is the trace of the original $U(3)_C$ symmetry:
\beq
        Q_{U(1)_C}^{(1)} ~=~ {1\over \sqrt{3}}\, (Q_1 +Q_2 +Q_3)~.
\label{U1Cgene}
\eeq
As before, we are properly normalizing all of the linear combinations
of charges so that massless states will have conformal dimension one.
Since the $L=1,2,3$ fermions all have identical
boundary conditions in every sector, the
squared charge polynomials corresponding to these factors
are then simply given by
\beqn
         Q_{SU(3)}^2 &=&Q_1^2~-~Q_1Q_2\nonumber\\
         Q_{U(1)_C}^2&=&Q_1^2~+~2\,Q_1Q_2~.
\label{SU3poly}
\eeqn
Likewise, the traceless $SU(2)$ generator is
\beq
               Q_{SU(2)}^{(1)}~=~{1\over{\sqrt2}}\,(Q_4-Q_5) ~,
\label{SU2gene}
\eeq
and the corresponding orthogonal $U(1)_L$ generator
is the trace of the original $U(2)_L$ symmetry:
\beq
         Q_{U(1)_L}^{(1)}~=~ {1\over \sqrt{2}}\,(Q_4 +Q_5)~.
\label{U1L}
\eeq
Because the $L=4,5$ fermions also have identical
boundary conditions in all sectors, the
polynomial insertions in the modified partition function are simply
given by
\beqn
   Q_{SU(2)}^2&=&Q_4^2~-~Q_4Q_5 \nonumber\\
   Q_{U(1)_L}^2&=&Q_4^2~+~Q_4Q_5~.
\label{SU2poly}
\eeqn

In this model, the electroweak  hypercharge is a combination of $U(1)_C$ and
$U(1)_L$,
\beq
          U(1)_Y~=~{1\over3}\,U(1)_C~+~{1\over2}\,U(1)_L~,
\label{U1Y}
\eeq
where the coefficients in this linear combination are appropriate for the
 {\it un}\/-normalized traces $U(1)_C$ and $U(1)_L$.
Thus, the properly normalized charge polynomial corresponding to $U(1)_Y$ is
simply
\beqn
    Q_{U(1)_Y}^2&=&{6\over5}
     \left({{Q_1+Q_2+Q_3}\over3}+{{Q_4+Q_5}\over2}\right)^2
                    \nonumber\\
             &=&{6\over5}\left({1\over3}\,Q_1^2+{2\over3}\,Q_1Q_2+2\,Q_1Q_4+
               {1\over2}\,Q_4^2+{1\over2}\,Q_4Q_5\right)~,
\label{U1poly278}
\eeqn
where in the second line we have again used the fact that the boundary
conditions within
each of the fermion sets $L=1,2,3$ and $L=4,5$ are identical.

\bigskip
\leftline{\it Threshold Contributions:  Results of Calculation}
\medskip

Given the charge polynomials $Q_{SU3)}^2$, $Q_{SU(2)}^2$,
and $Q_{U(1)_Y}^2$ for this model, we see that, as anticipated,
we need to calculate the contributions from all five of the ``basis'' charge
insertions
$\lbrace {Q_1}^2, Q_1 Q_2, Q_1 Q_4, {Q_4}^2, Q_4 Q_5\rbrace$.
Following the same procedure as outlined in the previous section, we
calculate the $b_{00}$ coefficients and
integrals $\Delta$ corresponding to each insertion, with
the results:
\beq
\begin{tabular}{c|cc}
 ~ & $b_{00}$ & $\Delta$  \\
\hline
${Q_1}^2$ &   $3.5$ &   $       9.51642$ \\
$Q_1 Q_2 $ &  $    4.5 $  &  $     1.53625 $  \\
$Q_1 Q_4 $ &  $    0.5 $  &  $      -0.291404$   \\
${Q_4}^2  $&  $     9.5$  &  $     12.951  $       \\
$Q_4 Q_5 $ &  $    4.5 $  &  $     1.53625 $  \\
\end{tabular}
\label{results2}
\eeq

As an aside, we note that
the modified partition function $B$ for the
insertion of $Q_1 Q_2$ is the {\it same}\/ as that
with the insertion of $Q_4 Q_5$.  This explains their
identical values of $b_{00}$ and $\Delta$.  Moreover,
this is also the same function $B$ which appeared for the $Q_1 Q_2$
insertion in the flipped $SU(5)$
model.  This is a reflection of the similarity of their
underlying free-fermionic structures.

\bigskip
\leftline{\it Self-Consistency Checks}
\medskip

Let us now verify the self-consistency of the above values of $b_{00}$.
Indeed, as we shall see, these checks become increasingly non-trivial
as the models become more complex.

For this model, the full massless spectrum was presented in
Ref.~\cite{EU}. Here we list only the states and their non-trivial charges
under $SU(3)_C\times SU(2)_L\times U(1)_C\times U(1)_L$.
The sectors $\b_1$, $\b_2$, and $\b_3$
produce three ${\bf 16}$ representations of $SO(10)$, decomposed under
$SU(3)_C\times SU(2)_L\times U(1)_C\times U(1)_L$.
The Neveu-Schwarz sector produces three pairs of electroweak
doublets with $(U(1)_C,U(1)_L)$ charges $(0,\pm1)$.
The sector $\b_1+\b_2+\alpha+\beta$ produces an additional
pair of electroweak doublets and a pair of color triplets
with $(U(1)_C,U(1)_L)$ charges $(0,\pm1)$
and $(\pm1,0)$ respectively.
Denoting the sector ${\bf 1}+\b_1+\b_2+\b_3$ as $I$,
and introducing the notation $(+I)$ to indicate the two
sectors with and without the vector $I$ added,
we find that the four sectors $\b_2+\b_3+\alpha\pm\gamma(+I)$ produce
an additional pair of color triplets with
$U(1)$ charges $(\pm{1\over4},\pm{1\over2})$,
and eight $SU(3)_C\times SU(2)_L\times U(1)_Y$ singlets with
$U(1)$ charges
$(\pm{3\over4},\mp{1\over2})$.
The sectors $\b_1+\b_3+\alpha\pm\gamma(+I)$
produce an additional pair of electroweak doublets with
$U(1)$ charges
$(\pm{1\over4},\pm{1\over2})$,
and fourteen $SU(3)_C\times SU(2)_L$ singlets with
$U(1)$ charges
$(\pm{3\over4},\mp{1\over2})$.
Finally,
the sectors $\b_1+\b_{2,3}+\alpha\pm\gamma$,
$\b_2+\b_{3}+\alpha\pm\gamma$,
and $\b_1+\b_2+\b_3+\alpha+\beta\pm\gamma(+I)$ produce a total
of 32 $SU(3)_C\times SU(2)_L$ singlets with
$U(1)$ charges
$(\pm{3\over4},\pm{1\over2})$.

Given this matter content, the one-loop
beta-functions coefficients are determined as follows.
In general, we have
\beqn
        b_{SU(3)}&=&-9~+~2\,N_g~+~N_3\nonumber\\
        b_{SU(2)}&=&-6~+~ 2\,N_g~+~N_2\nonumber\\
        b_{U(1)_Y}&=&{1\over k_1}\, {\rm Tr}\,Q_{U(1)_Y}^2
\label{betafun278}
\eeqn
where $N_g$ is the number of ${\bf 16}$
representations of $SO(10)$ in the massless spectrum,
$N_3$ in the number of color triplets in
vector-like representations, and $N_2$ is the number of electroweak doublet
pairs.
The trace in the $U(1)$ beta-function is taken over the entire massless
spectrum,
and the $U(1)_Y$ normalization is fixed by the standard $SO(10)$ embedding.
Since $U(1)_C$ and $U(1)_L$ have the standard $SO(10)$ embedding,
$U(1)_Y$ has the standard $SO(10)$ normalization, $k_1={5/3}$.
Using $N_g=3$, $N_3=2$, $N_2=5$, and the
$U(1)$ charges given above, we then find
\beqn
         &&b_{SU(3)}=-1~,~~~b_{SU(2)}=5~,\nonumber\\
         &&b_{U(1)_C}=12.5~, ~~~b_{U(1)_L}=14~,~~~b_{U(1)_Y}=14.6~.
\label{betafuncoef278}
\eeqn

It is now straightforward to compare these five results with
the values of $b_{00}$ listed in (\ref{results2}).
Indeed, given the expressions for the charge polynomials
in (\ref{SU3poly}), (\ref{SU2poly}), and (\ref{U1poly278}),
we find that in each case, the appropriate linear combinations of these values
of
$b_{00}$ agree with the values obtained from the massless spectrum.
This check is extremely non-trivial, essentially verifying
not only the known massless spectrum and our evaluation of the modified
partition
functions $B(\tau)$ as sums over all of the (thousands of) sectors, but also
verifying the $U(1)$ normalizations which we determined through other means.
This is therefore an important consistency check on our analysis.

\bigskip
\leftline{\it Final Threshold Corrections}
\medskip

Given the charge polynomials in (\ref{SU3poly}), (\ref{SU2poly}),
   and (\ref{U1poly278}),
we take the appropriate linear combinations of the five values of $\Delta$
listed in (\ref{results2}) in order to obtain
the relative string threshold corrections for the
$SU(3)_C$, $SU(2)_L$, and $U(1)_{\hat Y}$ group factors
(where $\hat Y$ refers to the {\it normalized}\/ weak hypercharge).
In this way we obtain the relative thresholds
\beq
     \Delta_{U(1)_{\hat Y}}-\Delta_{SU(3)}~=~  5.0483~,  ~~~~~~~
     \Delta_{U(1)_{\hat Y}}-\Delta_{SU(2)}~=~  1.6137~.
\label{DELTAS278}
\eeq
In Sect.~5 we will examine the effects of these threshold
corrections on the experimental parameters.

\subsection{Second SU(3) $\times$ SU(2) $\times$ U(1) model}

We now turn to a different $SU(3)\times SU(2)\times U(1)$
model \cite{GCU,custodial},
one whose analysis is substantially more complex due to the fact that portions
of its gauge group are realized as {\it enhanced}\/ symmetries.
In particular, this means that some of the corresponding
gauge bosons originate not in the Neveu-Schwarz sector of the theory,
but rather in various additional twisted sectors.
Therefore, for the sake of clarity, we shall first
discuss the origin and structure of the gauge symmetry
in this model before proceeding with the analysis.
The parameters defining this model are given in the Appendix.

\bigskip
\noindent{\it The Gauge Structure of the Model}
\medskip

Due to the enhanced gauge symmetry of this model, it turns out
that the gauge group is ultimately realized not only through the five
complex fermions listed in (\ref{firstfive}), but also through
the following additional five fermions denoted $L=6,...,10$:
\beqn
       L=6 ~&\Longleftrightarrow&~ \ell_i=(25,31)
    ~~~\longleftarrow ~{\rm fermion}~ \overline{y}^3 + i \overline{y}^6
                         \nonumber\\
       L=7 ~&\Longleftrightarrow&~ \ell_i=(21,30)
    ~~~\longleftarrow ~{\rm fermion}~ \overline{y}^4 + i \overline{\omega}^5
                         \nonumber\\
       L=8 ~&\Longleftrightarrow&~ \ell_i=(24,28)
    ~~~\longleftarrow ~{\rm fermion}~ \overline{\omega}^2 + i
       \overline{\omega}^4 \nonumber\\
       L=9 ~&\Longleftrightarrow&~ \ell_i=(41,57)
    ~~~\longleftarrow ~{\rm fermion}~ \overline{\phi}^1
                         \nonumber\\
       L=10 ~&\Longleftrightarrow&~ \ell_i=(48,64)
    ~~~\longleftarrow ~{\rm fermion}~ \overline{\phi}^8~.
\eeqn
As before, the indices $\ell_i$ refer to the ordering of the real
fermions presented in the Appendix, and the labels $\overline{y}^i$,
$\overline{\omega}^i$, and $\overline{\phi}^i$ refer to the discussion
in Sect.~2.
Three other complex fermions which we will also need to consider in our
discussion
(although not in our eventual calculations) are
\beqn
       L=11 ~&\Longleftrightarrow&~ \ell_i=(38,54) ~~~\longleftarrow ~{\rm
fermion}~ \overline{\eta}^1
                         \nonumber\\
       L=12 ~&\Longleftrightarrow&~ \ell_i=(39,55) ~~~\longleftarrow ~{\rm
fermion}~ \overline{\eta}^2
                         \nonumber\\
       L=13 ~&\Longleftrightarrow&~ \ell_i=(40,56) ~~~\longleftarrow ~{\rm
fermion}~ \overline{\eta}^3
        ~.
\eeqn

We begin by outlining the various $U(1)$ factors which will appear.
First, as before, there are the $U(1)_C$ and $U(1)_L$ factors
which are respectively the traces of the larger $U(3)_C$ and
$U(2)_L$ symmetries.  These correspond respectively  to the $L=\lbrace
1,2,3\rbrace$ and
$L=\lbrace 4,5\rbrace$ sets of complex fermions.
Next, there are $U(1)$ factors which correspond
to each of these additional complex fermions $L=6,...,13$ individually.
We shall follow the literature in naming these $U(1)$'s as follows:
\beqn
          &&(L=6) ~\Longrightarrow~ U(1)_4 ~, ~~~~~~~~(L=10) ~\Longrightarrow~
U(1)_9 ~, \nonumber\\
          &&(L=7) ~\Longrightarrow~ U(1)_5 ~, ~~~~~~~~(L=11) ~\Longrightarrow~
U(1)_1 ~,  \nonumber\\
          &&(L=8) ~\Longrightarrow~ U(1)_6 ~, ~~~~~~~~(L=12) ~\Longrightarrow~
U(1)_2 ~,  \nonumber\\
          &&(L=9) ~\Longrightarrow~ U(1)_7 ~, ~~~~~~~~(L=13) ~\Longrightarrow~
U(1)_3 ~.
\eeqn

In this model,
the observable gauge group formed by the gauge bosons from the Neveu-Schwarz
sector
alone is
\beq
       SU(3)_C \times SU(2)_L \times U(1)_C \times U(1)_L \times
U(1)_{1,2,3,4,5,6}~.
\label{originalgroup}
\eeq
However, in this model a new feature arises due to the appearance of two
additional gauge bosons from the twisted sector ${\bf1}+\alpha+2\gamma$
\cite{custodial},
and two corresponding new generators.
These new gauge bosons are singlets of the non-Abelian group, but carry $U(1)$
charges.
However, referring to these new generators as $T^\pm$, it turns out that we can
define the linear combination
\beq
    T^3 ~\equiv~ {1\over4}\biggl\lbrack U(1)_C+U(1)_4+U(1)_5+U(1)_6+U(1)_{7} -
U(1)_9\biggr\rbrack
\label{T3SU2C}
\eeq
in such a way that the three generators $\lbrace T^3,T^\pm\rbrace$
together form the enhanced symmetry group $SU(2)$.
Due to the custodial role played by this additional $SU(2)$ in this model,
we shall refer to this new factor as $SU(2)_{\rm cust}$.
Thus, we find that the original observable symmetry group (\ref{originalgroup})
of this model has been enhanced to
\beq
       SU(3)_C \times SU(2)_L \times SU(2)_{\rm cust} \times
             U(1)_{C'} \times U(1)_L
           \times U(1)_{1,2,3} \times U(1)_{4',5',7''}
\label{enhancedgroup}
\eeq
where (again following the nomenclature in the literature) we have chosen to
define
the following linear combinations of remaining $U(1)$ factors,
each of which is orthogonal to $T^3$:
\beqn
   U(1)_{C^\prime}&\equiv&{1\over3}\,U(1)_C-{1\over2}\,U(1)_{7} +{1\over
2}\,U(1)_9~\nonumber\\
   U(1)_{4^\prime}&\equiv&U(1)_4-U(1)_5\nonumber\\
   U(1)_{5^\prime}&\equiv&U(1)_4+U(1)_5-2\,U(1)_6\nonumber\\
   U(1)_{7^{\prime\prime}}&=&U(1)_C-{5\over3}\biggl\lbrack
U(1)_4+U(1)_5+U(1)_6\biggr\rbrack
           +U(1)_{7} -U(1)_9~.
\label{cuscom}
\eeqn

The final issue is the definition of the electroweak hypercharge in this model.
In fact, due to the extended symmetry, we now have the freedom to define the
weak
hypercharge in several ways.  One option is to define the weak hypercharge
just as in (\ref{U1Y}) for the previous $SU(3)\times SU(2)\times U(1)$ model.
However, in the present model, the $U(1)_C$ symmetry is now only {\it part}\/
of the extended custodial symmetry $SU(2)_{\rm cust}$.  Indeed, expressing
$U(1)_C$ in terms of the new linear combinations defined above, we have
\beq
    {1\over3}\,U(1)_{C}~=~{2\over5}\biggl\lbrace U(1)_{C^\prime}+
     {5\over{16}}\,\biggl\lbrack T^3+{3\over5}\,U_{7^{\prime\prime}}
\biggr\rbrack
        \biggr\rbrace~.
\label{U1Cin274}
\eeq
Thus $U(1)_Y$, by depending on the $T^3$ of the custodial $SU(2)$,
is no longer orthogonal to $SU(2)_{\rm cust}$.
We must therefore instead define the new linear combination with this term
removed,
\beqn
        U(1)_{Y'} &\equiv& U(1)_Y - {1\over 8}\,T^3 \nonumber\\
        &=& {1\over 2}\,U(1)_L + {5\over{24}}\,U(1)_C \nonumber\\
         &&~~~~~-{1\over8}\,\biggl\lbrack
U(1)_4+U(1)_5+U(1)_6+U(1)_7-U(1)_9\biggr\rbrack~,
\label{U1pin274}
\eeqn
so that the weak hypercharge is expressed in terms of $U(1)_{Y'}$ as
\beq
        U(1)_{Y} = U(1)_{Y'} + {1\over 2}\,T^3 ~~~~\Longrightarrow~~~~
           Q_{\rm e.m.} = T^3_L + Y = T^3_L + Y' + {1\over 2}\,T^3_{\rm
cust}~.\label{Qemin274}
\eeq
The final observable gauge group then takes the form
\beq
       SU(3)_C \times SU(2)_L \times SU(2)_{\rm cust}\times U(1)_{Y'} ~\times
        ~\biggl\lbrace ~{\rm seven~other~}U(1){\rm ~factors}~\biggr\rbrace~.
\label{finalgroup}
\eeq
Of course, these remaining seven $U(1)$ factors must be chosen
as linear combinations of the previous $U(1)$ factors so as to be orthogonal
to the each of the other factors in (\ref{finalgroup}).

\bigskip
\noindent{\it Determining the Charge Polynomials}
\medskip

Given the gauge structure outlined above,
we now determine the corresponding charge polynomials for the
physically relevant $SU(3)_C$, $SU(2)_L$, $SU(2)_{\rm cust}$, and $U(1)_{Y'}$
factors.

The charge polynomials for the color $SU(3)_C$ and the electroweak $SU(2)_L$
gauge groups are the same as in the previous $SU(3)\times SU(2)\times U(1)$
model:
\beqn
      Q^2_{SU(3)}&=& Q_1^2 ~-~ Q_1 Q_2 \nonumber\\
      Q^2_{SU(2)}&=& Q_4^2 ~-~ Q_4 Q_5~.
\eeqn
The charge polynomial for the custodial $SU(2)_{\rm cust}$ gauge group is
obtained from the expression (\ref{T3SU2C}) for its diagonal generator,
and, with proper normalization, is given by
\beqn
       Q^2_{SU(2)_{\rm cust}}&=&{1\over8}
    \biggl\lbrace
     3Q_1^2~+~6\,(Q_1Q_2+Q_1Q_6+Q_1Q_7+Q_1Q_8+Q_1Q_9-Q_1Q_{10})\nonumber\\
  &&~~+\, Q_6^2~+~2\,(Q_6Q_7+Q_6Q_8+Q_6Q_9-Q_6Q_{10})\nonumber\\
  &&~~+\, Q_7^2~+~2\,(Q_7Q_8+Q_7Q_9-Q_7Q_{10}) ~+~
            Q_8^2~+~2\,(Q_8Q_9-Q_8Q_{10})\nonumber\\
  && ~~+~Q_9^2 ~-~2\,Q_9Q_{10}~+~ Q_{10}\biggr\rbrace~.
\label{Q2ofcusSU2}
\eeqn
Likewise, the normalized charge polynomial combination for $U(1)_{Y^\prime}$ is
found to be
\beqn
      Q^2_{U(1)_{Y^\prime}}&=&
     {{24}\over{17}} \biggl\lbrace
    {{25}\over{192}}(Q_1^2+2Q_1Q_2)+{5\over4}Q_1Q_4 \nonumber\\
     &&~~~~-~{5\over{32}}(Q_1Q_6+Q_1Q_7+Q_1Q_8+Q_1Q_9-Q_1Q_{10})\nonumber\\
     &&~~~~+~ {1\over2}Q_4^2+{1\over2}Q_4Q_5-
      {1\over4}(Q_4Q_6+Q_4Q_7+Q_4Q_8+Q_4Q_9-Q_4Q_{10})\nonumber\\
     &&~~~~+~ {1\over{64}}\biggl\lbrack
     Q_6^2+2(Q_6Q_7+Q_6Q_8+Q_6Q_9-Q_7Q_{10}) + Q_7^2 \nonumber\\
     &&~~~~~~+~ 2(Q_7Q_8+Q_7Q_9-Q_7Q_{10})+ Q_8^2
    \nonumber\\
     &&~~~~~~+~ 2(Q_8Q_9-Q_8Q_{10})+ Q_9^2-2Q_9Q_{10}+ Q_{10} \biggr\rbrack
                    \biggr\rbrace~.
\label{Q2ofYp}
\eeqn
In assembling these charge polynomials we have made use of identical
boundary conditions wherever possible in order to simplify these expressions.

\bigskip
\leftline{\it Threshold Contributions:  Results of Calculation}
\medskip

Given the above polynomials, we see that we must now
calculate the contributions from a basis set of 30
different charge insertions corresponding to the ten complex worldsheet
fermions $L=1,...,10$.
Our results are as follows:
\beq
\begin{tabular}{c|cc||c|cc}
   ~ & $b_{00}$ & $\Delta$  & ~ & $b_{00}$ & $\Delta$  \\
\hline
$  {Q_1}^2   $&$   1.75   $&$  10.2704      $&$   {Q_6}^2   $&$
6.^\ast $&$     12.4133        $\\
$  Q_1 Q_2  $&$     1.75 $&$   1.57939         $&$     Q_6 Q_7   $&$      1.
 $&$  0.803486                $\\
$  Q_1 Q_4  $&$     -0.25 $&$     -0.0275848          $&$   Q_6 Q_8   $&$
-0.5^\ast  $&$   0.696872           $\\
$  Q_1 Q_6  $&$     -0.5 $&$    -0.401743            $&$     Q_6 Q_9   $&$
-2.5 $&$   -0.622422                  $\\
$  Q_1 Q_7  $&$    -0.5  $&$    -0.401743           $&$      Q_6 Q_{10} $&$
2.5 $&$  0.622422           $\\
$  Q_1 Q_8  $&$     -1.  $&$    -0.803486      $&$            {Q_7}^2   $&$
 6.^\ast $&$   12.4133      $\\
$  Q_1 Q_9   $&$    -1.75 $&$   -0.193094         $&$        Q_7 Q_8  $&$
-0.5^\ast $&$  0.696872        $\\
$  Q_1 Q_{10} $&$     1.5  $&$   0.165509       $&$           Q_7 Q_9   $&$
   -2.5  $&$   -0.622422            $\\
$  {Q_4}^2  $&$     11.75 $&$   12.7601      $&$             Q_7 Q_{10} $&$
    2.5$&$   0.622422          $\\
$  Q_4 Q_5  $&$   3.75 $&$    1.80007       $&$            {Q_8}^2  $&$
 6.^\ast  $&$   11.3147      $\\
$  Q_4 Q_6  $&$     0. $&$      0.               $&$    Q_8 Q_9   $&$      -1.
$&$   0.582808           $\\
$  Q_4 Q_7  $&$     0. $&$      0.              $&$     Q_8 Q_{10} $&$
1. $&$    -0.582808          $\\
$  Q_4 Q_8  $&$     0.  $&$     0.              $&$     {Q_9}^2     $&$
1.75$&$   10.2704         $\\
$  Q_4 Q_9  $&$    0.25 $&$   0.0275848        $&$            Q_9 Q_{10}  $&$
      -9.5  $&$   -2.43452          $\\
$  Q_4 Q_{10} $&$     -0.5 $&$    -0.0551696       $&$             {Q_{10}}^2
$&$       1.  $&$   10.1876         $\\
\end{tabular}
\label{results4}
\eeq

Note that unlike the simpler previous cases, some of these charge insertions
resulted in modified partition functions $B(\tau)$ containing contributions
from charged unphysical tachyons.  Specifically, in the $q$-expansion of
some of these functions $B(\tau)$, various coefficients $b_{mn}$ with $m<0$ or
$n<0$
are non-zero.  Those cases are indicated with an asterisk following the
corresponding value of $b_{00}$.
This occurrence is not unexpected, however, since such unphysical tachyons are
generically
present (and in fact unavoidable) in generic string models, and are
required for the consistency of the theory \cite{missusy}.
Indeed, they do not lead to any divergence in the corresponding threshold
integrals $\Delta$, since the fact that they are unphysical
({\it i.e.}, with $m\not = n$, $m-n\in \IZ$) implies that
they have no contributions from the region $\tau_2\geq 1$
of the fundamental domain ${\cal F}$
from which infrared divergences might arise.
We point out, however, that any unphysical tachyonic contributions
with $m+n<0$ will render incorrect the expression (\ref{Kapwrong}) which
was proposed in Ref.~\cite{Kaplunovsky}.
We will discuss some further implications of these unphysical
tachyons in Sect.~6.

Also note from the above results that the charges $Q_6$ and $Q_7$
appear interchangeable insofar as their effects
on the modified partition functions $B(\tau)$
are concerned.
Despite this fact, however, it can easily be verified that the
corresponding $L=6$ and $L=7$ worldsheet fermions
do {\it not}\/ share the same boundary conditions.

Finally, we observe that in the cases of the insertions $Q_4 Q_6$, $Q_4 Q_7$,
and $Q_4 Q_8$, the corresponding modified partition functions $B(\tau)$
actually vanish identically.
This occurs because each term in these expressions $B(\tau)$
contains a factor of $\tp{0}{0}$, $\tp{0}{1/2}$, or $\tp{1/2}{0}$.
As can be seen from their definition in (\ref{thetaprime}), these
particular singly primed $\Theta'$-functions vanish.

\bigskip
\leftline{\it Self-Consistency Checks}
\medskip

Given the complexity of this model, it is crucial now more than ever to verify
that our self-consistency checks are satisfied.
The complete massless spectrum of the model is given in Ref.~\cite{custodial}.
The model contains a total of eighteen color triplets, twelve from the
sectors $\b_1$, $\b_2$, and $\b_3$ (which produce the light generations),
and an additional three pairs from the sectors
$\b_{1,2}+\b_3+\beta\pm\gamma$ and ${\bf 1}+\alpha+2\gamma$. Thus,
 from (\ref{betafun278}), we see that the one-loop beta-function coefficient
 $b_{SU(3)_C}$ for the color $SU(3)$ gauge group factor vanishes.
Comparing this against the
above values of $b_{00}$ (in particular, the difference between those
for the $Q_1^2$ and the $Q_1 Q_2$ insertions), we see that agreement is
obtained.

Turning now to the $SU(2)_L$ electroweak symmetry, we recall \cite{custodial}
that this model contains 28 electroweak doublets. Five doublets are obtained
from each of the sectors
$\b_1$, $\b_2$, and $\b_3$. An additional doublet beyond the Minimal
Supersymmetric Standard Model arises
due to the transformation of the lepton left-handed doublet under
the custodial $SU(2)_{\rm cust}$ symmetry. The Neveu-Schwarz sector
and the sector $\b_1+\b_2+\alpha+\beta$ contribute
three and two additional pairs, respectively.
Finally, three additional electroweak doublets are
obtained from each sector, ${\bf 1}+\b_i+\alpha+2\gamma$ for $i=1,2,3$.
Thus, from (\ref{betafun278}), we find $b_{SU(2)_L}=8$.
Consulting the above charge polynomial expression for $SU(2)_L$
and the corresponding entries in (\ref{results4}),
we see that once again the values agree.

Finally, we compare the one-loop
beta-function coefficients
for the custodial $SU(2)_{\rm cust}$ symmetry.
It turns out that this model contains twelve massless doublets under the
custodial $SU(2)_{\rm cust}$ symmetry.
Thus, the one-loop beta-function
coefficient of the $SU(2)_{\rm cust}$ group vanishes. From the
polynomial expression (\ref{Q2ofcusSU2})
and the appropriate entries from (\ref{results4}),
it can be verified that the appropriate linear combination of $b_{00}$-values
vanishes as well.

\bigskip
\leftline{\it Final Threshold Corrections}
\medskip

First we must discuss the Ka\v{c}-Moody factors associated with
the $U(1)$ factors in this model.
In the class of string models we have been examining,
the Ka\v{c}-Moody level of the non-Abelian group factor is always one.
The situation is somewhat more complicated for the $U(1)$ factors, however.
In general, a given $U(1)$ current $U$ will be a combination of the simple
worldsheet
$U(1)$ currents $U_f\equiv f^*f$ corresponding to individual worldsheet
fermions $f$,
and will take the form $U=\sum_f a_f U_f$ where the $a_f$ are certain
model-specific coefficients.
The $U_f$ are each individually normalized to one, so that
$\langle U_f, U_f\rangle=1$. To produce the correct conformal dimension
for the massless states, each of the $U(1)$ linear combinations $U$ must also
be normalized to one.
The proper normalization coefficient for the linear combination
$U$ is thus given by $N=(\sum_f a_f^2)^{-1/2}$, so that
the properly normalized $U(1)$ current ${\hat U}$ is
given by ${\hat U}=N\cdot U$.

Now, in general the Ka\v{c}-Moody level of the $U(1)_Y$ generator
can be deduced from the OPE's between two of the $U(1)$ currents,
and will be
\beq
       k_1~=~ 2\,N^{-2}~ = ~ 2\,\sum_f \, a_f^2~.
\label{k1u1}
\eeq
For a weak hypercharge that is a combination
of several $U(1)$'s with {\it different}\/ normalizations,
the result (\ref{k1u1}) generalizes to
\beq
       k_1 ~=~ \sum_i\, a_i^2 \, k_i~
\label{ky}
\eeq
where the $k_i$ are the individual normalizations
for each of the $U(1)$'s.

Now, in the model analyzed in Sect.~4.2, the $U(1)_Y$ generator is given as a
combination of simple worldsheet
currents that produces the correct weak hypercharges for the Standard Model
particles. Thus, in that case, $k_1$ is simply given by (\ref{k1u1}).
For the weak hypercharges (\ref{U1pin274}) and (\ref{Qemin274})
that appear in this model, however, we instead use (\ref{ky}).
Thus, for this weak hypercharge, we see from (\ref{Qemin274}) and (\ref{ky})
that $k_1=(1/4)k_{2_C}+k_{Y^\prime}=1/4+17/12=5/3$, which is the same as the
standard
$SO(10)$ normalization.

 From (\ref{SU3poly}) and (\ref{SU2poly}),
we then obtain the following relative string heavy threshold corrections for
$SU(3)_C$, $SU(2)_L$, $SU(2)_{\rm cust}$, and $U(1)_{{\hat Y}^\prime}$:
\beqn
 && \Delta_{SU(2)_L} -  \Delta_{SU(3)_C}= 2.2690~,~~~
 \Delta_{U(1)_{\hat Y}} - \Delta_{SU(3)_C}= 5.3767 \nonumber\\
 && \Delta_{U(1)_{{\hat Y}^\prime}} - \Delta_{SU(2)_{\rm cust}}= 2.8587~,~~~
  \Delta_{U(1)_{{\hat Y}}} - \Delta_{SU(2)_{\rm cust}}= 2.4299~.
\label{DELTAS274}
\eeqn
Note that for future convenience [see (\ref{futureeq})], we have defined
\beq
     \Delta_{U(1)_{\hat Y}}~=~{3\over 5}\left({1\over 4}\Delta_{SU(2)_C}+
     {17\over 12}\Delta_{U(1)_{{\hat Y}^\prime}}\right)~,
\eeq
whereupon
\beq
    \Delta_{U(1)_{\hat Y}}-\Delta_{SU(2)_{\rm cust}}~=~
      {17\over 20}\,\left( \Delta_{U(1)_{{\hat Y}^\prime}}
     -\Delta_{SU(2)_{\rm cust}}\right)~ .
\eeq

Alternatively, we can define the weak hypercharge to be the combination
\beq
         U(1)_Y~=~{1\over2}\,U(1)_L~+~ U(1)_{C^\prime}
\eeq
where $U(1)_{C^\prime}$ is given in (\ref{cuscom}).
In this case the Ka\v{c}-Moody levels of
$U(1)_L$ and $U(1)_{C^\prime}$ are 4 and 5/3 respectively, so that $k_1=8/3$.
In this case we find that
$\Delta_{SU(3)_C}$ and  $\Delta_{SU(2)_L}$ are the same as above,
and that the relative threshold for the properly normalized $U(1)_{\hat Y}$ is
\beq
    \Delta_{U(1)_{\hat Y}}-\Delta_{SU(2)_{\rm cust}}~=~ 2.0939~.
\label{deltaywithk1=8/3}
\eeq

\subsection{SO(6) $\times$ SO(4) model}

Next we evaluate the threshold corrections in the $SO(6)\times SO(4)$
string model of Ref.~\cite{ALR}.
The parameters defining this model are also listed in the Appendix.

This string model realizes the Pati-Salam unification scenario, with
gauge group
\beq
       SO(6)_C\times SO(4) ~\simeq~ SU(4)_C \times SU(2)_L \times SU(2)_R~
\eeq
and three generations.
Here we have only listed that portion of the observable gauge group
which is of relevance to our analysis, neglecting both the hidden
symmetries [such as an $SU(8)$], and various $U(1)$'s.
In the construction of this string model,
a basis of nine boundary-condition vectors
with only periodic and anti-periodic boundary conditions is
used. In order to achieve the
reduction to three massless generations,
five of these vectors are taken to be the basis
vectors of the NAHE set, two others are identical to the vectors $\b_4$ and
$\b_5$ which
appear in the flipped $SU(5)$ model, another is similar to the vector $2\gamma$
of the flipped $SU(5)$ model,
and a final vector performs the required symmetry breaking from the larger
$SO(10)$.
Thus, within this construction,
the $SO(6)_C$ gauge symmetry is realized through the $L=1,2,3$ fermions
listed in (\ref{firstfive}), while the $SO(4)$ gauge symmetry
is realized through the $L=4,5$ fermions.

\bigskip
\noindent{\it Determining the Charge Polynomials}
\medskip

Given their similar worldsheet structures,
the charge polynomials corresponding to the group factors of the $SO(6)\times
SO(4)$
model are easily obtained from those of the first $SU(3)\times SU(2)\times
U(1)$ model
presented above in Sect.~4.2.
In particular, the three generators of the $SO(6)$ group are simply
the generators $Q_{SU(3)}^{(1)}$, $Q_{SU(3)}^{(2)}$, and $Q_{U(1)_C}^{(1)}$
which appear in (\ref{SU3gene}) and (\ref{U1Cgene}), from which we find
that
\beq
       Q_{SO(6)}^2 ~=~ Q_1^2~.
\label{polynomialofso6}
\eeq
Note that all of the non-diagonal terms $Q_L Q_M$ with $L\not=M$ have
cancelled in this case.
Likewise,
the generators of the $SO(4)$ symmetry
(or equivalently, of the $SU(2)_L$ and $SU(2)_R$ symmetry)
are respectively
the generators $Q_{SU(2)}^{(1)}$ and $Q_{U(1)_L}^{(1)}$
which appear in (\ref{SU2gene}) and (\ref{U1L}).
Their squares are therefore, as before,
\beqn
       Q_{SU(2)_L}^2 &=& Q_4^2 ~-~ Q_4 Q_5 \nonumber\\
       Q_{SU(2)_R}^2 &=& Q_4^2 ~+~ Q_4 Q_5~.
\eeqn

\bigskip
\leftline{\it Threshold Contributions:  Results of Calculation}
\medskip

Given these charge insertion polynomials, we see that
we have the basis set $\lbrace Q_1^2, Q_1 Q_2, Q_4^2,
Q_4 Q_5\rbrace$ of charge polynomials.
Our results are as follows:

\beq
\begin{tabular}{c|cc}
 ~ & $b_{00}$ & $\Delta$  \\
\hline
${Q_1}^2     $&$      3.    $&$     11.4457    $\\
$Q_1 Q_2   $&$        0.    $&$      0.    $\\
${Q_4}^2    $&$        13.  $&$     16.708    $\\
$Q_4 Q_5   $&$         -4.   $&$       -3.21395    $\\
\end{tabular}
\label{results5}
\eeq
Note that for $Q_1 Q_2$,
the modified partition function $B(\tau)$ vanishes identically.
As before, this occurs because every term in $B(\tau)$
contains a factor of either $\tp{0}{0}$, $\tp{0}{1/2}$, or $\tp{1/2}{0}$.
Each of these $\Theta'$-functions vanishes.

\bigskip
\leftline{\it Self-Consistency Checks}
\medskip

We now consider the massless spectrum of this model,
and focus on the representations of the $SO(6)\times SO(4)$
[or $SU(4)\times SU(2)_R\times SU(2)_L$] observable gauge symmetry.
The sectors $\b_{1,...,5}$ produce
three generations transforming as
$(4,2,1)\oplus({\bar4},1,2)$ and two pairs transforming as
$(4,1,2)\oplus({\bar4},1,2)$ under
$SU(4)\times SU(2)\times SU(2)$.
The Neveu-Schwarz sector produces a pair of Higgs doublets
transforming as  $(1,2,2)$ and four sextet fields transforming as $(6,1,1)$,
while the sector $S+\b_4+\b_5$ produces an additional pair of Higgs doublets
transforming as $(1,2,2)$. The sectors
$\b_{1,4}+\alpha$, $\b_1+\b_2+\b_4+\alpha$, $\b_2+\b_3+\b_5+\alpha$,
and $\b_1+\b_4+\b_5+\alpha$ produce a total of ten doublets of
$SU(2)_L$ and ten doublets of $SU(2)_R$. Finally, the sector
$S+\b_2+\b_4+\alpha$ produces an $SU(4)$ multiplet transforming as
$(4,1,1)\oplus({\bar4},1,1)$.

The one-loop beta-function coefficients of
$SU(4)\times SU(2)\times SU(2)$ are given by
\beqn
        b_{SU(4)}&=&-12~+~{1\over2}\,(n_4+n_{\bar4})~+~n_6\nonumber\\
        b_{SU(2)_{L,R}}&=&-6~+~{1\over2}\,n_{2_{L,R}}
\label{ALRbfc}
\eeqn
where $n_4$ and $n_{\bar4}$ are the total number of massless ${\bf 4}$
and $\overline{\bf 4}$ representations, where $n_6$ is the number of sextet
representations,
and where $n_{2_{L,R}}$ are the total number of doublets of $SU(2)_{L,R}$
respectively.
In this model, $n_4+n_{\bar4}=22$ (twenty from the
sectors $\b_{1,...,5}$ and two from the sector $S+\b_2+\b_4+\alpha$),
while $n_6=4$ (all sextets from the Neveu-Schwarz sector), and
$n_{2_L}=46$ and $n_{2_R}=30$.
Thus, we find that the one-loop beta-function coefficients are
\beq
b_{SU(4)}=3~,~~~~~b_{SU(2)_L}=17~,~~~~~b_{SU(2)_R}=9~.
\label{ALRbfcnv}
\eeq
With the values of $b_{00}$ given in (\ref{results5}), we see
that agreement is again obtained.

\bigskip
\leftline{\it Final Threshold Corrections}
\medskip

Given the above results, it is straightforward to calculate the
relative threshold corrections for each of the Planck-scale
group factors in this model,
yielding
\beq
      \Delta_{SU(2)_L}-\Delta_{SU(4)}~=~ 8.4763~, ~~~~~~
      \Delta_{SU(2)_R}-\Delta_{SU(4)}~=~ 2.0483~.\label{DELTASO64}
\eeq

We may also consider the threshold correction for
the low-energy weak hypercharge.
In the Pati-Salam unification scenario \cite{PS},
there is an intermediate scale at which a further symmetry-breaking
occurs, such that
\beqn
      SU(4) &\longrightarrow& SU(3)_C \times U(1)_{B-L}\nonumber\\
      SU(2)_R & \longrightarrow& U(1)_{R}~.
\eeqn
Here the $U(1)_{B-L}$ symmetry
is generated by the $(B-L)$ generator of the original $SU(4)$ symmetry,
while the remaining $U(1)_R$ symmetry
is generated by the diagonal generator $T^3_R$ of the
original $SU(2)_R$ symmetry.
The normalized low-energy weak hypercharge is then defined
as a linear combination of these two $U(1)$ factors,
\beq
      U(1)_{\hat Y}~\equiv~ \sqrt{2\over 5}\,U(1)_{B-L} ~+~ \sqrt{3\over
5}\,U(1)_{R}~.
\eeq
  From this relation we deduce that
\beq
           \Delta_{U(1)_{\hat Y}}~=~ {2\over5}\, \Delta_{SU(4)}~+~
         {3\over5}\,\Delta_{SU(2)_R}~,
\eeq
yielding
\beq
            \Delta_{U(1)_{\hat Y}} - \Delta_{SU(4)} ~=~
              {3\over 5}\,\left( \Delta_{SU(2)_R} - \Delta_{SU(4)}\right)~=~
1.229~.
\label{deltasforso64model}
\eeq

\subsection{Non-supersymmetric version of first
              SU(3) $\times$ SU(2) $\times$ U(1) model}

In order to assess the role that spacetime supersymmetry might play in
affecting the overall magnitude of these threshold correction expressions,
we have constructed a non-supersymmetric version of the first
$SU(3)\times SU(2)\times U(1)$ model analyzed above.  This was achieved by
altering the GSO projections of that model
in such a way that spacetime supersymmetry is broken at the Planck
scale ({\it e.g.}, the gravitinos are projected out of the spectrum),
but no physical tachyons are introduced.
Despite the absence of spacetime supersymmetry, the spectra of such
tachyon-free
models have a number of interesting properties,
such as the appearance of a
residual so-called ``misaligned supersymmetry'' \cite{missusy,missusyreview},
and the vanishing of certain mass supertraces \cite{supertraces}.
The complete parameters defining this new non-supersymmetric model are listed
in the Appendix.

Note that in performing our calculations for this non-supersymmetric model,
we have continued to use the threshold-correction expressions (\ref{Deltadef})
as originally given by Kaplunovsky \cite{Kaplunovsky}.
It is claimed in Ref.~\cite{Kaplunovsky} that these expressions hold for all
$(1,0)$ string vacua, regardless of whether or not spacetime supersymmetry
is present.   However, it has recently been shown \cite{Kiritsis}
that for non-supersymmetric vacua, there are also additional terms arising
from gravitational interactions such as, {\it e.g.}, dilaton
tadpoles.\footnote{
      We thank E.~Kiritsis for discussions on this point.}
Such terms are absent in the supersymmetric case, and do not
arise in a straightforward ``field-theoretic'' derivation in which such
gravitational interactions are not present, or in Kaplunovsky's
stringy generalization of such a field-theoretic calculation.
Nevertheless, they are present in the full string calculation,
and are found \cite{Kiritsis} to contribute to the
threshold corrections at the same order in $\alpha'$.
In this section, however, our goal is merely to construct a toy
non-supersymmetric model, and to determine how significantly the
original expression of Kaplunovsky changes when the
traces are evaluated over its non-supersymmetric spectrum.  Consequently
we are neglecting those additional terms which would need to be included if we
were performing a complete analysis of the gauge coupling unification in
such a model.  We hope to
perform such a calculation in the future.

\bigskip
\noindent{\it Determining the Charge Polynomials}
\medskip

Because the boundary conditions of the worldsheet fermions are unaltered
relative to those of the supersymmetric version of this model,
both the gauge group and its realization in terms of free fermions remain
intact.  Therefore the charge polynomials for the various group factors
are unchanged relative to the supersymmetric case in
(\ref{SU3poly}), (\ref{SU2poly}), and (\ref{U1poly278}).

\bigskip
\leftline{\it Threshold Contributions:  Results of Calculation}
\medskip

We then repeat the calculation with the same five ``basis'' charge insertions.
Since the boundary conditions for all fermions are the same for all sectors
as they were in the previous spacetime-supersymmetric case, all changes in the
results
relative to that case arise due to the changes in the GSO projection phases.
In particular, certain
particles which had previously been in the spectrum ({\it e.g.}, various
superpartners)
have been projected out, while new particles which had previously been
projected out
now appear.  Our results are as follows:
\beq
\begin{tabular}{c|cc}
 ~ & $b_{00}$ & $\Delta$  \\
\hline
$   {Q_1}^2     $&$      6.5^\ast   $&$      11.322   $\\
$   Q_1 Q_2     $&$        5.5^\ast   $&$      1.56759    $\\
$   Q_1 Q_4     $&$         1.5^\ast  $&$       -0.260057   $\\
$   {Q_4}^2     $&$         8.5^\ast  $&$      14.3152    $\\
$   Q_4 Q_5     $&$        5.5^\ast   $&$      1.56759    $\\
\end{tabular}
\label{results3}
\eeq
\smallskip

As in previous cases,
the asterisk indicates contributions from unphysical tachyons.
In fact, the existence of such contributions from unphysical tachyons
is {\it expected}\/ in non-supersymmetric string models, since the breaking
of spacetime supersymmetry in many
cases ensures that the contributions from such
unphysical bosonic tachyons will not be cancelled by those from any
unphysical fermionic tachyonic superpartners.
As before, however,
the existence of such contributions does not lead to any
divergence in the corresponding threshold integrals $\Delta$.
We will discuss the appearance of these unphysical tachyons in Sect.~6.

\bigskip
\leftline{\it Self-Consistency Checks}
\medskip

We now turn to the values of $b_{00}$.
Although the gauge group and charge polynomials are not
altered relative to the supersymmetric version of this model,
the breaking of spacetime supersymmetry
does modify the spectrum.
Nevertheless, calculating the one-loop beta-function coefficients
corresponding to the new non-supersymmetric spectrum,
we now obtain
\beqn
       && b_{SU(3)}=1~,~~~b_{SU(2)}=3~,\nonumber\\
       && b_{U(1)_C}=17.5~,~~~b_{U(1)_L}=14~,~~~b_{U(1)_Y}=19~.
\label{betafuncoef278nonsusy}
\eeqn
Comparing this to the appropriate linear combinations of the
above values of $b_{00}$, we see that once again agreement is obtained.

\bigskip
\leftline{\it Final Threshold Corrections}
\medskip

Using (\ref{SU3poly}), (\ref{SU2poly}), and (\ref{U1poly278}),
we obtain the relative string threshold corrections for
the $SU(3)_C$, $SU(2)_L$, and normalized $U(1)_{\hat Y}$ group factors,
\beq
      \Delta_{U(1)_{\hat Y}} - \Delta_{SU(3)} ~=~ 4.934 ~,~~~~~~
      \Delta_{U(1)_{\hat Y}} - \Delta_{SU(2)} ~=~ 1.9408~.
\label{DELTAS278nonsusy}
\eeq
We emphasize once again that these results are to be interpreted
as numerical evaluations of the expression given in (\ref{Deltadef}).
As discussed above, a complete evaluation of the threshold calculations
for such a toy non-supersymmetric model would require the inclusion
of additional terms as well \cite{Kiritsis}.

We see from these results, then, that breaking supersymmetry by
a GSO projection has no appreciable effect on the magnitude of
the expressions (\ref{Deltadef}).
Thus the assumption of spacetime supersymmetry
in these models is not what sets the magnitude of
these threshold correction contributions.
In Sect.~6 we shall discuss the underlying
reasons why these string-theoretic threshold corrections are
so small.

\input epsf.tex

\setcounter{footnote}{0}
\section{Confrontation with Experiment}

In the previous section we calculated the threshold corrections
$\Delta_G$ within a wide class of realistic free-fermionic string models.
In this section we will analyze the effects of these threshold
corrections on the experimentally observed parameters which are measured at low
energies.
In so doing, we shall also consider the other effects which might affect
the running of the couplings, including non-standard hypercharge
normalizations,
light SUSY thresholds, intermediate
gauge structure, and extra non-MSSM matter.

In order to study the effects of the heavy string threshold corrections
we have calculated in Sect.~4, one method of approach
might be to absorb their contributions $\Delta_G$
into the logarithms as in (\ref{su5xu1unificationscale}),
and thereby obtain a resulting ``effective'' string unification scale.
Indeed, this is the course followed in Ref.~\cite{AELN}.
One would then compare this modified string unification scale with
the usual value given for $M_{\rm GUT}$, which in turn is obtained
by extrapolating the low-energy data under certain field-theoretic
assumptions.
However, we shall instead choose a more general path which allows us
to take into account a number of different factors which might affect the
analysis over the whole energy range from the $Z$ scale to the string
scale.
In particular, starting from the string scale,
we shall solve the one-loop gauge-coupling renormalization
group equations (RGE's) for the measured parameters $\sin^2\theta_W(M_Z)$ and
$\alpha_3(M_Z)$
directly at the $Z$-scale, and thereby obtain the explicit dependence of these
parameters
on the string threshold corrections.
In these RGE's, we shall include
\begin{itemize}
\item  the corrections due to second-loop effects.
       We shall see, in fact, that the two-loop contributions
       are quite sizable, and can in some instances alter our results.
\item  the corrections due to Yukawa couplings.
\item  the corrections due to scheme conversion.   The so-called
       $\overline{DR}$-scheme is used in the definition of the
       string heavy threshold corrections, and in the supersymmetric RGE's.
       However, the $\overline{MS}$-scheme is the one used in extracting
       the low-energy parameters from experiments.
       One must therefore include explicit scheme-conversion terms in
       the RGE's.
\end{itemize}
Furthermore, by explicitly setting up the RGE's in this way, we
will also be able to include various additional field-theoretic
and string-theoretic
factors that can affect the analysis.
These include:
\begin{itemize}
\item  the effects due to non-standard values of $k_1$, as can occur
         in string theory.  The model-dependent parameter $k_1$ is the
         essentially the normalization of the
         weak hypercharge generator relative to the $U(1)$ generator
         in a unified theory.
         In $SU(5)$ and $SO(10)$ unified models, one has $k_1=5/3$.
         However, in string models, the normalization of the $U(1)$ generators
         is fixed by the requirement that the conformal dimension of
         the massless states be equal to one.  The value $k_1$ is thus the
         relative normalization between the properly normalized
         $U(1)$ generators, and the $U(1)$ generator which produces
         the correct weak hypercharge assignment for the
         Standard Model quarks and leptons.
	 Therefore, depending on how these $U(1)$
         generators are ultimately realized within a given string model,
         the value of $k_1$ may be different from $5/3$.
\item  the effects arising from light SUSY thresholds ({\it i.e.},
          the splittings of the sparticle mass spectrum).
\item  the effects arising from potential additional non-Abelian
	gauge structure at intermediate energy scale.
        Such additional gauge structure
         occurs, for example, in the Pati-Salam unification scenario,
         and in the corresponding $SO(6)\times SO(4)$ string model.
\item	the effects arising from possible
    additional matter thresholds at intermediate energy scales.
\end{itemize}
All of these factors contribute, along with the heavy string thresholds,
to the running of the gauge-couplings, and lead to similar correction terms
in the corresponding RGE's.  Thus, by
starting from the string scale and evolving our couplings directly down to
the $Z$-scale using such corrected RGE's, we can provide a serious
test of whether the predictions of string theory are truly in
accordance with the experimentally observed low-energy data.

\bigskip
\noindent{\it Low-Energy Experimental Inputs}
\bigskip

For our subsequent analysis, the input parameters are the tree-level string
prediction \cite{Kaplunovsky}
\beq
    M_S~\equiv~M_{\rm string}~\approx~g_{\rm
      string}\,\times\,5\,\times\,10^{17} ~{\rm GeV}~,
\eeq
the mass of the $Z$ \cite{expalpha}
\beq
    M_Z ~\equiv~ 91.161\pm 0.031~{\rm GeV}~,
\eeq
and the electromagnetic coupling at the $Z$-scale \cite{expalpha}
\beq
 a^{-1}~\equiv~\alpha_{\rm e.m.}(M_Z)^{-1}~=~ 127.9\pm0.1~.
\label{alphaem}
\eeq
 From the RGE's, we then obtain predictions for $\sin^2\theta_W(M_Z)$ and
$\alpha_{3}(M_Z)$.  The experimental values we wish to obtain
are as follows.
For the $\overline{MS}$-value of $\sin^2\theta_W$, extracted
from low-energy data \cite{exps2w}, we take
\beq
       \sin^2\theta_W(M_Z)\vert_{\overline{MS}}~=~0.2315\pm0.001~,
\label{leps2w}
\eeq
while for the $\overline{MS}$-value  of $\alpha_{3}(M_Z)$ we take
\cite{expalphas}
\beq
       \alpha_{3}(M_Z)\vert_{\overline{MS}}~=~0.120\pm0.010~.
\label{lepa3}
\eeq

\bigskip
\noindent{\it Renormalization Group Equations}
\bigskip

We seek to compare the above experimentally observed values with the
predictions from string theory.
To do this, we set up our calculation as follows.
Recall that in  general, string unification
implies that
each gauge coupling individually has a one-loop
RGE of the form
\beq
     {{16\pi^2}\over{g_i^2(\mu)}}~=~k_i{{16\pi^2}\over{g_{\rm string}^2}}~+~
     b_i\,\ln{ M^2_{\rm string}\over\mu^2}~+~\Delta_i^{\rm (total)}
\label{reminder}
\eeq
where $b_i$ are the one-loop beta-function coefficients, and where
the $\Delta_i^{\rm (total)}$
represent the combined corrections from each of
the effects outlined above.  From (\ref{reminder}) we
wish to obtain expressions for $\sin^2\theta_W(M_Z)$ and
$\alpha_{3}(M_Z)$.
To this end, we solve (\ref{reminder}) for $i=1,2,3$ simultaneously
in order to eliminate the direct dependence
on $g_{\rm string}$ from the first term on the right
sides of (\ref{reminder}).  A small dependence
on $g_{\rm string}$ remains through $M_{\rm string}$,
and in all subsequent numerical calculations we will
allow $M_{\rm string}$ to vary slightly
in order to account for this.
In each case we initially assume the MSSM spectrum
between the Planck scale and the $Z$ scale,
and treat all perturbations of this scenario through
effective correction terms.
The expression for $\alpha_{3}(M_Z)$
then takes the general form
\beqn
    \alpha^{-1}_{3}(M_Z)\vert_{\overline{MS}}&=&
      \Delta_{\rm MSSM}^{(\alpha)} +
      \Delta_{\rm h.s.}^{(\alpha)} +
      \Delta_{\rm l.s.}^{(\alpha)} +
       \Delta_{\rm i.g.}^{(\alpha)} +\nonumber\\
      &&~~~\Delta_{\rm i.m.}^{(\alpha)} +
      \Delta_{\rm 2-loop}^{(\alpha)} +
      \Delta_{\rm Yuk.}^{(\alpha)} +
      \Delta_{\rm conv.}^{(\alpha)},
\label{generalform}
\eeqn
and likewise for $\sin^2\theta_W(M_Z)\vert_{\overline{MS}}$ with
corresponding corrections $\Delta^{\rm (sin)}$.
Here $\Delta_{\rm MSSM}$ represents the one-loop
contributions from the spectrum of the Minimal Supersymmetric
Standard Model (MSSM) between the unification scale and the $Z$ scale, and
the remaining
$\Delta$ terms respectively correspond
to the second-loop corrections, the Yukawa-coupling corrections,
the corrections from scheme conversion,
the heavy string thresholds that were calculated in the previous
section, possible light SUSY threshold corrections,
corrections from possible additional intermediate-scale gauge structure
between the unification and $Z$ scales,
and corrections from possible extra intermediate-scale matter.
Each of these $\Delta$ terms has an algebraic
expression in terms of $\alpha_{\rm e.m.}$ as well
as model-specific parameters such as $k_1$,
the beta-function coefficients, and the
appropriate intermediate mass scales.
In particular, for $\sin^2\theta_W (M_Z)$ we find:
\beqn
    \Delta^{(\sin)}_{\rm MSSM} &=&
         {1\over{k_1+1}}\left\lbrack
     1-{a\over{2\pi}}(11-k_1)\ln{M_S\over{M_Z}}\right\rbrack
                \nonumber\\
    \Delta^{(\sin)}_{\rm l.s.} &=&
      {1\over{2\pi}}\,\sum_{\rm sp}\, {{k_1a}\over{k_1+1}}\,
              (b_{1_{\rm sp}}-b_{2_{\rm sp}})
                \ln{M_{\rm sp}\over{M_Z}}\nonumber\\
    \Delta^{(\sin)}_{\rm i.m.} &=&
      {1\over{2\pi}}\,\sum_{i}\,{{k_1a}\over{k_1+1}}\,
             (b_{2_i}-b_{1_i})\ln{M_S\over{M_i}}\nonumber\\
    \Delta^{(\sin)}_{\rm h.s.} &=&
      {1\over{2\pi}}\,{{k_1a}\over{k_1+1}}\,
          {{\Delta_2-\Delta_{\hat Y}}\over2}~.
\label{s2wmz}
\eeqn
where $M_S\equiv M_{\rm string}$ is the string unification scale,
$a\equiv \alpha_{\rm e.m.}(M_Z)$,
$M_{\rm sp}$ are the sparticle masses, $M_i$ are the intermediate gauge
mass scales,
and $\Delta_{3,2,\hat Y}$ are respectively
the heavy string threshold corrections for the strong,
electroweak, and properly normalized hypercharge gauge group factors.
Likewise, for $\alpha^{-1}_{3}(M_Z)$, we find:
\beqn
    \Delta^{(\alpha)}_{\rm MSSM} &=&
      {1\over{1+k_1}}\,\left\lbrack {1\over a}+{1\over{2\pi}}(-3 k_1-15)
       \ln{M_S\over{M_Z}}\right\rbrack\nonumber\\
    \Delta^{(\alpha)}_{\rm l.s.} &=&
       {1\over{2\pi}}\,\sum_{\rm sp}\,\left\lbrack
    \left({k_1\over{1+k_1}}\right)b_{1_{\rm sp}}+
         \left({1\over{1+k_1}}\right)b_{2_{\rm sp}}
       -b_{3_{\rm sp}}\right\rbrack \ln{M_{\rm sp}\over{M_Z}}\nonumber\\
    \Delta^{(\alpha)}_{\rm i.m.} &=&
        -~{1\over{2\pi}}\,\sum_{i}\,\left\lbrack
    \left({k_1\over{1+k_1}}\right)b_{1_{i}}+
       \left({1\over{1+k_1}}\right)b_{2_{i}}
     -b_{3_{i}}\right\rbrack\ln{M_{S}\over{M_i}}\nonumber\\
    \Delta^{(\alpha)}_{\rm h.s.} &=&
         -~{1\over{4\pi}}\,{1\over{1+k_1}}\left\lbrack
       k_1(\Delta_{\hat Y}-\Delta_3)+(\Delta_2-\Delta_3)\right\rbrack
\label{a3invmz}
\eeqn

Note that in the above solution of the one-loop RGE's
for $\sin^2\theta_W(M_Z)$ and $\alpha^{-1}_{3}(M_Z)$, only the {\it
differences}\/
of the RGE's (\ref{reminder}) for the separate gauge couplings are used.
Therefore, the expressions for
$\sin^2\theta_W(M_Z)$ and $\alpha_{3}(M_Z)$ depend only on the
 differences of the heavy string threshold
corrections $\Delta_{3,2,\hat Y}$ of the different group factors,
and not on their absolute values. Consequently, the group-independent
additive factors in the full expression for the one-loop string threshold
corrections do not affect the predictions for the experimentally
measured parameters $\sin^2\theta_W(M_Z)$ and $\alpha_3(M_Z)$.
This is important, for these extra factors are neglected in the original
definition for $\Delta_{G}$ in (\ref{Deltadef}).

Finally, before we can proceed, we must
estimate the second-loop and Yukawa-coupling corrections.
As discussed above, these are calculated assuming the MSSM spectrum
between the string unification and $Z$ scales.
To estimate the size of the second-loop
corrections, we run the one- and two-loop RGE's
for the gauge couplings and take the difference.
Likewise, to estimate the Yukawa-coupling corrections, we
evolve the two-loop RGE's for the gauge couplings coupled
with the one-loop RGE's for the heaviest-generation Yukawa couplings,
assuming $\lambda_t\approx 1$ and $\lambda_b=\lambda_\tau\approx 1/8$
at the string unification scale.
We then subtract the two-loop non-coupled result.
As indicated above, these differences are then each averaged for different
values
of $M_{\rm string}$ (in order to account for various values of $g_{\rm
string}$).
Numerically, this yields
the following results for
$\sin^2\theta_W(M_Z){\vert_{\overline{MS}}}$:
\beqn
          \Delta^{(\sin)}_{\rm 2-loop} &\sim& 0.00678  \nonumber\\
          \Delta^{(\sin)}_{\rm Yuk.} &\sim& -0.00091 \nonumber\\
          \Delta^{(\sin)}_{\rm conv.} &\sim& 0.00026~,
\label{s2wtwoloopYscc}
\eeqn
while for $\alpha_{3}^{-1}(M_Z)\vert_{\overline{MS}}$ we find:
\beqn
          \Delta^{(\alpha)}_{\rm 2-loop} &\sim& 0.80757  \nonumber\\
          \Delta^{(\alpha)}_{\rm Yuk.} &\sim& -0.16728  \nonumber\\
          \Delta^{(\alpha)}_{\rm conv.} &\sim& 0.0596831~.
\label{a3invtwoloopYscc}
\eeqn

In the next subsections, we will discuss and evaluate the various remaining
contributions to the experimentally measured parameters $\sin^2\theta_W(M_Z)$
and $\alpha_{3}(M_Z)$.

\subsection{MSSM spectrum between $M_Z$ and $M_{\rm string}$}

In this section we discuss the predictions for $\sin^2\theta_W(M_Z)$
and $\alpha_{3}(M_Z)$
assuming that the particle spectrum between the
string unification scale and the $Z$-scale is purely that of the Minimal
Supersymmetric Standard Model (MSSM).
The contributions of such a spectrum are
those given in the first lines
of (\ref{s2wmz}) and (\ref{a3invmz}) respectively.
We also include the two-loop and Yukawa-coupling
corrections to the one-loop RGE's,
and the corrections due to the conversion from the
$\overline{DR}$-scheme to the
$\overline{MS}$-scheme.  These are the
corrections which appear in (\ref{s2wtwoloopYscc})
and (\ref{a3invtwoloopYscc}).

Using the RGE's with these two terms as discussed in
Sect.~5.0, we find that the prediction for $\sin^2\theta_W(M_Z)$
deviates from the experimentally measured value by approximately 5\%,
and  that the prediction for $\alpha_3(M_Z)$
deviates from the experimentally measured value by approximately 100\%.
The discrepancy with $\alpha_{3}(M_Z)$ is, of course, expected, since
it is this discrepancy
which is ultimately the root of the factor of twenty that separates
the string scale $M_{\rm string}$ from the MSSM unification scale $M_{\rm
MSSM}$.
Indeed, in this way we can even see why the value of $\alpha_{3}(M_Z)$ should
be
off by 100\%.
Suppose, for example, that we were to run the RGE's not from
$M_{\rm string}\approx 20 M_{\rm MSSM}$ to $M_Z$, but rather from
$M_{\rm MSSM}$
to $M_Z/20$.  As expected from the running of the QCD coupling below
the $Z$ scale, we would then find that $\alpha_{3}(M_Z/20)$
deviates from $\alpha_{3}(M_Z)$ by approximately 100\%.
However, due to the additive nature of the logarithms involved,
this effect on $\alpha_{3}$ should be identical to that achieved by running
instead from $M_{\rm string}$ to $M_Z$.
Indeed, in the string unification scenario, we have simply
shifted this discrepancy from the low scale near $M_Z$ to the high scale
near $M_{\rm string}$.
Thus, we see that the simplest string unification scenario
assuming the MSSM spectrum below the string scale should indeed
predict a value of $\alpha_{3}(M_Z)$ that is off by 100\%.

%
\begin{figure}[thb]
\centerline{\epsfxsize 3.0 truein \epsfbox {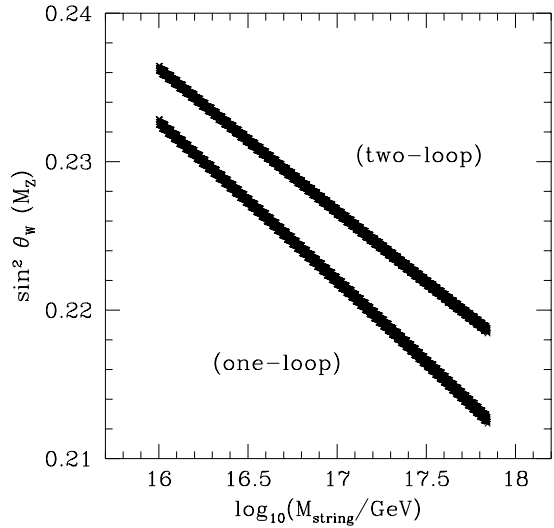}}
\caption{Dependence on the string scale:  $\sin^2\theta_W(M_Z)$
           vs.\ $M_{\rm string}$.
Results for both one-loop and two-loop running are plotted.  }
\label{sin_vs_Mstring}
\end{figure}

\begin{figure}[thb]
\centerline{\epsfxsize 3.0 truein \epsfbox {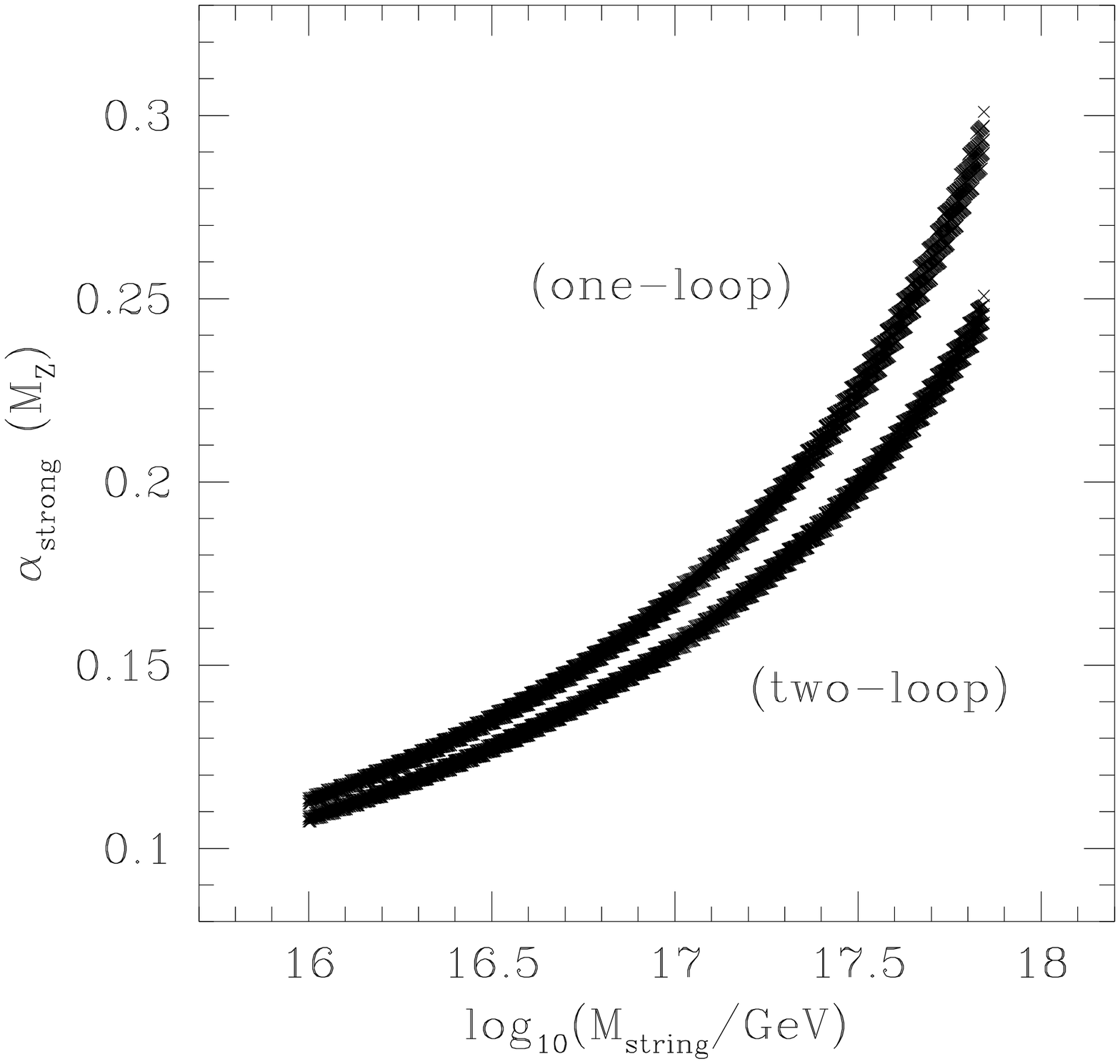}}
\caption{Dependence on the string scale:  $\alpha_{3}(M_Z)$
           vs.\ $M_{\rm string}$.
Results for both one-loop and two-loop running are plotted.  }
\label{alpha_vs_Mstring}
\end{figure}

\begin{figure}[htb]
\centerline{\epsfxsize 3.0 truein \epsfbox {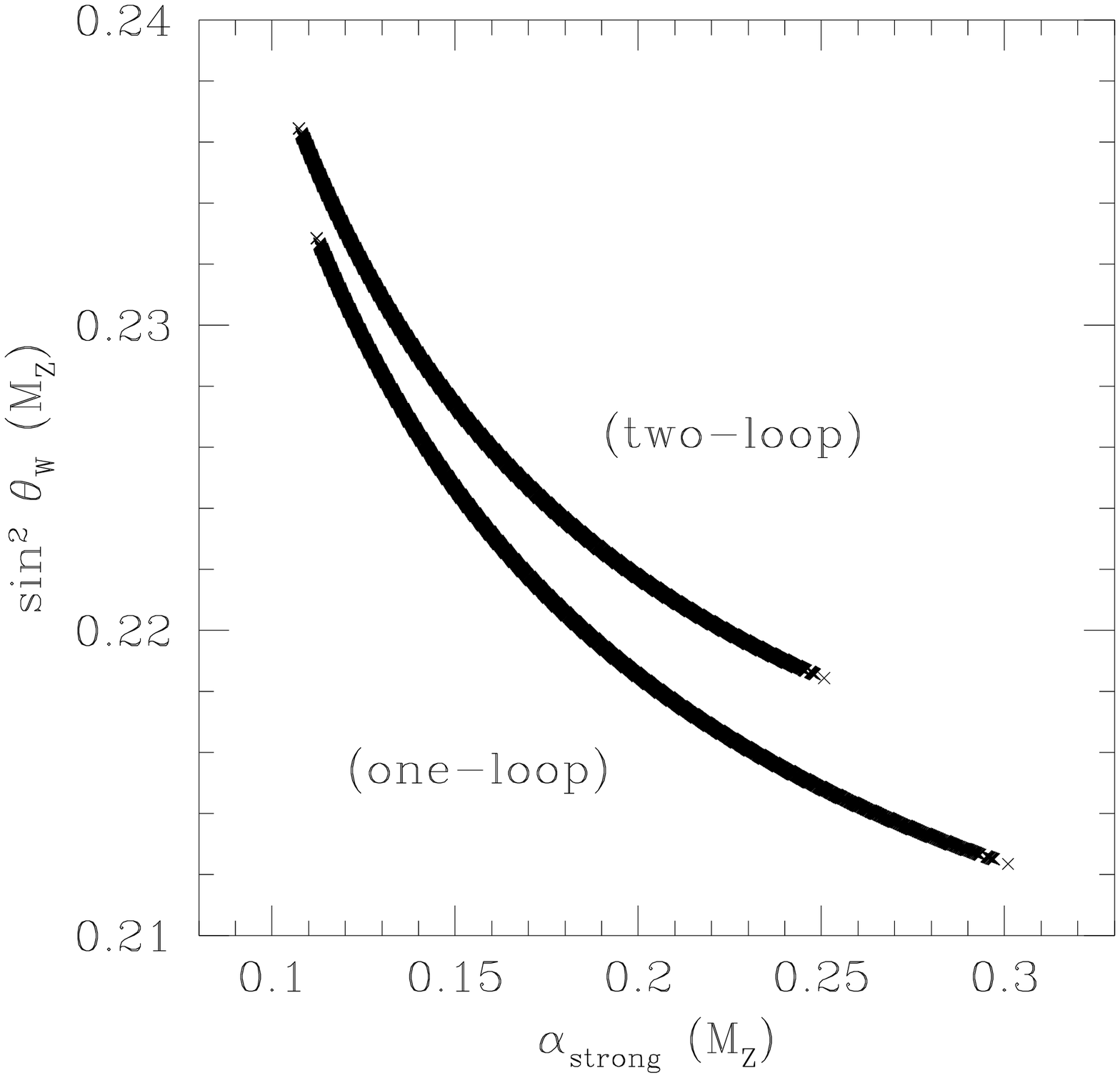}}
\caption{Dependence on the string scale:  $\alpha_{3}(M_Z)$ vs.\ $\sin^2
    \theta_W (M_Z)$, as $M_{\rm string}$ is varied.
Results for both one-loop and two-loop running are plotted.  }
\label{alpha_vs_sin}
\end{figure}

In Figs.~\ref{sin_vs_Mstring} and \ref{alpha_vs_Mstring},
we have plotted $\sin^2\theta_W(M_Z)$ and $\alpha_3(M_Z)$ versus
$M_{\rm string}$  in the range
$1\times 10^{16}\GeV\leq M_{\rm string}\leq 7\times 10^{17}\GeV$,
both with and without the two-loop corrections.
In the analysis for these figures,
rather than using the averaged second-loop corrections from
(\ref{s2wtwoloopYscc}) and
(\ref{a3invtwoloopYscc}), we have run the full
two-loop RGE's over
the entire range of $M_{\rm string}$.
In Fig.~\ref{alpha_vs_sin}, we have plotted the two low-energy variables
$\sin^2\theta_W(M_Z)$ and $\alpha_3(M_Z)$
versus each other as $M_{\rm string}$ is varied.  From these
figures we see that for
$3\times 10^{17}~{\rm GeV}\leq M_{\rm string}\leq 7\times 10^{17}~{\rm GeV}$,
both $\sin^2\theta_W(M_Z)$ and
$\alpha_{3}(M_Z)$ are approximately $\sim 0.22$. Thus,
as expected, assuming the MSSM spectrum between the $Z$ scale
and the string unification
scale results in significant disagreement with the low-energy data.

\subsection{Non-standard hypercharge normalizations}

\begin{figure}[hbt]
\centerline{\epsfxsize 3.0 truein \epsfbox {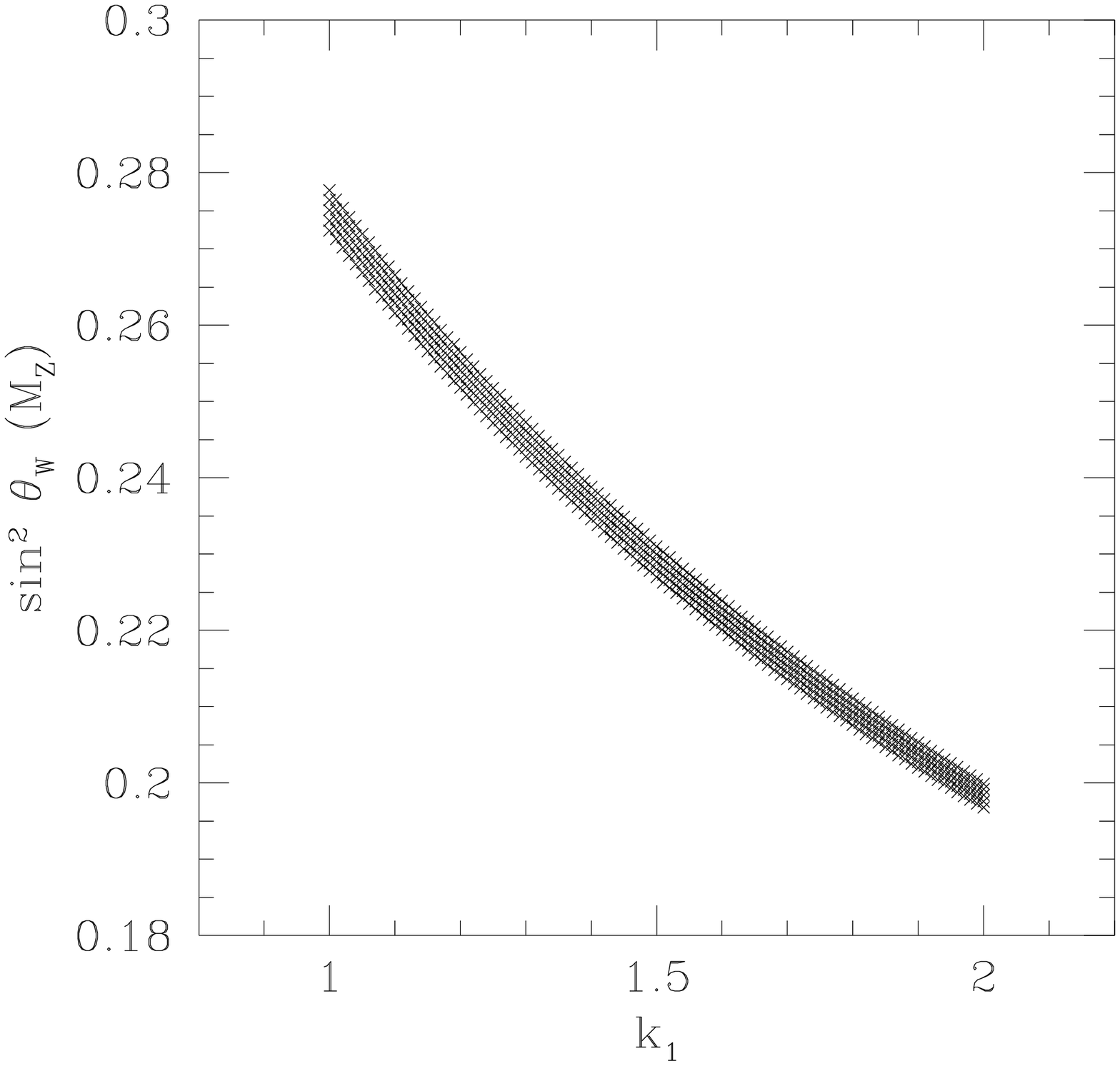}}
\caption{The effects of hypercharge normalization:  $\sin^2\theta_W(M_Z)$
     vs.\ $k_1$.
The width of the curve reflects the variations in $M_{\rm string}$. }
\label{sin_vs_kone}
\end{figure}

\begin{figure}[hbt]
\centerline{\epsfxsize 3.0 truein \epsfbox {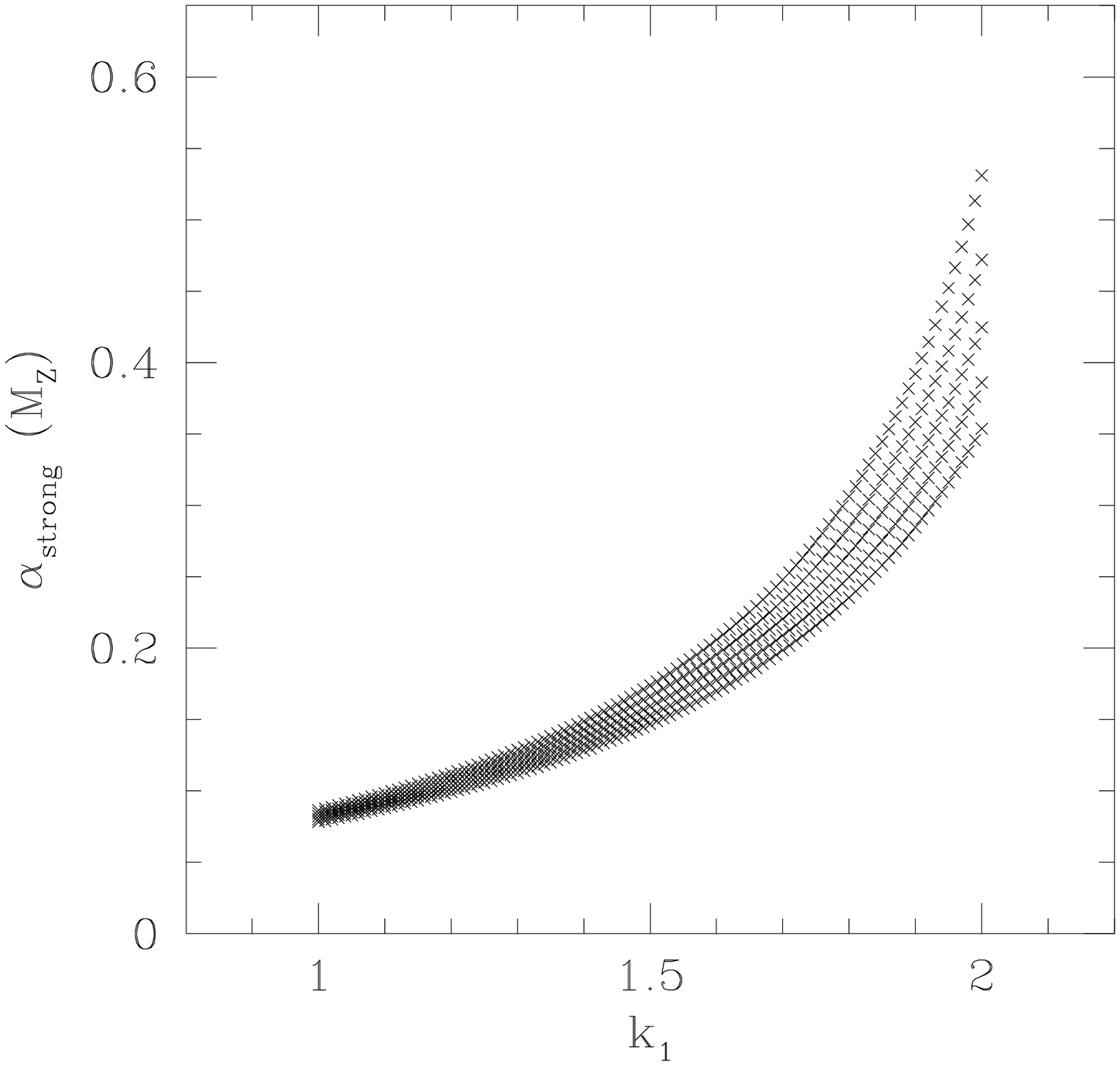}}
\caption{The effects of hypercharge normalization:  $\alpha_{3}(M_Z)$
       vs.\ $k_1$.
 The higher curves correspond to higher values of $M_{\rm string}$.}
\label{alpha_vs_kone}
\end{figure}

We can also examine the extent to which this MSSM string unification scenario
depends on the value of the hypercharge normalization $k_1$.
In the $SU(5)$ or $SO(10)$ unification schemes,
the value of $k_1$ is 5/3. However, in string unification, this assumption is
not
necessary, as other embeddings of the weak hypercharge into the
four-dimensional
string gauge group are possible.
This possibility is also discussed in Ref.~\cite{ibanez}.
In Figs.~\ref{sin_vs_kone}
and \ref{alpha_vs_kone},
we plot the predictions for $\sin^2\theta_W(M_Z)$ and
$\alpha_3(M_Z)$ versus $k_1$ in the range $1<k_1<2$.
We take $3\times 10^{17} \GeV \leq M_{\rm string}\leq
    7\times 10^{17} \GeV$.  From these figures we find that
the one-loop string prediction can be in
agreement with the measured values for a value of $k_1\approx 1.4$.

Unfortunately, none of the realistic models we have been analyzing
has a value of $k_1$ of this size, and indeed all of the models
have $k_1\geq 5/3$.  Similar results have also been found,
for example, in the case of semi-realistic orbifold models (see,
 {\it e.g.}, Ref.~\cite{FIQS}).
Moreover, in string theory, there are general reasons to expect
that $k_1 \ge5/3$ in realistic models.  The reason is as follows.\footnote{
     We thank J. March-Russell for discussions on this point \cite{DFM}.}
As discussed at the end of Sect.~4.3, the weak hypercharge is a combination
of simple worldsheet currents which are each normalized to one.
To produce the correct conformal dimension for the massless states,
every $U(1)$ generator (each of which is ultimately a combination of simple
worldsheet currents) must be normalized to one.  By contrast, the
$U(1)$ generator that produces the correct weak hypercharges
for the Standard Model particles is {\it not}\/ normalized to one, and
$k_1$ is essentially the normalization coefficient of the properly
normalized weak hypercharge generator. Since the weak hypercharge
generator is a combination of simple worldsheet currents, we can determine
the minimal number of simple worldsheet currents that must
be used in order to generate the correct hypercharges for all the
quark and lepton families and to satisfy the various anomaly cancellation
constraints. We then find that the minimal value of $k_1$ is
essentially $5/3$.
It is not likely, therefore, that this effect can explain the discrepancy
between the low-energy data and the suppositions of string unification.
Thus, in the analysis below, we shall take $k_1=5/3$ unless otherwise stated.

%

\subsection{Light SUSY thresholds}

We now study the effect of the light SUSY thresholds on the
predictions for $\sin^2\theta_W(M_Z)$ and $\alpha_3(M_Z)$.  These effects
correspond to the
second lines of (\ref{s2wmz}) and (\ref{a3invmz}) respectively.

Our purpose here is
not a detailed quantitative analysis of the sparticle threshold corrections,
but
rather a qualitative examination of whether the light SUSY thresholds
are capable of removing the discrepancy,
found in Sect.~5.1, between the
predicted and the experimentally observed values
of $\sin^2\theta_W(M_Z)$ and $\alpha_3(M_Z)$.
Consequently, to parametrize the sparticle thresholds, we shall consider the
spectrum of the
MSSM near the $Z$ scale,
and neglect the contributions from the Yukawa couplings and electroweak VEV's.
This implies that all of the supersymmetric scalar mass
matrices, including that of the stop quark,
will be diagonal, and that the $D$-term contributions to the sparticle masses
are
neglected.  With these assumptions, the sparticle masses can be obtained from
the
one-loop RGE's for the soft SUSY-breaking terms.
In terms of the usual soft SUSY-breaking parameters $m_0$, $m_{1/2}$, and
$\mu$,
the gluino, wino, and higgsino masses are then given by \cite{SUSYreviews}
\beq
  m_{\tilde g}={{\alpha_3(m_{\tilde g})}\over{\alpha_3}}\,m_{1/2}~,~~~~~
  m_{\tilde w}={{\alpha_2(m_{\tilde g})}\over{\alpha_3}}\,m_{1/2}~,~~~~~
  m_{\tilde h}=\mu~.
\label{gwhmasses}
\eeq
Likewise, the scalar sparticle masses are given by
\beq
      m_{\tilde p}^2~=~ m_0^2~+~c_{\tilde p}\, m_{1/2}^2
\eeq
where the coefficients $c_{\tilde p}$ for the different sparticles
are given in terms of their hypercharges $Y_{\tilde p}$ by
\beq
       c_{\tilde p}~=~ c_3(m_{\tilde p})~+~ c_2(m_{\tilde p})~+~
       Y^2_{\tilde p}\,c_y(m_{\tilde p})
\label{cpcoef1}
\eeq
with
\beqn
        c_3(m_{\tilde p})&=&-{8\over 9}\,\left\lbrack 1-
                      (1+ 3X)^{-2} \right\rbrack \nonumber\\
        c_2(m_{\tilde p})&=&{3\over 2}\,\left\lbrack 1-
                      (1- X)^{-2} \right\rbrack \nonumber\\
        c_y(m_{\tilde p})&=&{2k_1\over 11}\,\left\lbrack 1-
                      (1- 11X/k_1)^{-2} \right\rbrack
\label{cpcoef2}
\eeqn
and
\beq
     X~\equiv ~{1\over 2\pi} \,\alpha_{\rm string}\,\ln{M_{\rm
string}\over{m_{\tilde p}}}~.
\eeq
The corresponding corrections to $\sin^2\theta_W(M_Z)$
and $\alpha_3(M_Z)$
due to these light SUSY thresholds are then given by
\beq
     \Delta_{\rm l.s.}^{(\sin)}~=~ {1\over 2\pi}\, {k_1 \,a\over k_1+1}\,
\sum_R \,n_R\,C_R^{(\sin)}\,
          \ln {m_{R}\over M_Z}
\label{dlights2w}
\eeq
and
\beq
     \Delta_{\rm l.s.}^{(\alpha)}~=~ {1\over 2\pi}\,  \sum_R
\,n_R\,C_R^{(\alpha)}\,
          \ln {m_{R}\over M_Z}
\label{dlighta3}
\eeq
where the summations are over the particle representations $R$
with masses $m_R$.
Here $n_R$ is the number of times a particular generic
representation appears in the complete spectrum ({\it e.g.}, three
for three generations),
and the $C_R$-coefficients
are given in terms of the one-loop beta-function coefficients
$b_{3,2,\hat Y}(R)$ of the representation $R$ as
\beqn
       C_{R}^{(\sin)} &=& b_{\hat Y}(R)-b_2(R)\nonumber\\
       C_{R}^{(\alpha)} &=&{{k_1}\over{k_1+1}}\, b_{\hat
Y}(R)+{1\over{k_1+1}}\,b_2(R) -b_3(R)~.
\label{Ccoeffs}
\eeqn
Note that $b_{\hat Y}(R)\equiv b_Y(R)/k_1$.
For each representation $R$, these beta-function coefficients, degeneracies
$n_R$,  and $C_R$-coefficients are as follows:
\begin{equation}
\begin{tabular}{c||ccc|c|c|c}
       $R$ & $b_{\hat Y}(R)$ & $b_2(R)$ & $b_3(R)$  & $ C_R^{(\sin)}$  &
       $C_R^{(\alpha)}$ & $n_R$ \\
\hline
    ${\tilde g}$ & 0 &  0 & 2 & 0 & $-2$ & 1\\
    ${\tilde w}$ & 0 &  ${4\over3}$ & 0 & $-{4\over3}$ &
          ${4\over3}{1\over{k_1+1}}$ & 1\\
    ${\tilde\ell}_\ell$ & ${1\over{6k_1}}$ &  ${1\over6}$ & 0 &
${1\over{6k_1}}-{1\over6}$ &
          ${1\over3}{1\over{k_1+1}}$ & 3\\
    ${\tilde\ell}_r$ & ${1\over{3k_1}}$ &  0 & 0 & ${1\over{3k_1}}$ &
          ${1\over3}{1\over{k_1+1}}$ & 3\\
    ${\tilde Q}$ & ${1\over{18k_1}}$ &  ${1\over2}$ & ${1\over3}$ &
         ${1\over{18k_1}}-{1\over2}$   &
          ${5\over9}{1\over{k_1+1}}-{1\over3}$ & 3\\
    ${\tilde d}_r$ & ${1\over{9k_1}}$ &  0 & ${1\over6}$ &  ${1\over{9k_1}}$ &
           ${1\over9}{1\over{k_1+1}}-{1\over6}$ & 3\\
    ${\tilde u}_r$ & ${4\over{9k_1}}$ &  0 & ${1\over6}$ &  ${4\over{9k_1}}$ &
          ${4\over9}{1\over{k_1+1}}-{1\over6}$ & 3\\
    ${\tilde h}$     & ${1\over{3k_1}}$ &  ${1\over3}$ & 0 &
${1\over{3k_1}}-{1\over3}$ &
          ${2\over3}{1\over{k_1+1}}$ & 2\\
    $h$ & ${1\over{6k_1}}$ &  ${1\over6}$ & 0 &
          ${1\over{6k_1}}-{1\over6}$ & ${1\over3}{1\over{k_1+1}}$ & 1\\
    $t$ & ${{17}\over{18k_1}}$ &  1 & ${2\over3}$ & ${17\over{18k_1}}-1$ &
          ${{35}\over{18}}{1\over{k_1+1}}-{2\over3}$ & 1\\
\end{tabular}
\label{lstqnumbers}
\end{equation}

\begin{figure}[hbt]
\centerline{\epsfxsize 3.0 truein \epsfbox {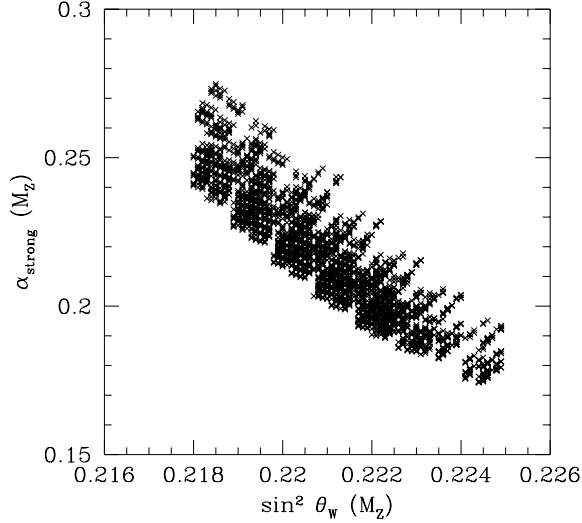}}
\caption{Scatter plot for $\lbrace\sin^2\theta_W(M_Z),
 \alpha_{3}(M_Z)\rbrace$ for various values of $\lbrace m_0,m_{1/2},\break
  m_h,m_{\tilde h},M_{\rm string}\rbrace$, assuming the MSSM spectrum.}
\label{lightsusyfigs}
\end{figure}

Assuming universal boundary conditions for the soft SUSY-breaking terms
({\it i.e.}, assuming that $m_{1/2}~{\rm and}~m_0$ are universal),
we have analyzed the possible light SUSY threshold contributions
for a wide range of points in
the parameter space $\lbrace m_0, m_{1/2}, m_h, m_{\tilde h}\rbrace$.
In general we find that the light SUSY thresholds are small, and
that the predictions for
$\sin^2\theta_W(M_Z)$ and $\alpha_3(M_Z)$ continue to disagree with the
measured values.
In Fig.~\ref{lightsusyfigs} we display
$\sin^2\theta(M_Z)_{\overline{MS}}$ versus $\alpha_3(M_Z)_{\overline{MS}}$
for a sampling of points in the SUSY-breaking parameter space.
Within this parameter space,
each free parameter $X$ is sampled in the interval $(X_{\rm min},X_{\rm max})$
with
spacing $\Delta X$ between consecutive points, as follows:
\begin{equation}
\begin{tabular}{c|ccc}
Parameter $X$ & $X_{\rm min}$  & $X_{\rm max}$  &   $\Delta X$  \\
\hline
$m_0$~(GeV) & 0 &  600 & 200 \\
$m_{1/2}$~(GeV) & 50 &  600 & 150 \\
$m_h$~(GeV) & 100 &  500 & 200  \\
$m_{\tilde h}$~(GeV)     & 100 &  500 & 200 \\
$M_{\rm string}$~(GeV) & ~~$3\times10^{17}$~~ &
     ~~$7\times10^{17}$~~ &  ~~$5\times10^{16}$~~   \\
$\alpha^{-1}_{\rm string}$ & $20$ &  32 & 4 \\
\end{tabular}
\label{sps}
\end{equation}
We have taken the top-quark mass $m_t=175$ GeV.

%

It is important to note that our assumption of universal boundary conditions
for
the soft SUSY-breaking terms is not necessary.
In fact, in string models, such boundary conditions are generally
expected to be {\it non}\/-universal \cite{sssbbc}.
We can then ask whether non-universal boundary conditions can
bring the string-scale unification into agreement with the experimentally
measured couplings at low energy.
By examining the different terms in (\ref{dlights2w})
and (\ref{dlighta3}), we can see that for this purpose, the
value of the light SUSY threshold corrections
must be positive. Now, the signs of the $C_R$-coefficients of
the different terms in these two equations are determined by the
one-loop beta-function coefficients of the different representations $R$.
Assuming that all the sparticle masses are larger than $M_Z$, we can maximize
the light SUSY corrections by setting the masses of the negative
contributions equal to the $Z$ mass.
For example, in (\ref{dlighta3}), we see from the signs of the
coefficients $C_{R}^{(\alpha)}$ that
the wino, top, sleptons, higgs and higgsino all give positive contributions.
Assuming a common
mass $M$ for all of these but the top, and summing over the resulting
coefficients,
we obtain $\Delta_{\rm l.s.}^{(\alpha)} =15/(16 \pi)\ln{(M/M_Z)}$.
Thus, with $\alpha_3^{\rm (predicted)}\sim0.2$ and
$\alpha_3^{\rm (measured)}\sim0.14$,
even the best-case scenario requires $M\approx100$ TeV.
For $\sin^2\theta_W$, on the other hand,
only ${\tilde u}_r$, ${\tilde d}_r$, and  ${\tilde e}_r$
contribute with positive coefficients. Assuming that all the other
thresholds are degenerate at $M_Z$, and that ${\tilde u}_r$, ${\tilde d}_r$,
and ${\tilde e}_r$
have a common mass ${\tilde M}$, we find that agreement with the measured value
of $\sin^2\theta_W$ requires ${\tilde M}\sim10$ TeV.
Alternatively, some of the sparticle thresholds may be below $M_Z$.
Assuming degenerate thresholds ${\tilde M}=45$ GeV
for the sparticles making negative contributions to $\sin^2\theta_W$,
and with $m_t=175$ GeV, we find that
the contribution for $\Delta_{\rm l.s.}^{(\sin)}$ from these thresholds is of
the
order of 0.001,
which still requires  ${\tilde M}\sim10$ TeV. Thus, even with these extreme
``best-case''
scenarios, light SUSY thresholds cannot by themselves bring the
string-unification predictions
into agreement with the experimentally observed values.

\subsection{Heavy string thresholds}

We now analyze the effects of the threshold corrections due
to the
infinite towers of heavy string states. These corrections were the focus of our
calculations
in Sect.~4, and contribute to the running of the gauge couplings as indicated
in the
fourth lines of (\ref{s2wmz}) and (\ref{a3invmz}).

The calculation of the heavy string threshold corrections in Sect.~4 was
performed for
normalized gauge group factors $G_i$ with $k_i=1$.
However, as indicated in (\ref{gaugedependence}),
the full string threshold corrections are in
general of the form $\tilde \Delta_i=\Delta_i +k_i Y$, where $Y$ is independent
of the gauge group. In Sect.~4 we calculated the gauge-dependent terms
$\Delta_i$ for the different gauge group factors. Therefore, by taking the
 {\it differences}\/ of the threshold corrections for two different gauge
groups,
the group-independent term cancels. Moreover, as we see from
(\ref{s2wmz}) and (\ref{a3invmz}), it turns out that the solutions
of the RGE equations for $\sin^2\theta_W(M_Z)$ and $\alpha_3(M_Z)$ depend
only on the differences $\Delta_i-\Delta_j$ for the different group factors
$G_i$ and $G_j$
in a given string model.
Therefore, the low-energy predictions for $\sin^2\theta_W(M_Z)$ and
$\alpha_3(M_Z)$
do not depend on the group-independent factor $Y$.
One might worry, of course, that the presence of a non-standard normalization
for $k_1$
in certain string models might render this claim incorrect.
However, the calculation of the threshold corrections is always done with
respect to the
properly normalized $U(1)$ generators. Similarly, only the threshold
corrections for the
properly normalized $U(1)$ generators appear in the RGE equations. This is
similar
to the usual practice in grand unified theories in which the coupling of the
properly normalized
$U(1)$ generator evolves below the GUT scale, and the matching with the weak
hypercharge
coupling is done at the $Z$ scale. Therefore the value of the
gauge-independent term $Y$ continues to be irrelevant for our analysis.

We now discuss the effects that these string threshold corrections
will have on the low-energy parameters.
Because the gauge structure and threshold corrections are highly
model-dependent,
we shall have to consider each realistic free-fermion model
in turn.

Considering first the string model analyzed in Sect.~4.2, we see that
the weak hypercharge has the standard $SO(10)$ embedding.
Therefore, we can directly insert the
relative values for $\Delta_{SU(3)}$,  $\Delta_{SU(2)}$ and $\Delta_{U(1)_{\hat
Y}}$ that
we found in (\ref{DELTAS278})
into (\ref{s2wmz}) and (\ref{a3invmz}).
For this model, we consequently obtain
\beq
   \Delta_{\rm h.s.}^{(\sin)}=-0.0006~~~~~{\rm and}~~~~~
      \Delta^{(\alpha)}_{\rm h.s.}=-0.3536~.
\eeq
Thus, the effect of the string threshold corrections in this model is to
slightly decrease the
values of $\sin^2\theta_W(M_Z)$ and $\alpha^{-1}_3(M_Z)$.  Alternatively, of
course,
this can be regarded
as effectively increasing the string unification scale. This can be seen by
absorbing
the string threshold corrections into $\ln(M_{\rm string}/M_Z)$ in
(\ref{s2wmz}) and (\ref{a3invmz}).  From (\ref{a3invmz}),
we see that the corrected string unification scale can be written as
\beq
    M^{\rm (corrected)}_{\rm string}~=~M^{\rm (uncorrected)}_{\rm string}\,
      \exp\left\lbrack {k_1(\Delta_1-\Delta_3)+(\Delta_2-\Delta_3)
          \over2(3k_1+15)}\right\rbrack~.
\label{correctedscale}
\eeq
Numerically, this yields
\beq
    M^{\rm (corrected)}_{\rm string}~\approx~6.72 \,\times\, 10^{17}~{\rm GeV}
\eeq
where we take $M_{\rm string}^{\rm (uncorrected)}\approx5\times 10^{17}$ GeV.
Thus, the effect of the string threshold corrections in this model is to
enhance
the disagreement with the experimentally observed values.

We next analyze the model of Sect.~4.3.
As we saw in Sect.~4, in this model the analysis is complicated due
to the presence of the enhanced symmetry.
The group factors relevant for the analysis
are $SU(3)_C\times SU(2)_L\times SU(2)_{\rm cust}\times U(1)_{Y^\prime}$.
Above the $SU(2)_{\rm cust}\times U(1)_{Y^\prime}$ breaking scale,
we have to consider the running of these four group factors.
The $SU(2)_{\rm cust}\times U(1)_{Y^\prime}$ symmetry can be broken down to
$U(1)_Y$ by, for example, the VEV of the right-handed neutrino,
and below this breaking scale we have the standard one-loop RGE's
for $SU(3)_C\times SU(2)_L\times U(1)_{\hat Y}$. The relation between the
weak hypercharge coupling and the $SU(2)_{\rm cust}\times U(1)_{Y^\prime}$
couplings is then
\beq
  {1\over{\alpha_Y}}~=~{1\over{4\,\alpha_{\rm cust}}}~+~ {17 \over
12\,\alpha_{\hat Y}^\prime}
\eeq
where $\alpha_{{\hat Y}^\prime}$ is the coupling of the properly normalized
$U(1)_{{\hat Y}^\prime}$.
The RGE for the $U(1)_{\hat Y}$ coupling then takes the form
\beqn
    {1\over{\alpha_{\rm string} }}&=&{3\over5}{1\over{\alpha_1{(\mu)}}}-
    {{3b_Y}\over{10\pi}}\ln{M_I\over{\mu}}-
    {1\over{2\pi}}{3\over 5}({1\over4}b_{2c}+{{17}\over{12}}b_{\hat
    Y^\prime})\ln{M_S\over{M_I}}\nonumber\\
     &&~~-
 {1\over{4\pi}}{3\over5}({1\over4}\Delta_{2c}+{{17}\over{12}}
     \Delta_{\hat Y^\prime})~.
\label{futureeq}
\eeqn

Let us assume for the moment that
the VEV of the right-handed neutrino
is of the order of $M_S$.  This implies that $M_I=M_S$.
Therefore, below the string scale we have
the three gauge couplings of the Standard Model,
and the threshold corrections for the properly normalized
weak hypercharge are given by (\ref{DELTAS274}). Inserting
the relative values of
$\Delta_3$,  $\Delta_2$, and $\Delta_{\hat Y}$ from
 (\ref{DELTAS274}) into (\ref{s2wmz}) and (\ref{a3invmz}), we then obtain
\beq
  \Delta_{\rm h.s.}^{(\sin)} =-0.0012~~~~{\rm and}~~~~~~
  \Delta_{\rm h.s.}^{(\alpha)} =-0.335~.
\eeq
As in the case of the previous model, we see that
the effect of the string threshold corrections is
to decrease the values of $\sin^2\theta_W(M_Z)$ and $\alpha^{-1}_3(M_Z)$,
or equivalently to increase the effective
string unification scale.  Using (\ref{correctedscale}) with
$M_{\rm string}^{\rm (uncorrected)}\approx 5 \times 10^{17}$ GeV, we now obtain
\beq
        M_{\rm string}^{\rm (corrected)}~\approx~6.62 \,\times\,10^{17} ~{\rm
GeV}~.
\eeq
We shall shortly discuss the modifications that arise if $M_I<M_S$.

Next, we analyze the effect of the threshold corrections in the
$SO(6)\times SO(4)$ string model which realizes Pati-Salam unification
scenario.  In this string model, there is an intermediate energy scale
at which the $SU(4)\times SU(2)_L\times SU(2)_R$ symmetry is broken to
$SU(3)_C\times SU(2)_L\times U(1)_Y$.
The low-energy parameters are consequently affected  by the presence of this
intermediate
energy scale. The weak hypercharge $\hat Y$ is
 a linear combination of $B-L$ and $T_{3_R}$:
\beq
  {\hat Y}~=~\sqrt{{2\over5}}\, T_{B-L}~+~\sqrt{{3\over5}} \,T_{3_R}~.
\eeq
The low-energy parameters are then obtained from the one-loop RGE's
via
\beqn
  {1\over{\alpha_3{(M_Z)}}}&=&{1\over{\alpha_{\rm string}}}+
        {{b_3}\over{2\pi}}\ln{M_I\over{M_Z}}+
        {{b_4}\over{2\pi}}\ln{M_S\over{M_I}}+
	       {\Delta_4\over{4\pi}}\nonumber\\
  {1\over{\alpha_2{(M_Z)}}}&=&{1\over{\alpha_{\rm string}}}+
    {{b_{2_L}}\over{2\pi}}\ln{M_I\over{M_Z}}+
    {{b_{2_L}}\over{2\pi}}\ln{M_S\over{M_I}}+
               {\Delta_{2_L}\over{4\pi}} \nonumber\\
  {1\over{\alpha_1{(M_Z)}}}&=&{1\over{\alpha_{\rm string}}}+
        {{b_1}\over{2\pi}}\ln{M_I\over{M_Z}}+
  {1\over{2\pi}}({2\over5}b_4+{3\over5}b_{2_R})\ln{M_S\over{M_I}}\nonumber\\
     &&~~~~~~~~~~~~~~~~~~
+{1\over{4\pi}}({2\over5}\Delta_4+{3\over5}\Delta_{2_R})
\label{1LOOPRGESSO64}
\eeqn
Assuming again that $M_I\approx M_S$, the threshold corrections for the three
properly normalized group factors are $\Delta_4$, $\Delta_{2_L}$ and
$\Delta_{\hat Y}=(2/5)\Delta_4+(3/5)\Delta_{2_R}$.
Inserting the values from (\ref{DELTASO64}) and (\ref{deltasforso64model})
into (\ref{s2wmz}) and (\ref{a3invmz}), we find
\beq
       \Delta_{\rm h.s.}^{(\sin)} =-0.0028 ~~~~~~{\rm and}~~~~~
       \Delta_{\rm h.s.}^{(\alpha)} =-0.3141
\eeq
Just as in previous models,
the effect of the string threshold corrections is
therefore to increase the effective string unification scale. Inserting the
values of
$\Delta_4$, $\Delta_{2_L}$ and $\Delta_{\hat Y}$ into
(\ref{correctedscale}),
we obtain
\beq
       M_{\rm string}^{\rm (corrected)} ~\approx~6.5 \,\times\,10^{17} ~{\rm
GeV}~.
\eeq

Thus, in all of the examples that we have explored,
we find that the effect of the string threshold corrections
is to {\it increase}\/ the effective string unification scale.
Therefore, the effect of the string threshold corrections in these examples
is always to enhance the disagreement with the low-energy observables.

It is an important observation that the sizes of
the heavy string threshold corrections are very small in
all of these realistic string models, and thus
do not greatly affect (either positively or negatively)
the magnitude of the string unification scale.
Although such threshold corrections receive contributions
from infinite towers of massive string states,
we have in fact been able to provide a general model-independent
argument which explains why these corrections
are naturally suppressed in string theory
(except of course for large moduli).
This will be discussed in Sect.~6.
Thus, we conclude that these threshold corrections cannot by
themselves resolve the experimental discrepancy.

\subsection{Intermediate gauge structure}

In the last two models that we studied at the end of Sect.~5.4, there exists an
intermediate
energy scale $M_I$ at which an extended gauge group is broken
to the gauge group of the Standard Model.  In each case, we assumed that the
extended symmetry is
broken at the string scale, and that therefore $M_I=M_S$. An obvious issue,
then,
is to determine the effect of the extended symmetry if it is broken at an
intermediate scale $M_I<M_S$.  In particular, might the breaking at the
intermediate
scale bring the string-scale predictions into agreement with experiment?

We claim that in the above examples, this cannot happen.
Our reasoning is as follows.
Since the larger discrepancy with experiment is for $\alpha_3$,
it is sufficient to focus on this observable. Now, in the first model we
considered,
the RGE for $\alpha_3$ does not depend on the intermediate scale breaking at
all.
Thus the disagreement with the experimentally observed value persists for all
$M_I$.
Of course, this argument is somewhat deficient due to the dependence
of the RGE for $\alpha_3$ on the group-independent contribution $Y$ to the
heavy string threshold corrections. However, unless (and contrary to
expectation \cite{KLNT})
this group-independent term $Y$ is very large, the argument will hold.
A more careful argument is provided by solving the RGE's, in the presence
if intermediate scale $M_I$, for $\sin^2\theta_W(M_Z)$ and $\alpha_3(M_Z)$.
The one-loop RGE for $\alpha_1$ takes the form
\beq
   {1\over{\alpha_1}}={1\over{k_1\alpha_Y}}-{1\over{2\pi}}
       {b_Y\over{k_1}}\ln{M_S\over{M_Z}} -{1\over{4\pi}}
    \Delta_{\hat Y}+{1\over{2\pi}}{1\over{k_1}}
      (b_Y-({b_c\over4}+{{17}\over{12}}
   b_{Y^\prime}))\ln{M_S\over{M_I}}
\label{a1withmi339}
\eeq
where
the first three terms in (\ref{a1withmi339}) are identical to the RGE without
an
intermediate scale, and the last term incorporates the effect of the
intermediate scale.
Thus, in the presence of the intermediate scale the solution of the RGE's
for $\sin^2\theta_W(M_Z)$ in (\ref{s2wmz}) has an additional term
\beq
      ...~+~{1\over{2\pi}}\,{\alpha\over{k_1+1}}\,\left\lbrack
b_Y-({b_c\over4}+{{17}\over{12}}
       b_{Y^\prime})\right\rbrack \,\ln{M_S\over{M_I}}~.
\label{s2wwithmi339}
\eeq
Since $\sin^2\theta_W$ appears divided by $a\equiv \alpha_{\rm e.m.}$
in the solution of the RGE's for $\alpha_3^{-1}$,
the effect of the intermediate scale on $\alpha_3(M_Z)$ is
that in (\ref{s2wwithmi339}) divided by $a$.
Now, for a spectrum consisting of three generations
and two Higgs doublets, the coefficient of the logarithm in
(\ref{s2wwithmi339})
is $-{301/48}$. The crucial point here is the {\it sign}\/ of this coefficient,
as it
decreases both $\sin^2\theta_W(M_Z)$ and $\alpha^{-1}_3(M_Z)$. Consequently,
$\alpha_3(M_Z)$
increases, and the disagreement with experiment is enhanced.

Similar considerations also apply to the
$SO(6)\times SO(4)$ model.  From (\ref{1LOOPRGESSO64}),
we see that the effects of the intermediate scale on $\sin^2\theta_W(M_Z)$
and $\alpha_3(M_Z)$ amounts respectively to the additive terms
\beq
     ...~+~{4\over 10\pi}\,{{k_1\alpha}\over{1+k_1}}\, \ln{{M_S}\over{M_I}}
\label{addtermone}
\eeq
and
\beq
    ...~+~{1\over 2\pi}\,\left(-1+{4\over 5}\,{k_1\over{k_1+1}}\right)\,
         \ln{{M_S}\over {M_I}}~=...~-~{1\over{4\pi}}\ln{{M_S}\over {M_I}}
\label{addtermtwo}
\eeq
in (\ref{s2wmz}) and (\ref{a3invmz}).  We assume that the spectrum
below $M_I$ is that of the MSSM, and that above $M_I$ it consists
of three {\bf 16}'s of $SO(10)$ that produce the three generations,
one (1,2,2) representation that produces the light Higgs, and
$(4,1,2)+(\bar4,1,2)$
representations that are used to break the $SU(4)\times SU(2)_R$ symmetry at
$M_I$.
The signs of the coefficients in (\ref{addtermone})
and (\ref{addtermtwo})
shows that the effect of
having $M_I<M_S$ is to increase $\sin^2\theta_W(M_Z)$ and
$\alpha_3(M_Z)$.
Note that in arriving at this conclusion, we have used only
the {\it differences}\/ of heavy string threshold corrections, which are
unambiguous.
Thus, the  effect of having an extended gauge structure broken at an
intermediate
energy scale is to enhance the disagreement with the experimental results for
$\alpha_3(M_Z)$.

Likewise, in the case of the flipped $SU(5)$ model,
the effect of the intermediate scale can be incorporated
via the following additional term in (\ref{s2wmz}):
\beq
        -{1\over 2\pi}\, {32\over 5}\, {k_1 a\over k_1+1}\,
         \ln\,{M_S\over M_I}~.
\eeq
Here we have assumed the spectrum below $M_I$ to be that of
the MSSM, and above $M_I$ to consist of three {\bf 16} representations
of $SO(10)$, one ${\bf 5}$ and $\overline{\bf 5}$
of $SU(5)$ that produces the light Higgs doublets,
and one {\bf 10} and $\overline{\bf 10}$ of $SU(5)$
that is used to break the $SU(5)\times U(1)$ symmetry
to $SU(3)\times SU(2)\times U(1)$.
Thus, the effect of the extended gauge structure in this model
is to reduce $\sin^2\theta_W(M_Z)$, and to enhance
the disagreement with the experimentally observed value.

It is remarkable that in all of these realistic string models,
the availability of an intermediate scale $M_I$ does not help in
removing the discrepancy between string-scale unification and
low-energy data.  {\it A priori}\/, one might have expected
that the presence of such an extra degree of
freedom would have enabled agreement to be reached.
However, we now see that within the context of the realistic
string models, this extra degree of freedom only worsens
the agreement.
Thus, we can effectively rule out intermediate gauge structure
as a potential explanation for the discrepancies.

\subsection{Intermediate matter thresholds}

We now turn to the effects induced by additional matter
below the string scale \cite{GCU,LNY,Gaillard}.
As we shall see, such matter appears naturally in
the string models we have examined.
It is evident from our analysis up to this point that such intermediate
matter now appears to be the {\it only}\/ way by which the disagreement
with experiment can be possibly be resolved.
Fortunately, as we shall find, in certain models exactly the required
matter appears, in just the right representations and with just the
right non-standard hypercharge assignments.
Hence, in these models, the disagreement with experiment can be resolved.

In the analysis up to this point, we have assumed that the matter spectrum
below the
string scale is simply that of the Minimal Supersymmetric
Standard Model, {\it i.e.},
that it consists of exactly three chiral generations with two Higgs doublets.
However, in the context of the realistic string models, this assumption
is {\it ad hoc}\/. In fact, in {\it all}\/ of the string models constructed to
date,
additional color triplets and electroweak doublets beyond the MSSM
appear in the massless spectrum, in vector-like representations.
The number of such additional color triplets and electroweak doublets is of
course
highly model-dependent, as are their mass scales.
However, the assumption that the massless spectrum below the string scale
is that of the MSSM is, in general, not justified.

Mass terms for these extra states beyond the MSSM may arise from
cubic or higher-order non-renormalizable terms in the superpotential.
In the
models studied to date, the mass scale of the additional
color triplets and electroweak doublets is not the string scale. In general,
the masses of the extra states are suppressed relative to the string scale,
the reason being that the mass terms arising from non-renormalizable terms are
suppressed relative to the cubic-level mass terms. For example, in
Ref.~\cite{FCP},
the mass scale of an additional pair of color triplets was
estimated to be of the order of $10^{11}$ GeV.

The contributions of such additional color triplets and electroweak doublets
to the low-energy parameters are given in the third
lines of (\ref{s2wmz}) and (\ref{a3invmz}).
Provided that these additional states exist at the appropriate scales,
we shall find that the presence of this additional matter results in agreement
between the hypothesis of direct unification at the string scale, and the
values of the low-energy parameters $\sin^2\theta_W(M_Z)$ and
$\alpha_3(M_Z)$.
Of course, there exist a large number of possible scenarios for the
mass scales of the extra states which will allow direct unification at the
string scale, and a classification of all these possibilities is beyond the
scope
of this paper.
Indeed,  such an analysis would require a detailed investigation of each
individual
string model, its full spectrum, and its renormalizable and
non-renormalizable superpotentials.
This is currently being investigated \cite{krdaef}.
However, the essential point that cannot be
overemphasized is that these string models,
in general, produce just the sort of
spectrum with additional matter as required to have the possibility of
direct string unification.

In the realistic free-fermionic models we have studied, the three generations
are obtained from the sectors $\b_1$, $\b_2$, and $\b_3$. The
electroweak doublets that couple to the states from these sectors, and
which therefore may correspond to the MSSM Higgs doublets, are obtained from
the Neveu-Schwarz sector and a sector which is a combination of the vectors
$\{\b_1,\b_2,\b_3,\alpha,\beta\}$. For example, in the models of
Sects.~4.2 and 4.3, the electroweak doublets are obtained from the
Neveu-Schwarz sector and the
$\b_1+\b_2+\alpha+\beta$ sector. The Neveu-Schwarz sector and the
$\b_1+\b_2+\alpha+\beta$ sector may, in general, also produce color triplets.
Color triplets from these sectors couple to the massless fermions from
the sectors $\b_1$, $\b_2$, and $\b_3$. Their mass scale is therefore
restricted from proton decay through Higgsino exchange.
However, for these two sectors there exist a superstring doublet-triplet
splitting mechanism through which the color triplets are removed from
the spectrum  via GSO projections \cite{DTSM}.

In addition to the massless spectrum from the above sectors,
the free-fermionic models may contain additional color triplets and
electroweak doublets form several other sectors.  In the $SO(6)\times SO(4)$
and the flipped $SU(5)\times U(1)$ models, additional ${\bf 16}$ and
$\overline{\bf 16}$
representations are obtained from the additional vectors beyond the NAHE set
that are used to reduce the number of generations to three.  Typically, there
are one or two
of such pairs.
These produce the $(3,2)_{1/6}$, $(\bar3,1)_{1/3}$, $(\bar3,1)_{-2/3}$, and
$(1,2)_{-1/2}$
representations of $SU(3)\times SU(2)\times U(1)_Y$. These states have the
usual
one-loop beta-function coefficients.
In the standard-like superstring models, additional color triplets and
electroweak
doublets may also appear from sectors which arise from combinations of the
vectors
$\{\b_1,\b_2,\b_3\}+\{\alpha,\beta,\gamma\}$. These states, in general, do not
have the
standard weak hypercharge assignments, and therefore their one-loop
beta-function coefficients will be different from those of the MSSM
representations.
For example, in the model of Ref.~\cite{GCU}, there exist color
triplets $\lbrace D_3,\overline{D}_3\rbrace$
and electroweak doublets $\lbrace \ell,\overline{\ell}\rbrace$
with the following beta-function coefficients:
\beq
      \pmatrix{ b_{SU(3)} \cr b_{SU(2)} \cr b_{U(1)}
\cr}_{D_3,\overline{D}_3}~=~
             \pmatrix{1/2\cr 0\cr 1/20\cr}~,~~~~~~~~~~
      \pmatrix{ b_{SU(3)} \cr b_{SU(2)} \cr b_{U(1)}
\cr}_{\ell,\overline{\ell}}~=~
               \pmatrix{0\cr 1/2\cr 0\cr}~.
\label{tripletswithoutstandardcharges}
\eeq

To estimate the effect of the intermediate thresholds induced by such
additional matter,
we shall examine in detail the spectrum of the model of Sect.~4.3.  In order to
be as
general as possible,
we will start from (\ref{s2wmz}) and (\ref{a3invmz}) and from the
experimental constraints on the low-energy observables, and derive
the corresponding constraints on these intermediate matter thresholds.
As our low-energy experimental constraints, we shall impose
\beqn
           0.230~<~\sin^2\theta_W(M_Z)~<~0.233\nonumber\\
           0.110~<~\alpha_3(M_Z)~<~0.135~.
\label{explimits}
\eeqn
As before, we set $M_{\rm string}=5\times 10^{17}$ GeV, and take
$k_1=5/3$ and $\alpha_{\rm e.m.}(M_Z) =1/127.9$.
We set all of the sparticle masses to be degenerate at $M_Z$.
These values, along with the corrections from two-loop contributions, Yukawa
couplings,
and scheme-conversion,
as well as the values for $\Delta_i$ given in (\ref{DELTAS274}),
are then inserted into the expressions for
$\sin^2\theta_W(M_Z)$ and $\alpha_3(M_Z)$
given in (\ref{s2wmz}) and (\ref{a3invmz}) respectively.
Using our experimental limits (\ref{explimits}),
we then obtain the constraints
\beqn
   & 10.29 ~<~ \sum_i \,(b_{2_i}-b_{1_i})\, \ln {\displaystyle {M_S\over{M_i}}}
    ~<~ 14.14~& \nonumber\\
   & 39.22 ~<~ - \sum_i \, \biggl\lbrack
   k_1b_{1_i}+b_{2_i}-(1+k_1)b_{3_i}\biggr\rbrack \,
                             \ln {\displaystyle {M_S\over{M_i}}}~<~ 67.40~.&
\label{imte1}
\eeqn
We can also combine these two equations
to obtain
\beq
            18.57~<~\sum_i\,(b_{3_i}-b_{1_i})\ln{M_S\over{M_i}}~<~ 30.58~.
\label{imte2}
\eeq

We can now use (\ref{imte1}) and (\ref{imte2}) to
examine the possible general scenarios for intermediate matter states.
Our first observation is that in order to accommodate the low-energy
parameters, both
intermediate color triplets and electroweak doublets are needed.
Let us now focus on the representations that are available
in the string model of Refs.~\cite{GCU,custodial}.
In this model, there are three color triplets in
vector-like representations, two with the one-loop
beta-function coefficients of (\ref{tripletswithstandardcharges}),
and one with the one-loop beta-function coefficients
of (\ref{tripletswithoutstandardcharges}).
There are five pairs
of electroweak doublets from the Neveu-Schwarz and $\b_1+\b_2+\alpha+\beta$
sectors with the quantum numbers
\beq
\pmatrix{ b_{SU(3)} \cr b_{SU(2)} \cr b_{U(1)} \cr}_{h,\overline{h}}~=~
               \pmatrix{0\cr 1/2\cr 3/10\cr}~,
\label{doubletswithstandardcharges}
\eeq
and three pairs of electroweak doublets
with the quantum numbers of (\ref{tripletswithoutstandardcharges}).
Let us suppose that all of the intermediate thresholds are
from states that fit into the ${\bf 5}$ and $\overline{\bf 5}$ representations
of $SU(5)$. For one such pair, we obtain from (\ref{imte1}) and (\ref{imte2})
the constraints
\beqn
        30.95~<~\ln{M_S\over{M_3}}-\ln{M_S\over{M_2}}~<~...\nonumber\\
        27.725~<~\ln{M_S\over{M_2}}-\ln{M_S\over{M_3}}~<~...
\label{only5bar5}
\eeqn
where the upper limits are not important.
It is clear that these two equations cannot be satisfied for any values of
$M_2$ and $M_3$, or for any number of doublets and triplets.
We therefore conclude that
in any string models containing only those states that fit into ${\bf 5}$ and
$\overline{\bf 5}$
representations of $SU(5)$,
intermediate matter thresholds cannot account for the disagreement with the
experimentally observed values of $\sin^2\theta_W(M_Z)$ and $\alpha_3(M_Z)$.
It is an important conclusion that some string models can actually be ruled out
in this way.

By contrast, in the model of Sect.~4.3, the massless spectrum contains
not only states that fit into the
${\bf 5}$ and $\overline{\bf 5}$ representations of $SU(5)$,
but also color triplets and electroweak doublets with exotic weak
hypercharge assignments.
These therefore cannot be fit into $SO(10)$ representations.
In this particular model, there are two pairs of color triplets
$\lbrace D_1, \overline{D}_1, D_2, \overline{D}_2 \rbrace$ with
one-loop beta-function coefficients
\beq
      \pmatrix{ b_{SU(3)} \cr b_{SU(2)} \cr b_{U(1)}
            \cr}_{D_1,\overline{D}_1,D_2,\overline{D}_2}~=~
       \pmatrix{ 1/2\cr 0\cr 1/5\cr}~
\label{tripletswithstandardcharges}
\eeq
in addition to one pair of color triplets
with the quantum numbers of (\ref{tripletswithoutstandardcharges}),
and three pairs of electroweak doublets with the quantum numbers of
(\ref{tripletswithstandardcharges}).  We shall set the masses of these
three doublet pairs to be degenerate at one scale, $M_2$, and examine possible
scenarios for the masses $M_3$ of the color triplets.
With one light color triplet, for example
$\{D_1,\overline{D}_1\}$, we obtain the limits
\beq
      {\rm experimental~limit}~<~M_3~<~ 18141~{\rm GeV}~.
\label{abcdefg}
\eeq
Setting $M_3$ at the upper limit of (\ref{abcdefg}),
we find
\beq
             7.2\,\times\,10^{13}~{\rm GeV}~<~ M_2 ~<~2.6\,\times\,10^{14}~{\rm
GeV}~,
\eeq
while for a lower limit of $M_3\sim500$ GeV we obtain
\beq
            3.6\,\times\,10^{5}~{\rm GeV}~<~M_2~<~1.7\,\times\,10^{6}~{\rm
GeV}~.
\eeq
By contrast, with {\it two}\/ triplet pairs degenerate at one mass scale $M_3$,
we
instead find
\beq
                4.3\,\times\,10^{6}~{\rm GeV}~<~ M_3 ~<~ 9.5\,\times\,
10^{10}~{\rm GeV}~,
\eeq
so that taking the upper limit for $M_3$ yields
\beq
           7.2\,\times\, 10^{13}~{\rm GeV}~<~ M_2 ~<~ 2.6\,\times\,
10^{14}~{\rm GeV}~
\eeq
while the lower limit on $M_3$ yields
\beq
               5\,\times\, 10^{12}~{\rm GeV}~<~ M_2
        ~<~1.8\,\times\,10^{13}~{\rm GeV}~.
\eeq
Finally, with all three color triplet pairs degenerate at the scale $M_3$,
we find
\beq
               2.4\,\times\,10^{11}~{\rm GeV}~<~ M_3~<~
7.2\,\times\,10^{13}~{\rm GeV}
\eeq
for which the upper and lower limits respectively yield
\beqn
    3.7\,\times\, 10^{14}~{\rm GeV}~&<~ (M_2)_{\rm upper}~&<~ 1.1\,\times\,
10^{15}~{\rm GeV}~\nonumber\\
    5.7\,\times\, 10^{13}~{\rm GeV}~&<~ (M_2)_{\rm lower}~&<~ 2\,\times\,
10^{14}~{\rm GeV} ~.
\eeqn

\begin{figure}[hbt]
\centerline{\epsfxsize 3.0 truein \epsfbox {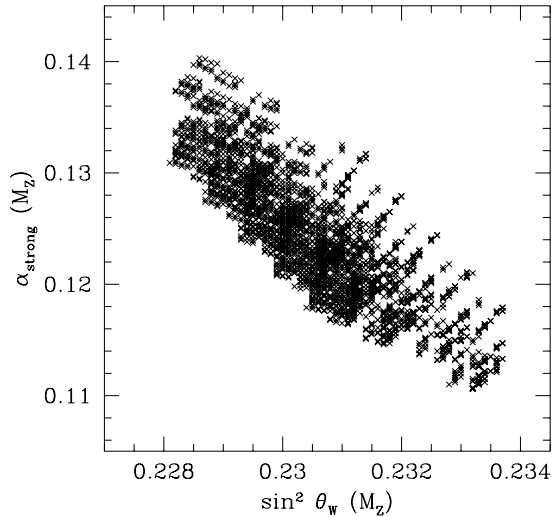}}
\caption{Scatter plot for
      $\lbrace\sin^2\theta_W(M_Z),\alpha_{3}(M_Z)\rbrace$,
      as in Fig.~\protect\ref{lightsusyfigs},
    except that the intermediate matter thresholds are now also included
     in the analysis.}
\label{scatterwithmatter}
\end{figure}

Clearly, many viable scenarios exist, and the above examples are
not exhaustive. In Fig.~\ref{scatterwithmatter} we plot $\sin^2\theta_W(M_Z)$
versus
$\alpha_3(M_Z)$, with $M_3=2\,\times\,10^{12}$ GeV and
$M_2=2\times 10^{14}$ GeV.   We have included in this analysis the two-loop,
Yukawa,
and scheme-conversion corrections given in (\ref{s2wtwoloopYscc}) and
(\ref{a3invtwoloopYscc}),
as well as the heavy string threshold corrections and the light-SUSY
corrections due to the
splitting of the sparticle spectrum. The unification scale is varied between
$3\times10^{17}~{\rm GeV}\leq M_{\rm string}\leq 7\times10^{17}~{\rm GeV}$, and
$0.03\leq \alpha_{\rm string}\leq 0.05$.
It is evident from the figure that in this model, the low-energy experimental
parameters can indeed be accommodated, provided that the
thresholds from this extra matter exist at appropriate scales.

%

Finally, we examine possible string models that contain additional
$(3,2)_{1/6}$ representations. These representations
are obtained, for example, in the flipped
$SU(5)\times U(1)$ model \cite{AEKN}
and in the $SO(6)\times SO(4)$ models \cite{ALT}, and arise
from the additional ${\bf 16}$ and $\overline{\bf 16}$ representations.
These may also arise in the standard-like string models.
For example, in the model of Ref.~\cite{FNY}, a change in the sign of the
GSO-projection phase $C\left\lbrack \matrix{\b_4\cr \bone\cr }\right\rbrack$
produces an additional $(3,2)_{1/6}$ representation from the
sector ${\bf b}_4$.
However, a single $(3,2)_{1/6}$ representation (or even several such
representations)
is not sufficient to accommodate the low-energy data, and
indeed at least one $(3,1)_{1/3}$ representation is needed.
With one pair of each, we  find
\beq
            4\,\times\,10^{13}~{\rm GeV}~ <~M_{32}~<~1\,\times\, 10^{15}~{\rm
GeV}~,
\eeq
with the corresponding constraints
\beqn
         3\,\times\, 10^{6}~{\rm GeV}~&<~(M_{31})_{\rm
upper}~&<~1.5\,\times\,10^{15}~{\rm GeV}
               ~\nonumber\\
         5\,\times\, 10^{2}~{\rm GeV}~&<~(M_{31})_{\rm
lower}~&<~2.6\,\times\,10^{11}~{\rm GeV}~.
\eeqn

In all of these cases,
it is an important observation that such ranges exist in which
string-scale gauge coupling unification is consistent with low-energy data.
It is of course obvious that the presence
of extra matter can in principle resolve the discrepancy between the GUT and
string unification scales.  What is highly non-trivial, however, is that
precisely the required sorts of extra states
naturally appear in some of the realistic free-fermion models
we have examined, with the necessary non-standard hypercharges to do the
job.
Indeed, the crucial representations, as we have seen, are those
for which the one-loop beta-function coefficients $b_2$ and $b_3$
are rather large, while $b_1$ is small.  It is for this reason
that this particular string-predicted extra
matter is able to modify the running of the
strong and electroweak couplings
without substantially affecting the $U(1)$ coupling.

\subsection{General conclusions}

It is apparent from the analysis presented in this section
that string gauge coupling unification imposes a strong constraint
on the allowed string models.
Models that would otherwise provide a very attractive low-energy phenomenology
can be ruled out on the basis that gauge coupling unification
at the string scale cannot be in agreement with the low-energy data.
The model of Ref.~\cite{EU} is an example of such a model.
In this model, the extra color triplets and electroweak doublets that appear
all have quantum
numbers that fit into ${\bf 5}$ and $\overline{\bf 5}$
representations of $SU(5)$. Consequently, as we have shown,
the extra states beyond the MSSM in these models
cannot bring the string unification prediction
into agreement with experiment.
We find it very encouraging that some otherwise very appealing string models
can be ruled out on this basis.

Perhaps even more importantly, however,
there exist realistic string models in which there naturally appears additional
matter that
does {\it not}\/ fit into ${\bf 5}$ or $\overline{\bf 5}$ representations of
$SU(5)$.
As we have shown, for such string models the hypothesis of gauge
coupling unification at the string scale can be in agreement with the
low-energy data.
The model of Refs.~\cite{GCU,custodial} is one explicit example of such a
string model
which contains all the needed representations to achieve string gauge coupling
unification,
in just the right combinations and with just the correct hypercharges.

We have seen, then, that intermediate
matter thresholds seem to be the only possible way in which string-scale gauge
coupling
unification can be achieved in realistic level-one string models.
By level-one string models,
we mean string models in which the non-Abelian gauge group content of the
Standard Model
is realized through a level-one Ka\v{c}-Moody algebra.
While our conclusions might be modified for models in which the gauge group
is realized by Ka\v{c}-Moody algebras at higher level, no realistic
three-generation models of this sort have been constructed to date.
Consequently, the need for (and appearance of) an additional matter spectrum
beyond the MSSM
seems to be a {\it prediction}\/ of realistic level-one string models.

It is remarkable that string theory, which predicts an unexpectedly high
unification scale $M_{\rm string}$, in many cases also simultaneously
predicts precisely the extra exotic particles needed to reconcile
this higher scale with low-energy data.
As we have shown, all of the other possible effects
are not likely to bridge the gap between the MSSM and string unification
scales.
This is a profound conclusion, and may have important experimental
consequences.


\setcounter{footnote}{0}
\section{Why are string-induced threshold corrections so small?}

In this section we address the general question of the {\it size}\/ of
the string-induced threshold corrections.  Our results in Sects.~4 and 5 for
the realistic
heterotic free-fermionic models, as well as similar previous calculations
for orbifold
models \cite{Kaplunovsky,MNS} and Type-II superstring models \cite{DL},
all consistently indicate that generic string-induced threshold corrections are
relatively
small.  This result is {\it a priori}\/ surprising, given the infinite towers
of massive states which could in principle affect the running of
the gauge couplings.
Indeed, we have already remarked that the calculation of the heavy
string threshold corrections $\Delta_G$ in (\ref{Deltadef}) is similar to
that of the one-loop cosmological constant $\Lambda$,
\beq
     \Lambda~\equiv~ \int_\calF {d^2\tau \over {\tau_2}^2}\, Z(\tau)~,
\label{Lambdadef}
\eeq
and typical values of this cosmological-constant integral (\ref{Lambdadef})
for non-supersymmetric string models are found \cite{KRDlambda} to
be in the range $\Lambda\sim {\cal O}(10^2)$.
Taking this as a typical scale for such one-loop string amplitudes,
the question then arises as to why a different particular amplitude, namely
the one-loop heavy string threshold correction, should be so highly suppressed.

\subsection{General argument}

This suppression of the string-induced threshold correction
is independent of any particular class of string model-construction,
and is in fact a general property of string theory.  We therefore
seek a model-independent explanation of the underlying reason for
this suppression.

We begin by focusing on the behavior of the modified partition function
$B_G(\tau)$
which serves as the string integrand in the calculation of the threshold
correction
$\Delta_G$ in (\ref{Deltadef}), and in particular let us compare it with
the ordinary partition function $Z(\tau)$ which plays the same role in the
calculation of
the cosmological constant $\Lambda$.  In analogy to (\ref{Bexpand}), we can
always expand the partition function $Z$ in the form
\beq
	Z ~=~ {\tau_2}^{-1}\, \sum_{m,n} a_{mn}\,\qbar^m\,q^n~
\label{Zexpand}
\eeq
so that, in analogy with (\ref{Deltasum}), we have
\beq
      \Lambda ~=~ \sum_{m,n} \, a_{mn}\,I^{(3)}_{mn}~.
\label{Lambdasum}
\eeq
There are therefore two potential sources of difference between the values
of $\Delta$ and $\Lambda$:  the integrals themselves are different due to the
different
powers of $\tau_2$, {\it i.e.}, $I^{(1)}_{mn}\not= I^{(3)}_{mn}$,
and the coefficients may also be different, {\it i.e.}, $b_{mn}\not= a_{mn}$.
Now, while it is true that
$|I^{(1)}_{mn}|$ is usually less
than $|I^{(3)}_{mn}|$, this is not a major source of difference
for two reasons:  first,
$|I^{(1)}_{mn}|$ is  numerically smaller than $|I^{(3)}_{mn}|$
by only a few percent, and second, the signs of these integrals
tend to alternate with different values of $m$ and $n$.
Thus, the major difference between the values of $\Delta$ and $\Lambda$
must involve the relative behavior of the
coefficients $\lbrace a_{mn}\rbrace $ and $\lbrace b_{mn}\rbrace$.

In order to pin down the essential
difference in the behavior of these coefficients,
let us first recall their physical interpretations.
In the ordinary partition function $Z$,
each coefficient $a_{mn}$ represents simply the net number of
states with spacetime squared masses $(M_R^2,M_L^2)=(m,n)$, where ``net''
refers to the difference between the numbers of spacetime bosonic and
fermionic states.  By contrast, in the modified partition functions $B$,
the coefficients $b_{mn}$ represent
the {\it charges}\/ of these states relative to the gauge group in question.
Thus, states only contribute to $b_{mn}$ if they carry a non-zero gauge charge.

Now, the general behavior of the ordinary state degeneracies $a_{mn}$
in string theory is well-known (see, for example, Ref.~\cite{missusy}).
In particular, there are two generic features which concern us.  The first
is the appearance of so-called ``unphysical tachyons'' in non-supersymmetric
heterotic string theories --- {\it i.e.}, states which contribute to
coefficients $a_{mn}$ with $m+n<0$
(``tachyonic'') but $m\not= n$ (``unphysical'').
More specifically, it can be shown \cite{missusy}
that {\it any}\/ non-supersymmetric heterotic string theory which contains
gravitons will have $a_{0,-1}\not=0$.  (Note that this does not imply that
the corresponding spacetime spectrum contains physical tachyons; indeed,
the absence of {\it physical}\/  tachyons
requires only that $a_{nn}=0$ for all $n<0$.)
By contrast, physically sensible string models
will never have {\it charged}\/ unphysical tachyons
with energy configuration $(m,n)=(0,-1)$.
This occurs because any state with energy
configuration $(m,n)=(0,-1)$
must arise as the vacuum state
in a Neveu-Schwarz sector,
and since such a vacuum state is
necessarily a gauge singlet,
it cannot carry a non-zero gauge charge.
We therefore find that although $a_{0,-1}\not=0$
for all non-supersymmetric heterotic string models,
we must have $b_{0,-1}=0$  regardless of whether spacetime
supersymmetry is present.

This is a crucial distinction, because the potential contributions from
the $(m,n)=(0,-1)$ unphysical tachyons
are typically larger than those from any other state.  Indeed, the integral
$I^{(s)}_{0,-1}$ is typically an order of magnitude greater than any
other integral $I^{(s)}_{m,n}$.
This feature alone, therefore, is responsible for a
sizable reduction in the total threshold correction.

Of equal importance, however, is the second generic difference between
the coefficients $a_{mn}$ and $b_{mn}$:  their behavior as $m,n\to\infty$.
As is well-known, the number of physical states in string
theory grows {\it exponentially}\/ with energy, so that
\beq
	a_{nn} ~\sim~ A \,n^{-B}\,e^{C\sqrt{n}} ~~~~~{\rm as}~ n\to\infty~
\eeq
where $A,B,C$ are constants.
This exponential growth in the number of states is the
famous Hagedorn phenomenon
which signals the existence of either a maximum (Hagedorn) temperature,
or string phase transition.
Indeed, as discussed in \cite{missusy}, the existence of such exponential
growth in the number of string states at
high energy is directly related, through
modular invariance, to the existence of physical and/or unphysical tachyons at
the low energy of the string spectrum, and is hence unavoidable.
Of course, this rapid growth in the number of physical states does not lead
to a divergent cosmological constant $\Lambda$, for there is an even stronger
corresponding suppression for the integrals $I_{nn}^{(s)}$:
\beq
	 I_{nn}^{(s)}~\sim~ e^{-C' n}~ ~~~~~{\rm as}~ n\to\infty~
\eeq
where $C'$ is a positive constant.
Nevertheless, the fact that the $a_{nn}$ grow so quickly
opens up a significant range of values of energy $n$
for which the contributions
of massive states of energy $n$ to the cosmological
constant are still sizable.
Thus, the cosmological constant receives important
contributions not only from the unphysical
tachyonic states (and the physical massless states),
but also from the first several
massive states.

By contrast,
for the threshold corrections,
this second source of contributions is often removed as well,
for in the case of certain threshold corrections it can be shown
that the coefficients $b_{nn}$ will exhibit growth which
is at most {\it polynomial}\/ rather than exponential \cite{missusy}.
This observation, which is ultimately related to the modular
properties of the modified partition function $B(\tau)$,
will be discussed below.  In particular, we shall see that
this suppressed polynomial rate of growth occurs for those functions $B(\tau)$
which arise solely from gauge-charge insertions of the form $Q_L Q_M$
with $L\not= M$, and for which
there are no contributions from charged unphysical tachyons.
Indeed, a quick scan
of the intermediate results listed in Sect.~4 for the appropriate
tachyon-free models verifies
that the values of $\Delta$ for those cases with $L\not= M$ are further
suppressed by a sizable amount
relative to those with $L=M$.
Thus, in these cases, not only are the contributions
from the unphysical tachyonic
states absent (as are the contributions from the massless states),
but even the contributions from the massive states are
extraordinarily suppressed.
Therefore, in these cases, there are {\it two}\/
features which combine to produce
the unusual suppression of the $L\not= M$ string-induced threshold
corrections relative to the corresponding cosmological constant.
Indeed, for certain free-fermionic embeddings of the gauge group (such as
that of the flipped $SU(5)$ models), these $L\not=M$ insertions are the only
ones which are relevant.  This double-suppression mechanism is then directly
responsible for the diminished size of total relative threshold corrections
such as $\Delta_{SU(5)}-\Delta_{U(1)}$.

Let us now briefly discuss the cases in which charged unphysical tachyons do
appear.
In these cases, although there will be charged unphysical tachyons making
contributions
to the threshold corrections $\Delta_G$,
we still must have $b_{0,-1} =0$ (for the general reasons indicated above).
Thus, the unphysical $(m,n)=(0,-1)$ tachyons which would have led to the
largest contributions
to the threshold corrections must still be absent.
For example, the charged unphysical tachyons that end up
contributing to the
threshold corrections in the non-supersymmetric
$SU(3)\times SU(2)\times SU(1)$ string model discussed in Sect.~4.5
are all of a more harmless variety,
with energies $m$ and $n$ never more negative than $-1/4$:
\beq
     {\tau_2}^{-1}\,B(\tau) ~=~
      (4 \,\qbar^{3/4} ~+~ 120\, \qbar^{7/4} ~+~ ...)\,q^{-1/4}
                 ~+~ {\rm non-tachyonic}~.
\eeq
Indeed, these charged unphysical tachyonic states do not contribute nearly
as much as the $(0,-1)$ states, since their corresponding
integrals are highly suppressed, with
$I^{(1)}_{3/4,-1/4}\approx -1.44\times 10^{-3}$
as compared to $I^{(1)}_{0,-1}\approx -11.04$.
Furthermore, even though these milder charged unphysical tachyonic states
may lead to exponentially growing values of $b_{mn}$ even
for the $L\not= M$ basis insertions,
modular transformations can be used to show that
the resulting {\it rate}\/ of exponential growth is
also highly suppressed;  this suppression is ultimately due to
the fact that these tachyons are never as tachyonic as the
$(0,-1)$ tachyons which would otherwise control the
rate of growth \cite{missusy}.
Thus, even in the cases that charged unphysical tachyons appear,
the above arguments remain intact, for its two main features,
namely that
\beq
        b_{0,-1}=0 ~~~~~~~{\rm and}~~~~~~~
         b_{mn}\ll a_{mn} ~~{\rm as}~~m,n\to\infty~,
\label{argument}
\eeq
remain unaltered.
Thus the string-induced threshold corrections
continue to be suppressed in spite of the appearance of such unphysical
tachyons.

Finally, we remark that this type of analysis can also incorporate
the one case in which string-theoretic threshold corrections are {\it known}\/
to be large \cite{louis,ILR}:  namely, as the values of particular
moduli (such as radii of compactification) are taken to infinity.
Indeed, we shall see below that taking such a limit changes the modular
properties of the modified partition function $B(\tau)$ in so fundamental a way
that the above suppression mechanisms (\ref{argument}) no longer apply.
Thus, even in this case, the sizes of the threshold corrections
can still be understood as a consequence of the modular properties
of the modified partition functions $B(\tau)$ and the behavior of
their coefficients $b_{mn}$.

\subsection{The modular properties of the modified partition function}

We now turn to the modular properties of the modified
partition functions $B(\tau)$, and the resulting behavior of their coefficients
$b_{mn}$ as $m,n \to \infty$.

As discussed above, if a given string model is devoid of charged physical
or unphysical tachyons, then
the corresponding
modified partition function $B(\tau)$ must have
coefficients $b_{mn}$ which always vanish for all $m+n<0$,
and which often grow at most {\it polynomially}\/ as $m,n\to\infty$.
However, this is behavior for the coefficients is certainly unusual, given the
expectations based
on analyzing ordinary partition functions $Z(\tau)$.
Indeed, it can be shown \cite{missusy} that any partition function $Z(\tau)$
with $a_{mn}=0$ for all $m+n<0$ must in fact have {\it all}\/ coefficients
vanishing, so that $Z=0$ identically;  otherwise, if these unphysical tachyonic
coefficients
are non-zero, then the partition function $Z$ must have coefficients $a_{mn}$
which grow {\it exponentially}.
It turns out that this is a general theorem which applies to all partition
functions
$Z$ which correspond to string theories in $D>2$ uncompactified spacetime
dimensions
(or equivalently, for all modular functions with modular weights $k<0$).
The question then arises as to how the {\it modified}\/ partition functions
$B_G(\tau)$ manage
to evade these constraints, and survive to exhibit non-zero coefficients
$b_{mn}$ at
higher levels despite having no tachyonic contributions in many cases.
Moreover, we also wish to determine the underlying reason why this growth is
fundamentally different depending on the form of the $Q_L Q_M$ insertion,
polynomial if $L\not= M$, and  exponential otherwise.
We also wish to understand why the limit of large moduli changes these
results, and allows the corresponding threshold corrections to grow large.

To answer these questions, we must first recall some facts about modular
functions.
In general, modular functions $f_i(\tau)$ (such as conformal-field-theoretic
characters $\chi_i$) transform covariantly under
modular transformations,
\beq
	     f_i\left({a\tau+b}\over{c\tau+d}\right) ~=~
	 (c\tau+d)^k \,\sum_{j}\,M_{ij}\,f_j(\tau)~,
\label{modcovariant}
\eeq
where the exponent $k$ is called the {\it modular weight}\/ and where the
matrix $M_{ij}$
is a mixing matrix which represents the particular modular transformation in
the space
of functions $f_i$.
One then builds a full holomorphic/anti-holomorphic modular invariant function
(such as a partition function) by combining two such sets of characters
$f_i$ and $g_i$, each with modular weight $k$, in the form
\beq
	   Z(\tau,\overline{\tau})~=~
	  {\tau_2}^k\, \sum_{ij} \,N_{ij}\,\overline{g_i(\tau)}\, f_j(\tau)
\eeq
where $\tau_2\equiv {\rm Im}\,\tau$ and where $N_{ij}$ is a matrix chosen
to satisfy $\tilde M^\dagger N M = N$ (where $\tilde M$ and $M$ are
respectively the modular
transformation representation matrices
in the spaces of functions $g_i$ and $f_i$).
For example, for the classes of free-fermionic string models we have been
examining,
our fundamental modular functions are the $\eta$ and $\Theta$ functions which
appear in
(\ref{trace1}) and (\ref{trace2}), and from (\ref{trace1}) we see that
the total combined holomorphic and anti-holomorphic functions $f_i$ and $g_i$
take the
general schematic forms
\beqn
	    f &\sim& \eta^{-24}\,\Theta^{22} \nonumber\\
	    g &\sim& \eta^{-12}\,\Theta^{10} ~.
\eeqn
Since the $\eta$ and $\Theta$ functions are modular functions with weights
$k=1/2$,
we see that the total partition functions $Z$ for these models all have the
total
modular weight $k= -1$.  Note that this is consistent with the factor of
${\tau_2}^{-1}$ which appears in (\ref{trace1}).
Indeed, the general relation between the total modular
weight $k$ of the partition function $Z$
and the spacetime dimension of the corresponding string theory is
\beq
	 k ~=~ 1 -D/2~.
\eeq

For the {\it modified}\/ partition functions $B$,
however, this is no longer the
case, for the effect of the charge insertions into
the trace is to {\it increase}\/ the modular weight
from $k$ to $k+2$.  We can see this easily as follows.
Let us first define the more general $\Theta$ function of
two variables $z$ and $\tau$,
\beq
	\t{\alpha}{\beta}(z|\tau)~\equiv~
	  \sum_{n= -\infty}^\infty\,
         e^{2\pi i(z+\beta)(n+\alpha)}\,q^{(n+\alpha)^2/2}~,
\eeq
so that our usual $\Theta$ functions of a single variable $\tau$
can be obtained from these more general functions by projecting to $z=0$:
\beq
	\t{\alpha}{\beta} (\tau)~=~ e^{-2 \pi i \alpha \beta}\, \biggl\lbrace
	  \t{\alpha}{\beta}(z|\tau)\biggr\rbrace\bigg|_{z=0}~.
\eeq
Then, in terms of this generalized function,
the singly primed function $\Theta'$ which results from the
insertion of a single charge operator $Q$ into the trace can be written
as the result of a single derivative with respect to $z$:
\beq
	\tp{\alpha}{\beta} (\tau)~=~ e^{-2 \pi i \alpha \beta}\,
        \biggl\lbrace {1\over 2 \pi i} \,{d\over dz}\,
          \t{\alpha}{\beta}(z|\tau)\biggr\rbrace\bigg|_{z=0}~.
\label{onederivative}
\eeq
Likewise, the doubly primed function $\Theta''$, which results from the
insertion
of {\it two}\/ charge-operator insertions, can be obtained from
$\Theta(z|\tau)$ via a
second derivative with respect to $z$:
\beqn
	\tpp{\alpha}{\beta} (\tau)&=& e^{-2 \pi i \alpha \beta}\, \biggl\lbrace
    \left({1\over 2\pi i}\right)^2\,{d^2\over dz^2}\,
       \t{\alpha}{\beta}(z|\tau) \biggr\rbrace\bigg|_{z=0}~\nonumber\\
	  &=& {1\over i\pi}\,{d\over d\tau}\,\t{\alpha}{\beta}(\tau)~.
\label{twoderivative}
\eeqn

Now, the general $\Theta(z|\tau)$ functions have modular transformations of
the form
\beq
	 \Theta\left({z\over c\tau+d}\bigg|{a\tau+b\over c\tau+d}\right)
       ~\sim~ (c\tau+d)^{1/2}\, \exp\left( {i\pi c z^2\over c\tau+d}\right)\,
\Theta(z|\tau)~
\label{Thetatrans}
\eeq
where we have neglected overall $\tau$- and $z$-independent phases and
mixing matrices.  From this result [in particular the
exponent of the $(c\tau+d)$ factor],
we easily see that the modular weight of each $\Theta$-function is $k=1/2$.
However, using (\ref{Thetatrans}) and taking a derivative with respect to $z$
before
the projection to $z=0$,
we find that $\Theta'$ transforms just like $\Theta$ except with exponent
$k+1=3/2$.
Likewise, a second derivative with respect to $z$ further increases the
exponent
to $k+2=5/2$.
Now, recall from (\ref{modtracegen}) that the modified partition functions $B$
contain a total of two (helicity) charge insertions for the right-moving
anti-holomorphic sector, and two corresponding (gauge) charge
insertions for the left-moving holomorphic sector.
Thus, the total modular weight of the modified partition function $B$
is not the negative value $k= -1$ which we would have expected for four
dimensional theories, but rather $k'=k+2=+1$.
Note that this is also consistent with the two extra factors of $\tau_2$
which were inserted along with the charge insertions into the trace in
(\ref{modtracegen}).

It is for this reason that $B$ can be non-zero even though it may contain no
contributions
from physical or unphysical tachyons.  Indeed, the theorem mentioned above,
which
would have forced such functions to vanish, holds only for functions with
negative
modular weights.  By contrast, as discussed in Ref.~\cite{missusy}, the
coefficients for
tachyon-free functions with {\it positive}\/ modular weights are expected
to grow polynomially.

Given that the coefficients $b_{mn}$ are expected to grow polynomially for
tachyon-free theories, we still
must explain why this growth is in fact exponential for the cases of gauge
charge
insertions $Q_L Q_M$ with $L=M$, or for the limit of large moduli.
Indeed, such exponential growth is more
typical for
partition functions of negative modular weight.

It turns out that for the cases of insertions $Q_L Q_M$ with $L=M$, this
behavior
is caused by a modular anomaly which prevents the modified partition functions
$B(\tau)$
from being truly modular-invariant.
We can see how this anomaly arises as follows.
We have already shown above in (\ref{onederivative}) that a single charge
insertion $Q_L$
is equivalent to differentiation with respect to $z$ followed by projection to
$z=0$,
so that if a certain trace $f_i(\tau)$ without any charge insertion transforms
modular-covariantly
as in (\ref{modcovariant}), then the same trace with a single
charge insertion $f'_i(\tau)$ will
transform modular covariantly with increased modular weight:
\beq
	     f'_i\left({a\tau+b}\over{c\tau+d}\right) ~=~
	 (c\tau+d)^{k+1} \,\sum_{j}\,M_{ij}\,f'_j(\tau)~.
\eeq
However, this is not the case for the traces $f''_i(\tau)$ with {\it two}\/
identical
charge insertions.  As shown in (\ref{twoderivative}), the insertion of two
identical
charges $Q_L$ is tantamount to two $z$-derivatives
or a single $\tau$-derivative, yet
these derivatives are not covariant with respect to modular transformations.
Indeed, it is easy to see that if $f_i(\tau)$ transforms as in
(\ref{modcovariant}) under
$\tau\to \tau'\equiv(a\tau+b)/(c\tau+d)$, then
\beq
    {d\over d\tau}f_i(\tau)\to {d\over d\tau'}f(\tau') ~=~
      (c\tau+d)^{k+2}\,\sum_j M_{ij} {d\over d\tau}f_j(\tau)
           ~+~ ck (c\tau+d)^{k+1} \,\sum_j M_{ij} f_j(\tau)~.
\eeq
While the first term is of the proper covariant form, the second term is not.
Rather, the true ``covariant derivative'' on the space of modular functions of
weight $k$ is actually
\beq
	 D ~\equiv~ {d\over d\tau} ~-~ {ik\over 2\,\tau_2}~;
\eeq
here the contribution from the second term cancels the anomaly caused by the
first.
Thus, double gauge charge insertions of the form $Q_L Q_M$ with $L=M$
destroy the modular covariance of the holomorphic or left-moving sector,
transforming a modular function of weight $k=-1$ into not only a
modular function of weight $k+2=+1$,
but also an anomaly term consisting of
a modular function of the original weight $k= -1$ divided by $\tau_2$.

It is this feature which ultimately explains why exponential growth of the
coefficients
$b_{mn}$ can occur for these double gauge-charge insertions.
Recall that regardless of the particular gauge-charge insertion for the
left-moving
sector, there is always a double charge insertion
for the right-moving sector --- this is the helicity
insertion ${\overline{Q}_H}^2$.
Thus, for the $L=M$ cases, the anomaly term from the right-moving sector can
combine
with the anomaly term from the left-moving sector to recreate a modular
function
with equal left- and right-moving modular weights $k= -1$.
Indeed, such a term will appear divided by two powers of $\tau_2$, just as
required
for reproducing a proper modular function of weight $k= -1$.
Thus, for the $L=M$ cases, a remnant of the original
$k= -1$ behavior survives despite
the charge insertions.  It is this which is ultimately responsible for the
exponential
growth rate for the coefficients $b_{mn}$ in these cases.

A similar mechanism occurs in the cases of models for which a modulus
(such as a radius of compactification) is taken to infinity.
In these cases, we are effectively changing the dimensionality
of the theory, so that once again the modular weight $k$ of
the modified partition function is decreased below zero.\footnote{
     We thank I. Antoniadis for discussions on this point.}
Indeed, the act of taking a radius of compactification to infinity
amounts to replacing
\beq
       {\overline{\Theta}\,\Theta \over \overline{\eta}\, \eta}~~
      \longrightarrow ~~ {1\over \sqrt{\tau_2}} \,{1\over
\overline{\eta}\,\eta}~
\eeq
in the modified partition functions $B(\tau)$.
This occurs because the $\Theta$-functions, which had previously represented
the lattice sums
over discrete momentum- and winding-mode vacua, now become continuous integrals
which
can be evaluated, yielding mere extra inverse factors of $\tau_2$ in the
decompactification limit.
Thus, the effective modular weight of the remaining function $B(\tau)$ is
decreased in this limit of large moduli,
and the above arguments then lead
to larger values of threshold corrections.

Finally, we remark that since the helicity
insertion ${\overline{Q}_H}^2$ always takes the form of
a double-charge insertion (or equivalently a single $\tau$-derivative, with its
associated modular anomaly), the
modified partition functions $B(\tau)$ are never modular-invariant in any case.
However, an improved derivation of the
string-theoretic gauge coupling threshold corrections
which is manifestly modular-invariant has recently appeared in
Ref.~\cite{Kiritsis}.
In this analysis, manifestly modular-invariant results are achieved
by explicitly introducing an infrared regulator
(provided by introducing a curved spacetime), and by taking into account
the back-reaction from both gauge and gravitational interactions.
Moreover, these modifications result in expressions
for threshold corrections which yield
their {\it absolute}\/ sizes, rather than merely their relative differences.
These expressions also incorporate the additional contributions (such as those
due to dilaton tadpoles) which must be included in the cases of string
models without spacetime supersymmetry.
We hope to repeat our threshold analysis using these expressions in a future
work.

\setcounter{footnote}{0}
\section{Conclusions}

In this paper we examined in detail the problem
of gauge coupling unification in realistic heterotic string models.
The class of models that we studied are all
constructed in the free-fermionic formulation, and are among
the most realistic string models constructed to date.
We believe that this fact is not accidental, but may reflect
deeper properties of string compactification that are at present unknown.
Indeed, the free-fermionic models are constructed at
a highly symmetric point in
the string compactification space, and the $Z_2\times Z_2$ orbifold structure
that is realized in these realistic free-fermionic models may be deeply
connected to
the existence of only three light generations in nature.

Despite their many attractive properties, however, the realistic free-fermionic
models
(and string theory in general) predict that the gauge couplings unify
at the string scale, which is of approximately $5\times 10^{17}$ GeV.
This implies that if the string spectrum below
the string scale is assumed to be that of the MSSM, then
the string-predicted values of the low-energy parameters
$\sin^2\theta_W(M_Z)$ and $\alpha_3(M_Z)$
will be in disagreement with the experimentally observed values.
We explicitly illustrated this disagreement by evaluating the associated
renormalization group equations including two-loop, Yukawa-coupling,
and scheme-conversion corrections.

The question then arises as to whether this disagreement with the low-energy
observables can be used to rule out this class of realistic string models,
or whether other effects may arise to alter this conclusion.
To answer this question, therefore, it is necessary
to examine all of the possible effects that can modify this result.

One possibility is that the tree-level string predictions may
be modified by the
one-loop heavy string threshold corrections due to the infinite tower of
heavy string modes.  We developed a method for evaluating
these threshold corrections in the free-fermionic string models,
and were able not only to evaluate the string threshold corrections to any
desired accuracy, but also to perform various non-trivial consistency checks
in our analysis.
We evaluated these string threshold corrections within a range of realistic
free-fermionic models, and in general we found that the string threshold
corrections are small and cannot explain the disagreement with
the experimentally observed values.  In fact, in all the cases that we
studied, we found that the string threshold corrections
tend to {\it elevate}\/ the string unification scale
by approximately 20\%,
and consequently enhance the disagreement with experiment.
Moreover, we found that the string threshold corrections
are, in general, not significantly affected by the choice of the gauge group,
the existence of spacetime supersymmetry, or the presence of charged unphysical
tachyons.
We were also able to provide a model-independent argument which explains
why such threshold corrections are naturally suppressed in string theory,
except at points with large moduli.
Our argument relied on only the modular properties of the modified
partition functions $B(\tau)$
which enter the calculation of the threshold corrections,
and hence should have validity beyond the class of free-fermionic
string models studied here.
Hence, we conclude that string threshold corrections cannot resolve the
disagreement with the experimentally observed values in these models.

Other possible corrections may arise due to stringy modifications
of the $U(1)_{\hat Y}$
hypercharge normalizations, light SUSY thresholds, additional gauge structure
at intermediate energy scales, or additional matter thresholds at intermediate
energy
scales. We found that smaller values of the $U(1)_{\hat Y}$ normalization $k_1$
can result in agreement with experiment, but we argued on general
grounds that such smaller values of $k_1$ are not possible in self-consistent
string models.
To analyze the possible contributions from light SUSY thresholds,
we parametrized the low-energy SUSY spectrum assuming universal
soft SUSY-breaking terms, and  found that light SUSY thresholds
also cannot bring the string predictions into agreement with experiment.
This conclusion is unaltered even if the assumption of universality is relaxed.
We also examined the possibility of additional gauge structure at intermediate
energy scales,
and showed that within the context of these realistic free-fermionic string
models,
this also cannot resolve the discrepancy.

Finally, we examined the effect of intermediate matter thresholds.
Because we were able to rule out each of the other effects in the realistic
string models, only the presence of additional non-MSSM matter could
possibly reconcile string-scale unification with low-energy data.
In the realistic free-fermionic string models, additional matter beyond
the MSSM generically appears in the massless spectrum.
This matter takes the form of
color triplets and electroweak doublets, in vector-like representations.
While some of these extra matter representations have the weak hypercharge
assignments that are common in grand-unified theories, some
have weak hypercharge assignments that are unique to the string models
and do not arise in regular GUT's. The ultimate mass scales
of these extra states will be determined by the renormalizable
and non-renormalizable terms in the superpotential, but
in general the mass scale of the additional states will not be at
the string scale. Therefore,
in general it is in unjustified to assume that the
spectrum below the string scale is that of the MSSM.
We showed that in some models, these appearance of these additional states
does not resolve the discrepancy between string-scale unification
and low-energy data;  hence these models can be ruled out.
More interestingly, however,
we found that certain other realistic string
models provide just the right combinations of extra matter representations,
with just the right gauge quantum numbers,
to allow string-scale unification to be consistent with the low-energy data.
Indeed, within these models,
we found that a significant window exists for the additional mass scales
so that the predicted values of the low-energy
parameters $\alpha_{\rm strong}(M_Z)$
and $\sin^2\theta_W(M_Z)$ will be in agreement with the
experimentally observed values.
In some of these models ({\it e.g.}, the model of Refs.~\cite{GCU,custodial}),
these extra states have a uniquely stringy origin,
and may have profound experimental implications.
It is then imperative to examine whether the mass scales of these additional
thresholds can be derived from the string models, and take the desired values.
Such work is currently in progress, and will be reported in future
publications.

\bigskip
\medskip
\leftline{\large\bf Acknowledgments}
\medskip

We are pleased to thank S. Chaudhuri, S.-W. Chung,
L. Dolan, J. Louis, J. March-Russell, J. Pati, M. Peskin,
F. Wilczek, E. Witten, and especially I. Antoniadis and
E. Kiritsis for discussions.
This work was supported in part by DOE Grant No.\ DE-FG-0290ER40542.

\setcounter{section}{0}   

\Appendix{List of Models Considered}

Here we provide the explicit $\V_i$ and $k_{ij}$ defining parameters
for each of the models we consider in the text.

As we discussed in Sect.~2, four-dimensional string models
in the free-fermionic construction are defined by a set
of boundary conditions for the 64 worldsheet Majorana-Weyl fermions,
as well as a set of phases which ultimately describe how the generalized
GSO projections are to be performed in each sector.  In the notation
of Ref.~\cite{KLST} which we shall use below, these boundary-condition
vectors are denoted $\V_i$, $i=0,...,N$ (where $N+1$ is the model-dependent
number of vectors which are necessary);  likewise, the phases are defined
through
an $(N+1)\times (N+1)$ matrix denoted $k_{ij}$.  However, most of the
models we analyze originally appeared in the literature in the different
notation of Ref.~\cite{ABK}.  In this latter notation, boundary-condition
vectors
are denoted $\b_i$ and the generalized GSO phases are denoted $\C{\b_i}{\b_j}$.
For completeness, we now give below the explicit mapping between these
two notations:
\beqn
       \b_i &=& -2\, \V_i \nonumber\\
      \C{\b_i}{\b_j} &=& \exp\left\lbrack
	  2\pi i \left(
      \V_i^1 + \V_j^1 + k_{ji} - \V_i \cdot \V_j \right)\right\rbrack
\label{mappingKLTABK}
\eeqn
where $\V_\ell^1$ denotes the first component of the vector $\V_\ell$
(\ie, the component corresponding to the worldsheet fermion $\psi^\mu$ carrying
spacetime
Lorentz indices).
Furthermore, inner products such as $\V_i \cdot \V_j$ are defined
with opposite signs in the two notations, so that
\beq
	\b_i\cdot\b_j ~=~ -4 \,\V_i\cdot \V_j~.
\eeq
Finally, however, we point out that an important issue is the adoption
of a self-consistent scheme for handling the zero-modes of real
Ramond fermions.
In the notation of Ref.~\cite{KLST}, an explicit convention
is established and we have adopted this convention in our calculations;
this is the origin, for example, of the crucial phase contribution
$\Gamma^{\overline{\alpha\V}}_{\overline{\beta\V}}$
which appears in (\ref{phases}).
Without this phase contribution,
our expressions for the partition function $Z(\tau)$ would not
have been modular-invariant.
However, in the literature (and, in particular, in the
papers in which these models first appeared), other conventions
have been implicitly adopted for handling the zero-modes.  Thus,
it is necessary to take these changes of convention into account
when translating between the two notations.  Fortunately,
for the case of all of the present models, these
extra changes of convention can be absorbed through
changes in the $k_{ij}$-phases [beyond
those $k_{ij}$ values implied by (\ref{mappingKLTABK})].
It is only with these additional effective phase changes that the
correct particle spectrum for each model is produced.

Thus, in terms of the exact $(\V_i,k_{ij})$ notation and zero-mode conventions
of Ref.~\cite{KLST},
the models which we have analyzed in this paper
can be listed as follows.
First, as discussed in Sect.~2, they all share the same
first boundary-condition vectors which comprise the so-called
NAHE set.  In the present $\V$-notation appropriate for 64 Majorana-Weyl
fermions, the NAHE set takes the following form:
$$
\begin{tabular}{c|c|c}
 ~ & right-movers & left-movers \\
\hline
$\V_0$ & ${\tt  11111111111111111111 }$& ${\tt
11111111111111111111111111111111111111111111
     }$ \\
$\V_1$ & ${\tt  11100100100100100100 }$& ${\tt 00000000000000
      0 0 0 0 0 0 0 0 0 0 0 0 0 0 0 0 0 0 0 0 0 0 0 0 0 0 0 0 0 0
	 }$ \\
$\V_2$ & ${\tt   1 1 1 0 0 1 0 0 0 1 0 0 1 0 0 1 0 0 1 0 }$& ${\tt  0 0 0 0 1 0
1 0 1 0 1 0
    1 1 1 1 1 1 0 0 0 0 0 0 0 0 0 0 1 1 1 1 1 1 0 0 0 0 0 0 0 0 0 0
	 }$ \\
$\V_3$ & ${\tt   1 1 0 1 0 0 1 0 1 0 0 1 0 0 0 0 1 0 0 1 }$& ${\tt 1 0 1 0 0 0
0 0 0 1 0 1
    1 1 1 1 1 0 1 0 0 0 0 0 0 0 0 0 1 1 1 1 1 0 1 0 0 0 0 0 0 0 0 0
	}$ \\
$\V_4$ & ${\tt   1 1 0 0 1 0 0 1 0 0 1 0 0 1 1 0 0 1 0 0 }$& ${\tt 0 1 0 1 0 1
0 1 0 0 0 0
    1 1 1 1 1 0 0 1 0 0 0 0 0 0 0 0 1 1 1 1 1 0 0 1 0 0 0 0 0 0 0 0
      }$ \\
\end{tabular}
$$
where the entry `$1$' is shorthand for `$-1/2$'.
Here the $\V_0$ through $\V_4$ vectors correspond to the vectors
 $\lbrace \bone, S,  {\bf b}_1, {\bf b}_2, {\bf b}_3 \rbrace$ described
previously.

The individual models are then each derived from this underlying NAHE
structure through the addition of extra boundary-condition vectors $V_i$,
and the specific choice of GSO phases $k_{ij}$.
These are listed below.

\subsection{Flipped SU(5) model}

This is the model we considered in Sect.~4.1.
This model has the additional boundary-condition vectors
$$
\begin{tabular}{c|c|c}
 ~ & right-movers & left-movers \\
\hline
$\V_5  $& ${\tt    1 1 1 0 0 1 0 0 0 1 0 0 0 1 0 0 1 0 1 0 }$& ${\tt  0 0 0 0 1
0 0 1 0 1 1 0
    1 1 1 1 1 1 0 0 0 0 0 0 0 0 0 0 1 1 1 1 1 1 0 0 0 0 0 0 0 0 0 0
     }$\\
$\V_6  $& ${\tt  1 1 0 0 1 0 1 0 1 0 0 1 0 0 0 0 1 0 1 0 }$&$ {\tt  0 1 1 0 0 0
0 0 0 1 1 0
    1 1 1 1 1 0 1 0 0 0 0 0 0 0 0 0 1 1 1 1 1 0 1 0 0 0 0 0 0 0 0 0
     }$\\
$\V_7 $& ${\tt   0 0 0 0 0 0 0 0 0 0 0 0 1 1 0 0 0 0 1 1 }$&$ {\tt 0 0 0 1 0 1
1 1 0 0 1
    1}$\m\m\m\m\m\m\m\m\m\m\m\m${\tt 1 1 0 0}$\m\m\m\m\m\m\m\m\m\m\m\m${\tt 1 1
0 0 }$\\
\end{tabular}
$$
where `$-$' is shorthand for `$-1/4$'.
Furthermore, $k_{00}$ and the following phases $k_{ij}$ with $i>j$ are equal to
$1/2$:
$k_{2 0}$, $ k_{2 1}$, $ k_{3 0}$, $ k_{3 1}$, $ k_{3 2}$,
$ k_{4 0}$, $ k_{4 1}$, $ k_{4 2}$, $ k_{4 3}$, $ k_{5 1}$, $ k_{5 2}$,
$ k_{5 3}$, $ k_{5 4}$, $ k_{6 1}$, $ k_{6 2}$, $ k_{6 3}$, $ k_{6 4}$,
$ k_{6 5}$, $ k_{7 0}$, $ k_{7 6}$.
The remaining phases with $i>j$ are vanishing, and those with $i\leq j$ are
determined from those with $i>j$ using the constraint equations
given in Ref.~\cite{KLST}.

Note that starting from this model as it was originally published and
following the strict mapping given in (\ref{mappingKLTABK}), we would have
instead found that $k_{50}$ and $k_{60}$ should also be non-zero (as well as
corresponding changes in those $i\leq j$ phases
which are related to these through the constraint equations of
Ref.~\cite{KLST}).
However, as discussed above,
we must set $k_{50}$ and $k_{60}$ to zero in order to account for
the difference in the implicit conventions
for dealing with the Ramond zero-modes.
Only with such modifications is the correct model with the same
particle spectrum reproduced.

\subsection{First SU(3) $\times$ SU(2) $\times$ U(1) model}

This is the model we considered in Sect.~4.2.
This model has the additional boundary-condition vectors
$$
\begin{tabular}{c|c|c}
 ~ & right-movers & left-movers \\
\hline
$ \V_5   $&${\tt  0 0 0 0 0 0 0 0 0 1 1 0 0 0 0 0 0 0 1 1  }$&${\tt    1 0 0 1
0 1 0 1 0 1 0 1
	 1 1 1 0 0 0 0 0 1 1 1 1 0 0 0 0
	 1 1 1 0 0 0 0 0 1 1 1 1 0 0 0 0
	}$\\
$ \V_6   $&${\tt   0 0 0 1 1 0 0 0 0 0 0 0 0 0 0 1 1 0 0 0  }$&${\tt  0 1 0 1 1
0 0 1 1 0 1 0
	 1 1 1 0 0 0 0 0 1 1 1 1 0 0 0 0
	 1 1 1 0 0 0 0 0 1 1 1 1 0 0 0 0
	}$\\
$ \V_7  $&${\tt    0 0 0 0 0 0 1 1 0 0 0 0 1 1 0 0 0 0 0 0  }$&{\tt
     101010100110\m\m\m\m\m\m\m\m\m011\m\m\m0\m\m\m\m\m\m\m\m\m011\m\m\m0  }\\
\end{tabular}
$$
where again `$-$' is shorthand for `$-1/4$'.
In this model, $k_{00}$ and the following phases $k_{ij}$ with $i>j$ are equal
to $1/2$:
 $k_{2 0 }$, $k_{2 1 }$, $k_{3 0 }$, $k_{3 1 }$, $k_{3 2 }$, $k_{4 0 }$,
 $k_{4 1 }$, $k_{4 2 }$, $k_{4 3 }$, $k_{5 0 }$, $k_{5 2 }$, $k_{6 0 }$,
 $k_{6 3 }$,  $k_{6 5 }$, $k_{7 4 }$, $k_{7 5 }$.
All remaining $i>j$ phases vanish, and
those with $i\leq j$ may be determined from those with $i>j$ using the
constraint equations
given in Ref.~\cite{KLST}.

In this case, no phase adjustments are necessary
in order to account for the difference in Ramond zero-mode conventions.

\subsection{Non-supersymmetric version of first SU(3) $\times$ SU(2) $\times$
U(1) model}

This is the model we considered in Sect.~4.5.
This model has the same additional boundary-condition vectors and phases
as the model above, except that $k_{51}$, $k_{61}$, and $k_{71}$
are changed from $0$ to $1/2$.
(Of course, given the constraint equations of Ref.~\cite{KLST},
this induces corresponding changes in the
coefficients $k_{ij}$ with $i\leq j$.)
As required, these changes have the effect of
breaking spacetime supersymmetry {\it without}\/ introducing physical
tachyons into the spectrum.

\subsection{Second SU(3) $\times$ SU(2) $\times$ U(1) model}

This is the model we considered in Sect.~4.3.
This model has the additional boundary-condition vectors
$$
\begin{tabular}{c|c|c}
    ~ & right-movers & left-movers \\
\hline
$\V_5 $ & ${\tt   0 0 0 1 1 0 1 1 0 1 1 0 1 1 0 1 1 0 1 1    }$ & ${\tt    0  1
1 0 0 1 1 0 1 0 0 1
   1 1 1 0 0 0 0 0 1 1 1 1 0 0 0 0 1 1 1 0 0 0 0 0 1 1 1 1 0 0 0 0
      }$\\
$\V_6 $ & ${\tt    0 0 0 0 0 0 1 1 0 0 0 0 1 1 0 0 0 0 0 0    }$ & ${\tt    1 0
1 0 1 0 1 0 0 1 1 0
   1 1 1 0 0 0 0 0 1 1 1 1 0 0 0 0 1 1 1 0 0 0 0 0 1 1 1 1 0 0 0 0
      }$\\
$\V_7 $ & ${\tt    0 0 0 1 1 0 0 0 0 0 0 0 0 0 0 1 1 0 0 0   }$ &
      {\tt 010110011010\m\m\m\m\m\m\m\m\m011\m\m\m0\m\m\m\m\m\m\m\m\m011\m\m\m0
 }\\
\end{tabular}
$$
In this model, $k_{00}$ and the following phases $k_{ij}$ with $i>j$ are equal
to $1/2$:
 $k_{2 0 }$, $k_{2 1 }$, $k_{3 0 }$,  $k_{3 1 }$, $k_{3 2 }$, $k_{4 0 }$,
 $k_{4 1 }$, $k_{4 2 }$, $k_{4 3 }$, $k_{5 0 }$, $k_{5 2 }$,  $k_{5 3 }$,
 $k_{5 4 }$, $k_{6 0 }$, $k_{6 5 }$, $k_{7 0 }$, $k_{7 3 }$.
All remaining $i>j$ phases vanish, and
those with $i\leq j$ may be determined from those with $i>j$ using the
constraint equations
given in Ref.~\cite{KLST}.

For this model, it is necessary to set
$k_{64}$ to zero in order be consistent with our
Ramond zero-mode conventions.

\subsection{SO(6) $\times$ SO(4) model}

This is the model we considered in Sect.~4.4.
This model has the additional boundary-condition vectors
$$
\begin{tabular}{c|c|c}
    ~ & right-movers & left-movers \\
\hline
$\V_5 $ & ${\tt   1 1 1 0 0 1 0 0 0 1 0 0 0 1 0 0 1 0 1 0    }$&${\tt      0 0
0 0 1 0 0 1 0 1 1 0
   1 1 1 1 1 1 0 0 0 0 0 0 0 0 0 0 1 1 1 1 1 1 0 0 0 0 0 0 0 0 0 0
     }$\\
$\V_6 $ & ${\tt   1 1 0 0 1 0 1 0 1 0 0 1 0 0 0 0 1 0 1 0   }$&${\tt   0 1 1 0
0 0 0 0 0 1 1 0
   1 1 1 1 1 0 1 0 0 0 0 0 0 0 0 0 1 1 1 1 1 0 1 0 0 0 0 0 0 0 0 0
     }$\\
$\V_7 $ & ${\tt   0 0 0 0 0 0 0 0 0 0 0 0 0 0 0 0 0 0 1 1   }$&${\tt   0 0 0 0
0 0 0 0 0 0 1 1
   1 1 1 1 1 1 1 1 1 1 1 1 0 0 0 0 1 1 1 1 1 1 1 1 1 1 1 1 0 0 0 0
     }$\\
$\V_8 $ & ${\tt   0 0 0 0 0 0 0 0 0 0 0 0 1 1 0 0 0 0 1 1   }$&${\tt    0 0 0 1
0 1 1 1 0 0 1 1
   1 1 1 0 0 1 1 0 0 0 0 1 1 0 0 0 1 1 1 0 0 1 1 0 0 0 0 1 1 0 0 0
     }$\\
\end{tabular}
$$
In this model, $k_{00}$ and the following phases $k_{ij}$ with $i>j$ are equal
to $1/2$:
 $k_{2 0 }$, $k_{2 1 }$, $k_{3 0 }$, $k_{3 1 }$, $k_{3 2 }$, $k_{4 0 }$,
 $k_{4 1 }$, $k_{4 2 }$, $k_{4 3 }$, $k_{5 0 }$, $k_{5 1 }$, $k_{5 2 }$,
 $k_{5 3}$,
 $k_{6 0 }$, $k_{6 1 }$,  $k_{6 5 }$,  $k_{7 0 }$,  $k_{7 2 }$,
 $k_{7 3 }$,  $k_{7 4 }$,  $k_{8 0 }$,  $k_{8 4 }$, $k_{8 6 }$.
All remaining $i>j$ phases vanish, and
those with $i\leq j$ may be determined from those with $i>j$ using the
constraint equations
given in Ref.~\cite{KLST}.

For this model,
in order be consistent with our Ramond zero-mode conventions,
it is necessary to set $k_{63}$ and $k_{83}$ to zero,
and $k_{5 3}$ and $k_{8 6}$ to $1/2$.


\vfill\eject

\bigskip
\medskip

\bibliographystyle{unsrt}

\vfill\eject

\end{document}